\documentclass[aps,pre,twocolumn,showpacs,superscriptaddress]{revtex4-1}
\usepackage{graphicx,epsfig}
\usepackage{times}
\usepackage{graphics,dcolumn,bm,fleqn,epic,eepic}
\usepackage{amssymb,amsmath,multirow,rotate,color}
\usepackage[caption=false]{subfig}
\usepackage{natbib}
\usepackage{color}
\usepackage{soul}
\usepackage{tikz}
\usetikzlibrary{positioning}

\begin{document}

\title{Hierarchical benchmark graphs for testing community detection algorithms}

\author{Zhao Yang}
\email{zhao.yang@business.uzh.ch}
\affiliation{URPP Social Networks, University of Z\"urich, Andreasstrasse 15, CH-8050 Z\"urich, Switzerland}

\author{Juan I.~Perotti}
\email{juanpool@gmail.com}
\affiliation{IMT School for Advanced Studies Lucca, Piazza San Francesco 19, I-55100 Lucca, Italy}

\author{Claudio J.~Tessone}
\email{claudio.tessone@business.uzh.ch}
\affiliation{URPP Social Networks, University of Z\"urich, Andreasstrasse 15, CH-8050 Z\"urich, Switzerland}
\affiliation{IMT School for Advanced Studies Lucca, Piazza San Francesco 19, I-55100 Lucca, Italy}

\date{\today}

\begin{abstract}
Hierarchical organization is an important, prevalent characteristic of complex systems; in order to understand their organization, the study of the underlying (generally complex) networks that describe the interactions between their constituents plays a central role. Numerous previous works have shown that many real-world networks in social, biologic and technical systems present hierarchical organization, often in the form of a hierarchy of community structures.
Many artificial benchmark graphs have been proposed in order to test different community detection methods, but no benchmark has been developed to throughly test the detection of hierarchical community structures. 
In this study, we fill this vacancy by extending the Lancichinetti-Fortunato-Radicchi (LFR) ensemble of benchmark graphs, adopting the rule of constructing hierarchical networks proposed by Ravasz and Barab\'asi. We employ this benchmark to test three of the most popular community detection algorithms, and quantify their accuracy using the traditional Mutual Information and the recently introduced Hierarchical Mutual Information.
The results indicate that the Ravasz-Barab\'asi-Lancichinetti-Fortunato-Radicchi (RB-LFR) benchmark generates a complex hierarchical structure constituting a challenging benchmark for the considered community detection methods.
\end{abstract}

\maketitle

\section{Introduction}

Hierarchical organization \cite{hayek1964theory, pattee1973hierarchy, simon1996sciences} is a typical trait of complex systems, appearing in many biological, social (corporations, education systems, governments, and organized religions) or technological (internet and other infrastructure) arrangements whose different scales are apparent. The interactions between the constituents of those systems are correctly described as networks of interconnected modules nested hierarchically  \cite{corominas2013origins, gross2005graph}. Typical hierarchical networks include food webs, protein interaction networks, metabolic networks, gene regulatory networks, social networks, etc. \cite{clauset2008hierarchical}. While interactions ultimately occur between the basic or {\em microscopic} constituents of the systems, effective coarse-grained elements and interactions between them emerge at the different levels of organization which should be characterized and understood at their own scale. Because of this, finding the appropriate hierarchical and modular structure of complex networks is of great interest for the understanding of complex systems \cite{clauset2008hierarchical, newman2012communities}.

Community detection helps to unveil the non-trivial organization of complex systems at the mesoscopic scale~\cite{lancichinetti2008benchmark, fortunato2010community, fortunato2016community}. 
Many algorithms have been developed to identify the community structure in networks \cite{girvan2002community, clauset2004finding, newman2006finding, raghavan2007near, rosvall2008maps, blondel2008fast, reichardt2006statistical, lancichinetti2011finding}. 
Some of them are also able to reveal the hierarchical community structure within.
Without the intention of being exhaustive, the most widely used are: \textit{Infomap} \cite{rosvall2008maps}, which uses the probability flow of random walks on the network under consideration as a proxy for information diffusion in the real system; it then proceeds by decomposing the network into modules by compressing a specific description of probability flow.
\textit{Louvain}~\cite{blondel2008fast}, which employs a computationally efficient greedy-algorithm for the optimization of Newman's modularity~\cite{newman2006modularity}. \textit{Spinglass}~\cite{reichardt2006statistical}, which uncovers the community structure of networks by minimizing the energy of a Hamiltonian whose spin-states represent the community indices. \textit{OSLOM}~\cite{lancichinetti2011finding}, which detects clusters by using the local optimization of a fitness function expressing the statistical significance of a community  with respect to random fluctuations. And \textit{hierarchical stochastic block model}~\cite{peixoto2014hierarchical}, which seeks to fit a hierarchy of stochastic block models to the different levels of organization of networks. 

Comparing the accuracy of different community detection algorithms is a non-trivial problem.
Commonly, two separately, intricate tools are required for the task~\cite{fortunato2010community}.
The first one are~\textit{benchmark graphs}. These can be either real networks with known community structure (i.e.~ground truth) or ensembles of artificial graphs with built-in community structure \cite{zachary1977information,girvan2002community,danon2006effect,lancichinetti2008benchmark,bagrow2008evaluating,lancichinetti2009benchmarks, orman2010effect,yang2016comparative}. The second tool required is a measure quantifying the similarity between different allocations of nodes into communities for the same network. This enables the comparison between the known community structure and the identified by the algorithms under study.
Recently, to cover the need of the second requirement, a similarity measure for the comparison of hierarchical community structures has been introduced -- the so-called \textit{Hierarchical Mutual Information (HMI)}~\cite{perotti2015hierarchical} -- which is a generalization of the \textit{Mutual Information (MI)}, a standard measure for the comparison of non-hierarchical community structures~\cite{danon2005comparing}. 
As we show in this paper the HMI can be further combined with the more traditional approach, where a level-by-level comparison of the hierarchies is performed with the standard MI~\cite{lancichinetti2009detecting,lancichinetti2011finding}.

The development of a benchmark graph model mimicking the hierarchical community structure of real  complex networks -- i.e. to cover the need of the first tools previously mentioned -- is the central topic of the present paper.
Namely, in this work, we introduce the \textit{Ravasz-Barab\'asi LFR benchmark (RB-LFR)}.
Broadly speaking, in its simplest incarnation, the RB-LFR is obtained combining the complex community structure of the standard LFR benchmark \cite{lancichinetti2008benchmark} with the celebrated Rabasz-Barab\'asi mechanism of constructing hierarchies \cite{ravasz2003hierarchical}. 
While we develop the  benchmark as a stylized representation of real-world networks, given that data about properties of the  hierarchical organization with multiple levels is scarce, we reproduce properties of well-established artificial models that are also inspired in real data.
We argue that only after solid hierarchical community detection methods have been developed and pass tests posed by artificial benchmarks, a proper understanding of  hierarchical organizations in real world will be possible.
As we show in this paper, the RB-LFR benchmark poses challenging detection problems for the most popular hierarchical community detection methods and it allows us to show that the HMI is a superior tool for the comparison of hierarchical community structures as compared to the traditional MI. 

The outline of the paper is the following. In section II, the construction of the benchmark is presented. In section III, three community detection algorithms have been tested on the RB-LFR benchmark graphs with different setups: in subsection III A, the benchmark graphs have two levels, while in subsection III B, the benchmarks have three levels. Finally, the discussions and conclusions are summarized in section IV.

\section{The RB-LFR hierarchical benchmark}

In this section, we provide a detailed description for the construction of the networks in the ensemble defined by the RB-LFR benchmark.
By performing a topological analysis, we also show that the resulting networks exhibit both: power-law degree and community size distributions.

\begin{figure}[h]
\captionsetup[subfloat]{farskip=-2pt,captionskip=0pt}
\begin{center}
\subfloat[]{\includegraphics[width=.22\textwidth]{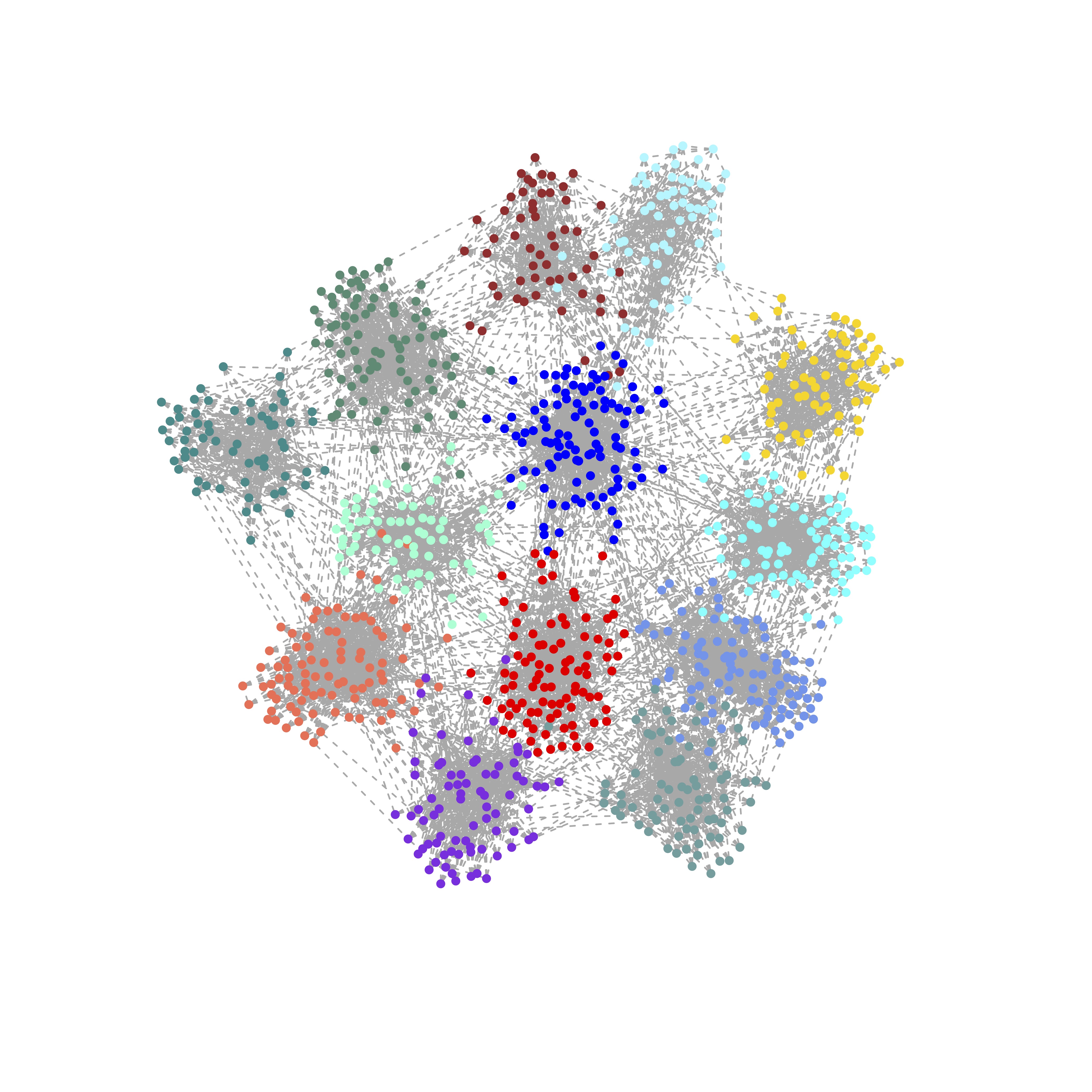}} 
\subfloat[]{\includegraphics[width=.22\textwidth]{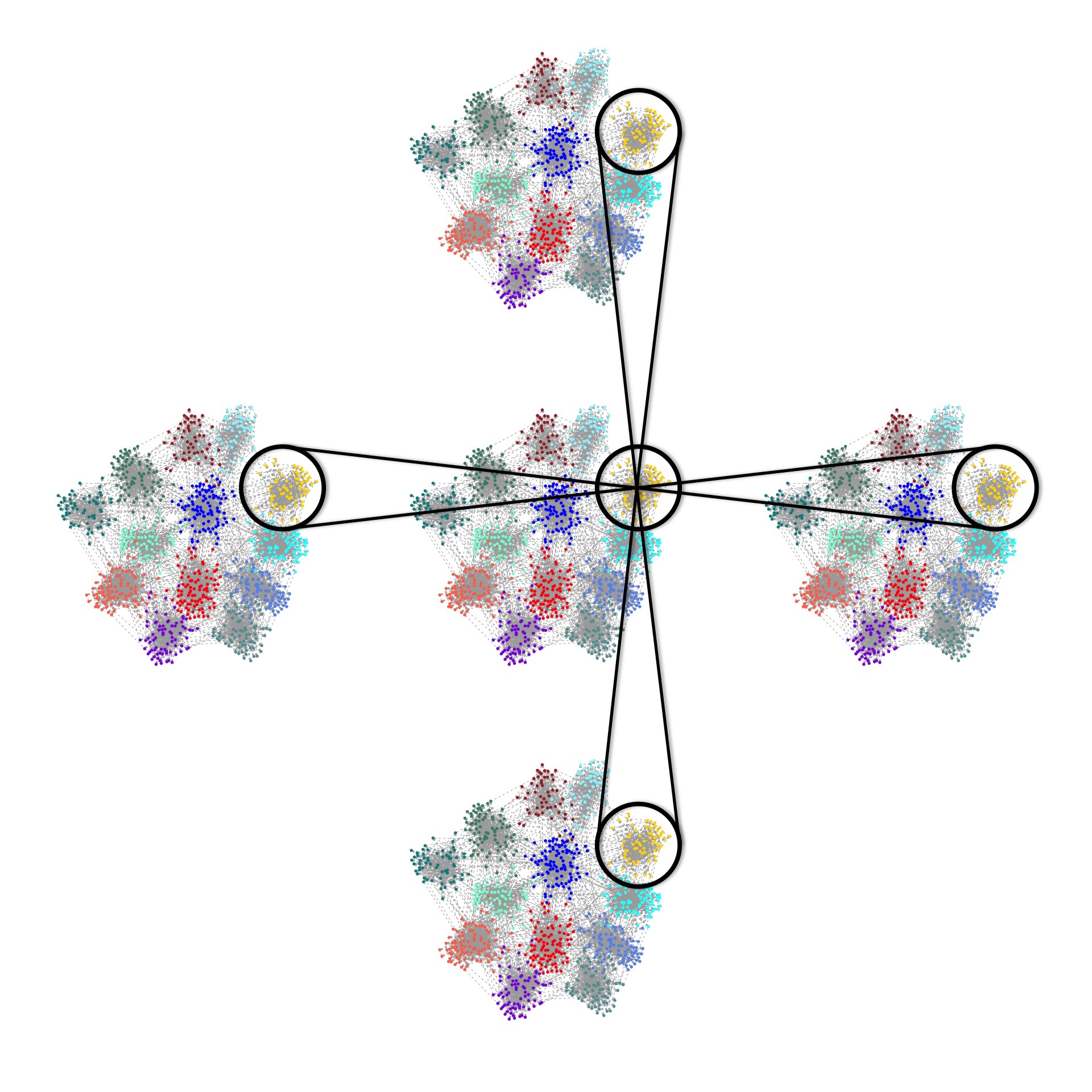}} \\
\subfloat[]{\includegraphics[width=.22\textwidth]{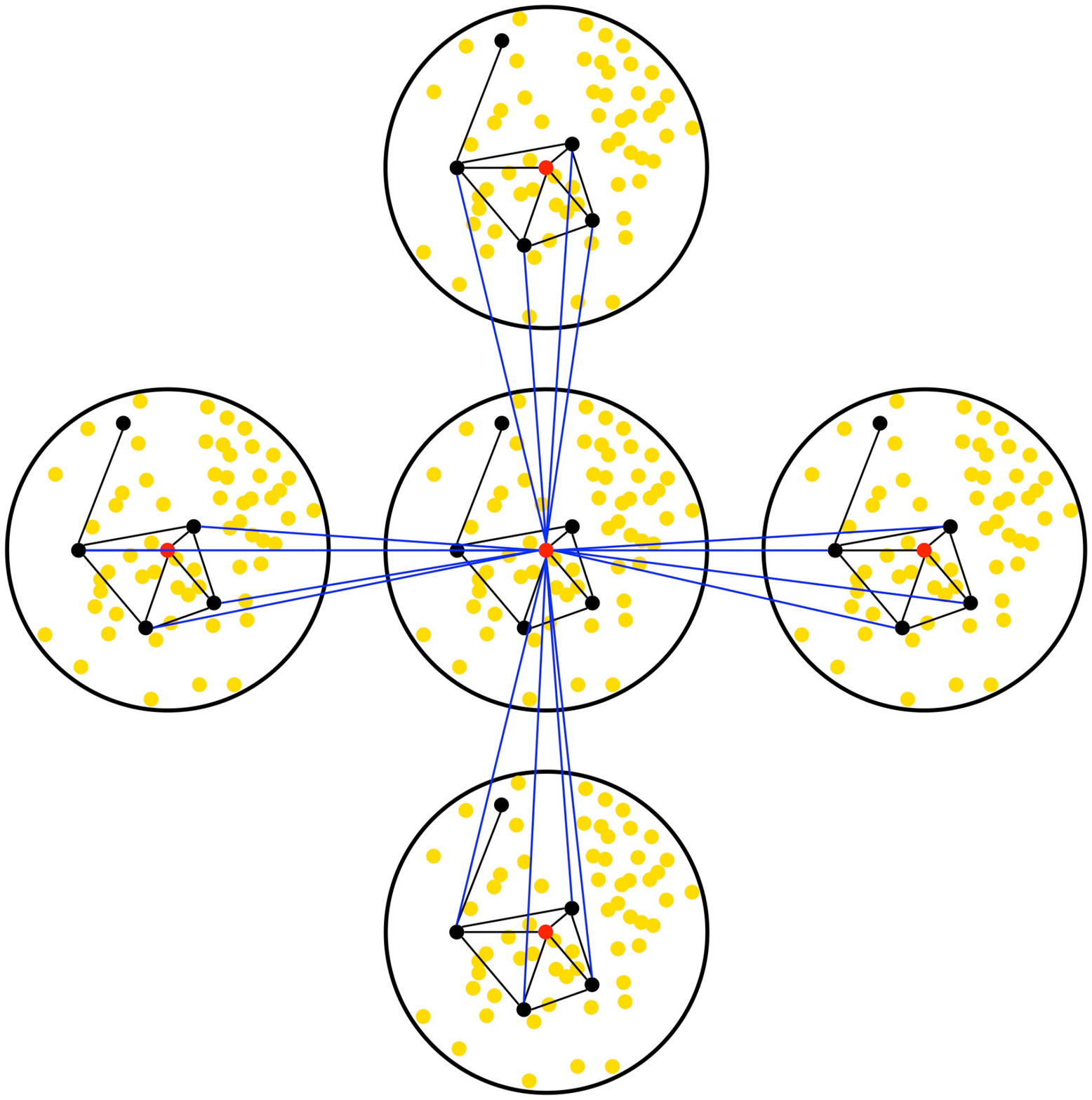}} 
\subfloat[]{\includegraphics[width=.22\textwidth]{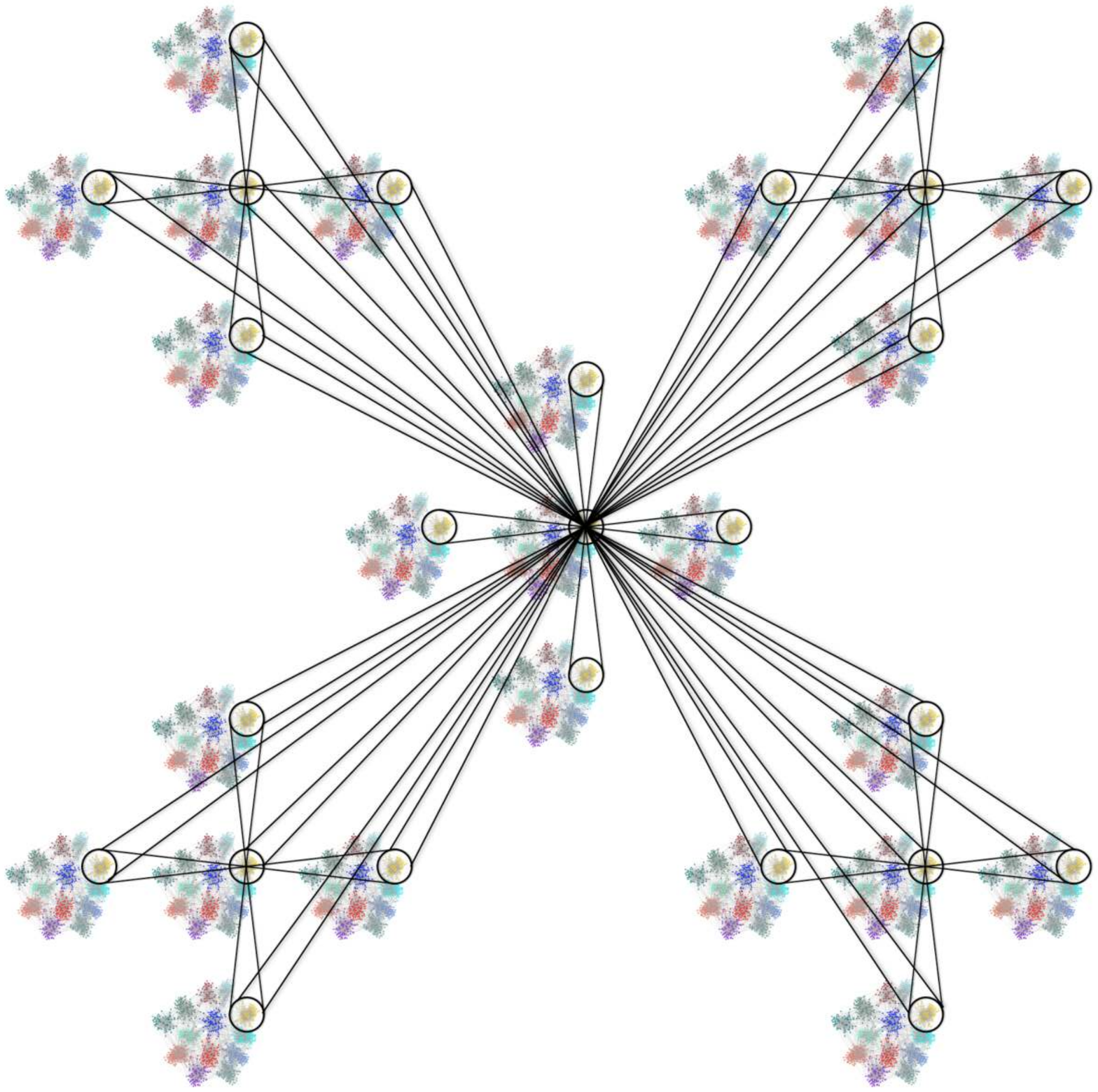}} \\
\end{center}
\caption{(a) An example of the LFR benchmark taken as original building block of the benchmark. (b) Four replicated LFR benchmarks are generated and connected to the original or seed LFR benchmark, community by community. (c) An schematic diagram of the connections between the seed community and the replicated ones; the red node is the hub, i.e.~the node with the largest degree in the community. We have only shown the links between the black nodes and the hub. The other links are not visible. (d) A realization of a three-level RB-LFR benchmark. Links of the other communities are not visible.} 
\label{figure1}
\end{figure}

Before we go into the construction details, let us first motivate the convenience of the RB-LFR benchmark as compared to other existing alternatives.
Different hierarchical network structures have been already proposed in the literature.
For instance, the Sierpinski gasket \cite{sierpinski1915courbe}, the hierarchical planted partition model \cite{lancichinetti2009detecting},
the hierarchical stochastic block model and its variants~\cite{peixoto2014hierarchical,
clauset2008hierarchical,
herlau2012detecting}, the Ravasz-Barab\'asi model \cite{ravasz2003hierarchical} and a hierarchically nested version of the LFR benchmark \cite{lancichinetti2011finding}. 
While some of these network structures have been already employed in the problem of community detection, they display certain limitations when considered as benchmark graphs.
For example, the Sierpinski gasket and the standard Ravasz-Barab\'asi models have an excessively regular structure, while real networks have more complex hierarchical community structures (see for example, the political blog network of Adamic and Glance and the IMDB film-actor network~\cite{adamic2005political, peixoto2014hierarchical}).
The hierarchical planted partition model contains disorder, but it has exceedingly simple communities and connection structures which fail to reflect the properties of the communities found in real networks where largely varying community sizes and node degrees are found.
The hierarchical stochastic block model admits generalized communities of different sizes and approximately arbitrary degree distributions, improving over the hierarchical partition model.
In practice however, at least to the extent of our knowledge, it has never been used to construct benchmark graphs with hierarchical structures and power-law distribution of community sizes. The most promising alternative is the hierarchical version of the LFR benchmark, since it presents complex and realistic degree and community structures like the standard LFR does, and a hierarchical community structure. 
However, although the general idea is given, a precise definition of the hierarchical LFR is still missing, nor its properties have been systematically tested.
Only realizations with two levels have been considered -- so-called  \textit{fine} or micro-community level and \textit{coarse} or macro-community level -- and, according to the given specifications, it is not clear how the macro-communities should be obtained by merging micro-communities, something that is required to generate networks with more than two levels. In other words, a guiding principle or mechanism is required to combine LFR networks into hierarchies with an arbitrary number of levels. 
The straightforward way is to appropriately extend the definition of the hierarchical LFR, recursively building LFR networks within the modules of other LFR networks.
However, this approach presents two important disadvantages, affecting the computational cost required to analyze and generate the networks.
Firstly, the number of nodes in the network quickly grows with the number $L$ of levels as $N_{L}\sim C^L N_0$ where $N_0$ is the number of nodes and $C$ the number of communities in a non-hierarchical LFR playing the role of a seed-network. 
Secondly, in order to be conceptually consistent, the algorithm devised to generate the non-hierarchical LFR networks should be appropriately modified in order to preserve the power-law community-size and degree distributions across every level of the resulting hierarchy.

Since we want to develop a computationally accessible benchmark combining well studied ideas, we propose a different approach. Namely, we introduce the \textit{Ravasz-Barab\'asi LFR benchmark (RB-LFR)}, an extension of the LFR benchmark \cite{lancichinetti2008benchmark} obtained by combining it with a construction procedure inspired in the work by Ravasz and Barab\'asi \cite{ravasz2003hierarchical}.
Compared to  previous alternatives, the RB-LFR benchmark has a complex and realistic network degree and community-size distributions -- like the LFR benchmark does --: its hierarchy can  have an arbitrary number of levels and the RB procedure can be generalized in a straightforward manner even further.
In the standard RB method, the hubs of different network motifs are connected to  the nodes of corresponding replicas~\cite{ravasz2003hierarchical} but, in a more general setup, these restrictions can be relaxed by allowing alternative inter-replica connections by combining different ways or modes of doing so~\cite{song2006origins}.
In the present work, in order to simplify the analysis, we restrict ourselves to study the case of the original RB procedure, leaving for future work the study of the alternative generalizations of the RB-LFR benchmark.

Our starting point is a standard non-hierarchical LFR benchmark network (Fig.~\ref{figure1}a), which we consider as the seed network motif for an adapted Rabasz-Barab\'asi procedure for constructing hierarchical networks. The parameters used to generate this LFR benchmark network are indicated in Table \ref{table1}. The number of nodes in the seed network is $N_0 = 1000$. Each node is given a degree taken from a power-law distribution with exponent $\gamma = -2$. We have fixed the average degree $\langle k \rangle = 20$,  and the maximum degree to $k_{max} = 0.1 N_0$. Community size is taken from a power-law distribution with exponent -1 and the upper bound and lower bound of community size are $0.1N_0$ and $\langle k \rangle$, respectively. The mixing parameter, $\mu$, which represents the fraction of links with the other nodes outside of its community, is defined as
$$
\mu = \frac{\sum_i k^{\mathrm{ext}}_i}{\sum_i k_i^{\mathrm{tot}}}, 
$$ 
where $k^{\mathrm{ext}}_i$ stands for the external degree of node $i$ and $k^{\mathrm{tot}}_i$ is the total degree of $i$. 
In this study, the values of $\mu$ are taken from an arithmetic sequence from 0.01 to 0.89 with step 0.04.

\begin{table}[h]
\begin{center}
\begin{tabular}{ l | l }
\hline
Parameter & Value \\ 
\hline
Number of nodes, $N_0$ & 1000 \\  
\hline
Average degree, $\langle k \rangle$ & 20\\
\hline
Maximum degree & $0.1 N_0$\\
\hline
Maximum community size & $0.1 N_0$\\
\hline
Minimum community size & $\langle k \rangle$\\
\hline
Degree distribution exponent, $\gamma$ & -2\\
\hline
Community size distribution exponent, $\beta$ & -1\\
\hline
Mixing parameter, $\mu$ & [0.01, 0.05, ..., 0.89]\\
\hline
\end {tabular}
\caption{Parameters defining the ensemble of seed LFR benchmark graphs. To deal with possible discrepancies in the network properties, we have  generated 10 independent networks for every set of parameters.}
\label{table1}
\end{center}
\end{table}

Next, following the constructing RB procedure, we generate $R$  replicas of the seed LFR network in this context, it means that we generate $R$ replicas of each seed community and connect each seed community to their corresponding replica communities (Fig.~\ref{figure1}b) \cite{lancichinetti2008benchmark, ravasz2003hierarchical}. 
We denote \textit{community hubs}, the node with the largest degree in that community.
Then, the connections between the seed and the replica communities are always between the hub of the seed community and nearest neighbors of the replicated hub (Fig.~\ref{figure1}c). 
This replication and connection procedure can be repeated up to the desired number of levels. 
Each replication increases the number of nodes of the benchmark graph by a factor $R+1$, so the number of nodes of a RB-LFR network with $L$ levels scales as $N_L \sim (R+1)^L N_0$, a number that can be considerably smaller than the analogous for the hierarchical LFR since, in practice, $R+1$ can be chosen to be significantly smaller than $C$.
In Figure~\ref{figure1}d we show a three-level RB-LFR benchmark graph. 
Importantly, by assuming that each node in the network chooses to join the community to which the maximum number of its neighbors belong to \cite{raghavan2007near}, introducing the inter-community connections does not cause vanishing, merging, or generation of communities. For instance, in the most stylized case, the hub node has the same amount of links to the seed community and to the replica communities. As we will show later by introducing a non-zero probability of removing connections between the seed communities and the replicas, we can guarantee that the hubs will always belong to the seed communities.   
Hence, a power-law community structure is preserved at the bottom level (or top level, depending of the benchmark parameters) of the hierarchy, while a uniform community-structure is generated at the other levels.

\begin{figure}[h]
\begin{minipage}{0.48\textwidth}
\begin{tikzpicture}
  \node (img)  {\includegraphics[scale=0.48]{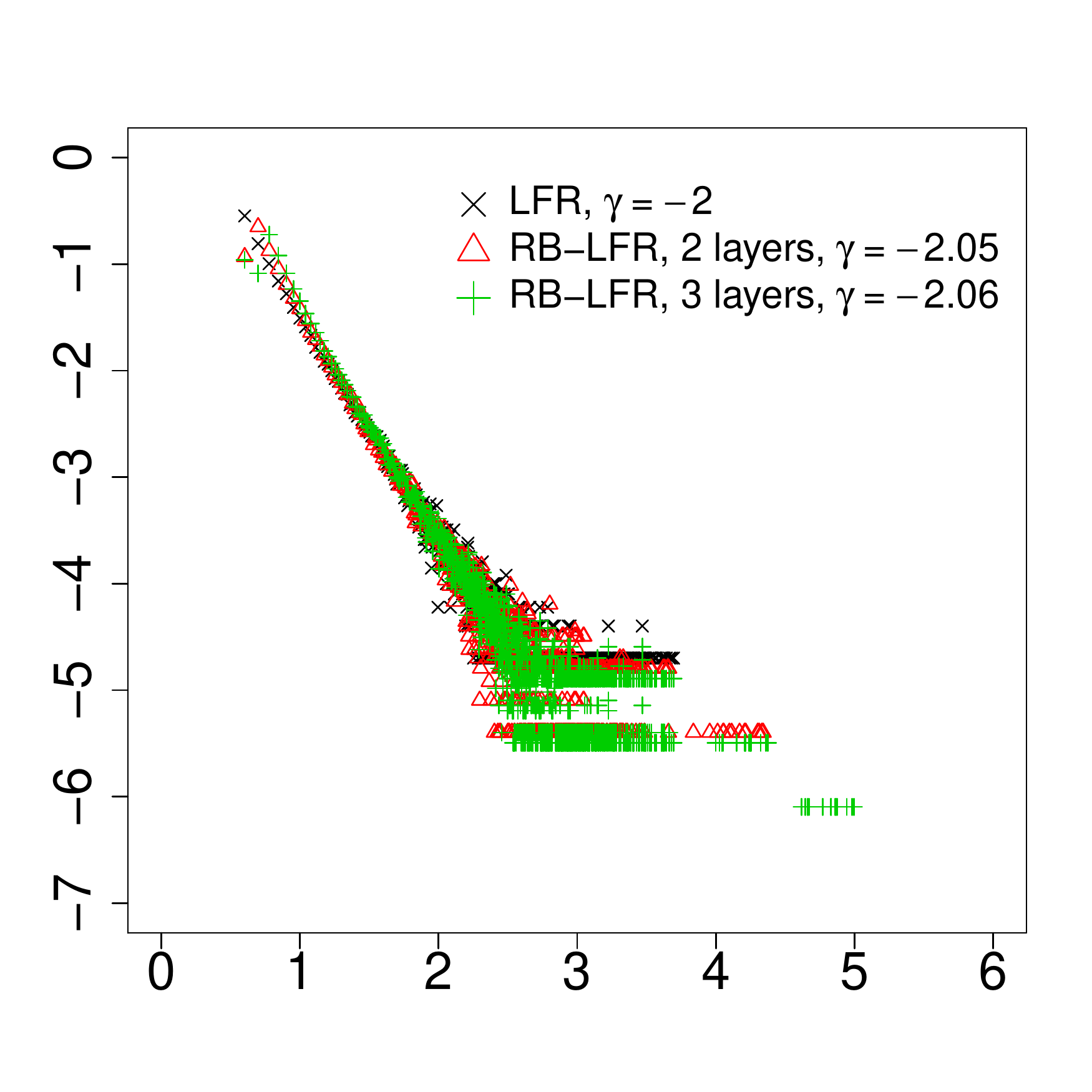}};
  \node[below = of img, node distance=0cm, yshift=1.8cm,font=\large, align = center] {degree, $ \log_{10} k$};
 \node[left = of img, node distance=0cm, rotate=90, anchor=center,yshift=-1.2cm,font=\large , align = center] {degree distribution $\log_{10} P(k)$};
 \end{tikzpicture}
\end{minipage} 
\caption{\textit{log-log} plot of the degree distribution for different RB-LFR benchmark graph samples with two levels (red triangles) and three levels (green pluses), constructed from a seed LFR with $N_0=50000$ nodes and mixing parameter $\mu = 0.05$ (black crosses).} 
\label{figure5}
\end{figure}

In Fig.~\ref{figure5}, the degree distributions of 2 and 3 layers networks generated by the RB-LFR benchmark are plotted, always starting with a seed LFR graph with the same set of parameters. We have fitted the degree distributions and reported the exponent of the fitted power-law distribution. As it can be seen, the added inter-community connections produce minor changes to the exponent of the degree distribution. In other words, an RB-LFR benchmark network approximately preserves the degree distribution of the seed LFR.

\begin{figure}[h]
\captionsetup[subfloat]{farskip=-2pt,captionskip=0pt}
\begin{center}
\subfloat[]{\includegraphics[width=.2\textwidth]{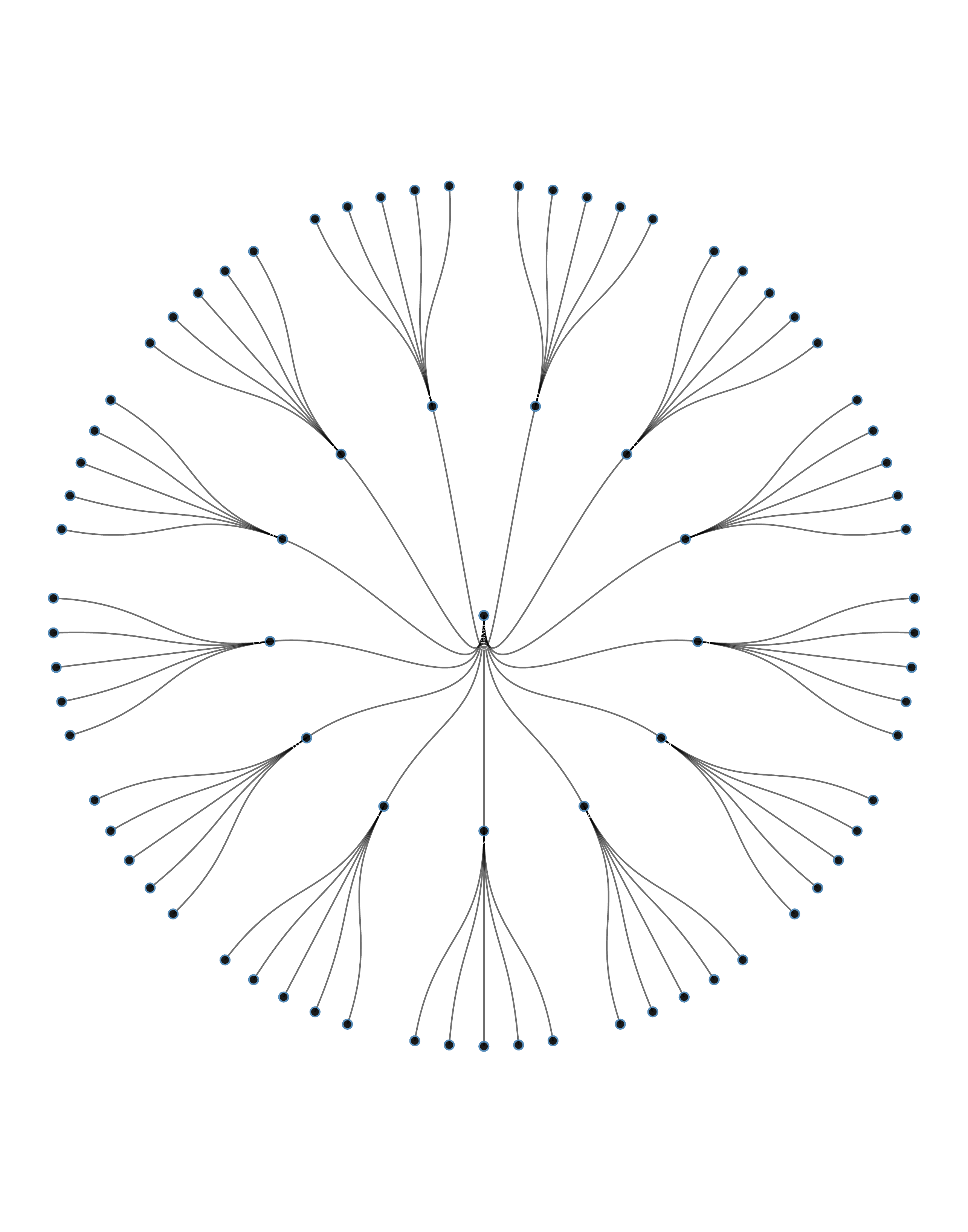}} 
\subfloat[]{\includegraphics[width=.2\textwidth]{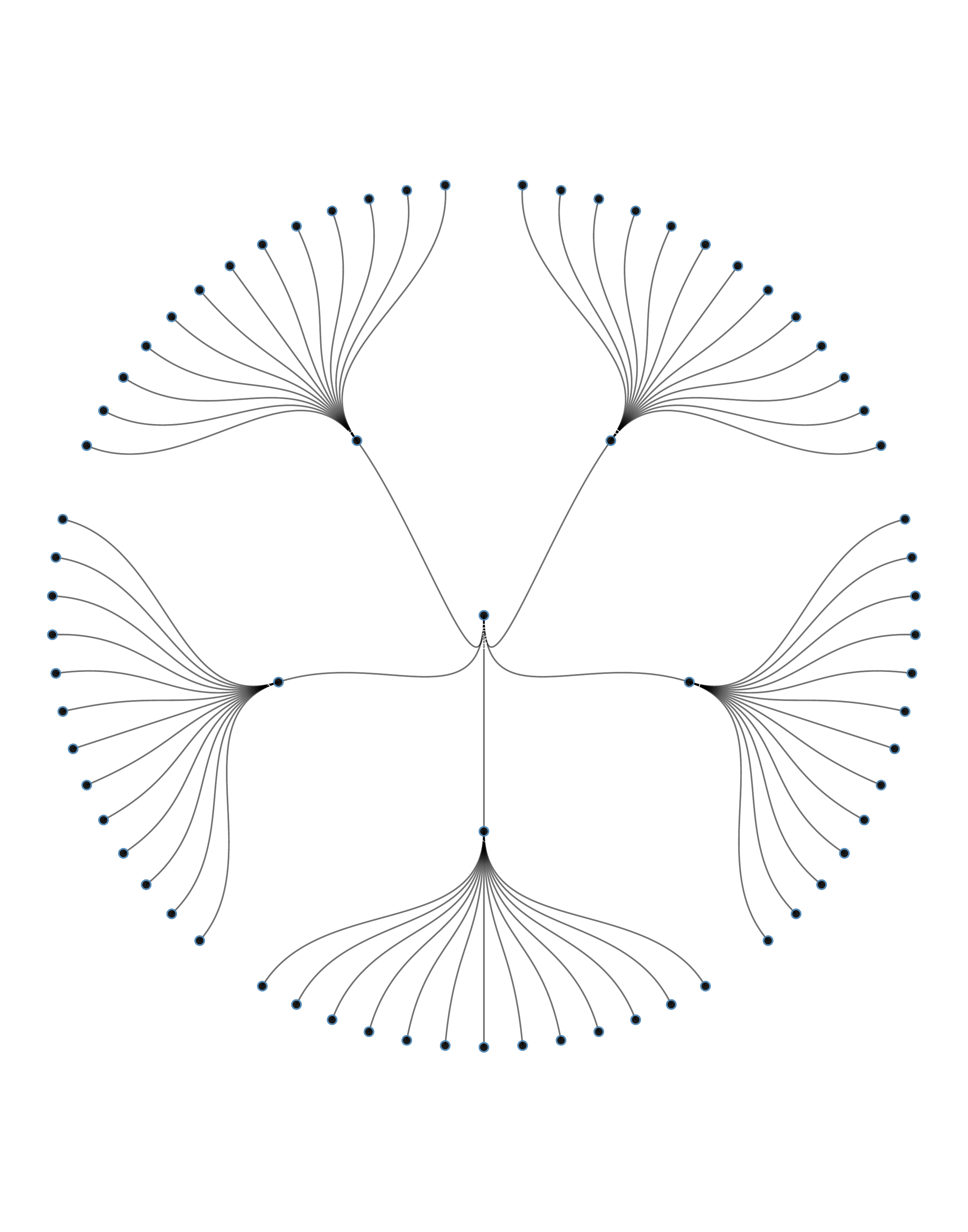}} \\
\subfloat[]{\includegraphics[width=.2\textwidth]{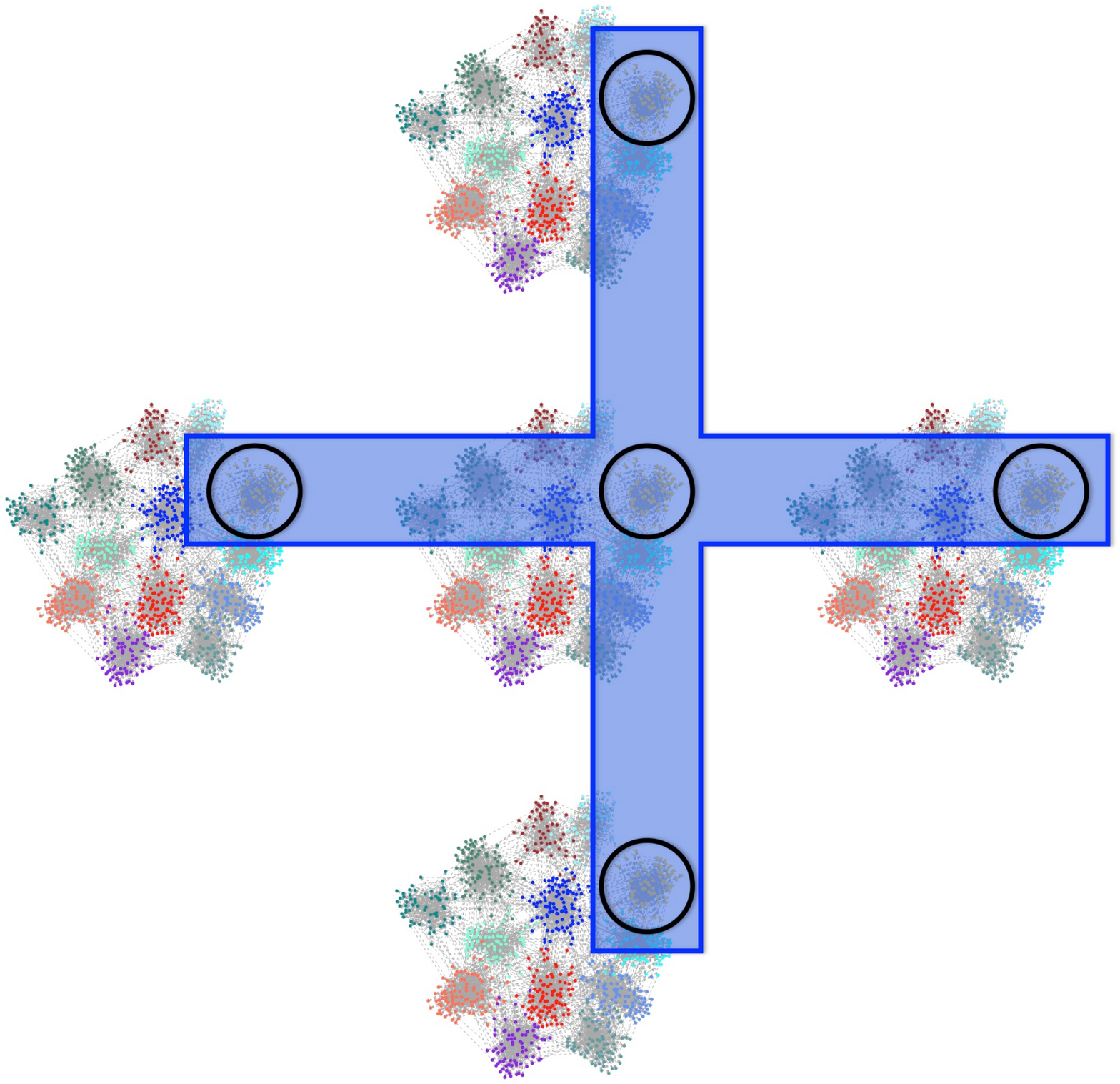}} 
\subfloat[]{\includegraphics[width=.2\textwidth]{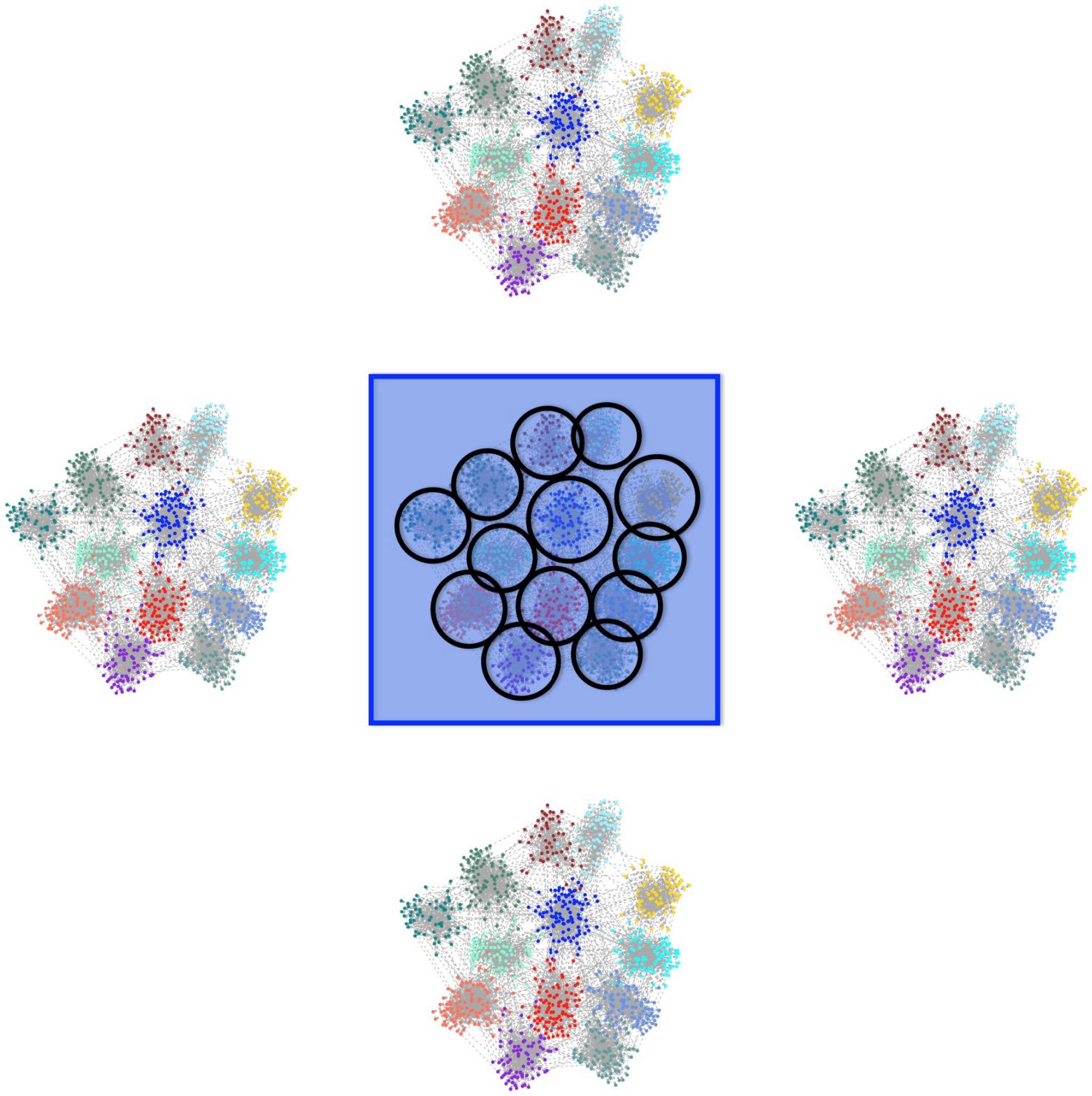}} 
\end{center}
\caption{(a) and (b) are the circular representations of the hierarchical structure of an RB-LFR benchmark with $R=4$ replicas. The center represents the whole network at level 0. 
In the example, LFR seeds with $N_0=1000$ nodes, $C=13$ communities and varying mixing parameter$\mu$ are used. 
In (a), the mixing parameter of the seed LFR benchmark is small, and the RB-LFR has $C$ communities on the first level and each of them has $R+1$ sub-communities on the second level. A larger mixing parameter for the seed LFR is used in (b), where the RB-LFR benchmark has $R+1$ communities on the first level, each having $C$ sub-communities on the second level. In panels (c) and (d) schematic network representations corresponding to the hierarchies in (a) and (b) are shown, respectively. The shaded (blue) areas represents a community on the first level, and the black circles represent sub-communities on the second level. Communities might have different sizes. For clarity reasons, links between the seed LFR and the replicas are not shown.} 
\label{figure4}
\end{figure}

Depending on the value of the mixing parameter $\mu$ for the seed LFR benchmark, the process described above can generate hierarchical graphs with two different well-defined ground truths. 
Taking the two-level RB-LFR benchmark graphs as an example, when the mixing parameter of the seed LFR benchmark is small, its community structure and that of its replicas are well-defined. 
First, on the first level, the RB-LFR benchmark displays as many communities as the seed LFR has, i.e.~$C$ communities. 
Each community in this first layer contains one community of the seed LFR together with all its replicas.
At the second level, each community of the first one contains $R+1$ sub-communities  (Fig.~\ref{figure4}a \& c) -- one for each replica plus the seed one -- summing a total of $C\times(R+1)$ sub-communities in the complete network\.
Notice, this occurs because there are no connections between each of the seed communities and the replicas of other seed communities.
This sort of inter-replica connections could be added and studied in future works, an interesting aspect showing how much richer in possible variations is the hierarchical case as compared to the non-hierarchical one.

When the mixing parameter $\mu$ is increased, the community structure of the seed LFR becomes more fuzzy and harder to detect.
Therefore, the seed and the replica communities within the RB-LFR benchmark become harder to detect, too. 
This obviously occurs to all replicas, while the number of inter-layer links remain the same regardless of $\mu$. 
Therefore, the seed LFR and the replicas may be interpreted as $R+1$ communities at the first layer.
Each of them has as many sub-communities at the second level as the seed LFR had, i.e.~$C$ (see Figs.~\ref{figure4}b \& d).
Again, the total number of sub-communities at the second level is $(R+1)\times C$ but, this time, such number is reached through different means, as you can see by comparing Figs.~\ref{figure4}a \& b.

If the mixing parameter of the seed LFR becomes too large, then the communities become impossible to detect and the community structure of the RB-LFR benchmark network becomes mono-level; i.e.~no second level arises and only $R+1$ communities exist at the first level, one for the seed LFR and others for the replicas.

\section{Test}
In the previous section, we have given the intuition that the RB-FLR benchmark is compatible with different ground truths for the hierarchical community structure. 
In this section, we verify that this topological transition occurs. But the main result of this section is the use the RB-LFR benchmark to test the performance of three hierarchical community detection algorithms: \textit{Infomap}~\cite{rosvall2008maps}, a recursive application of \textit{Louvain} method for the generation of hierarchies~\cite{blondel2008fast,perotti2015hierarchical} and
the Minimum Description Length implementation of the \textit{Hierarchical Stochastic Block Model (HSBM)}~\cite{peixoto2014hierarchical}.
\textit{Spinglass} algorithm~\cite{reichardt2006statistical} is not tested because is computationally slow and \textit{OSLOM}~\cite{lancichinetti2011finding} is not employed because we  focus on networks with non-overlapping communities. 

As we already mentioned, we compare the similarity of the ground truth and detected community structures, employing the \textit{Normalized Mutual Information (NMI)}~\cite{danon2005comparing} and the \textit{Normalized Hierarchical Mutual Information (NHMI)}~\cite{perotti2015hierarchical}. In addition, we calculate the difference between the \textit{Hierarchical Mutual Information (HMI)} and the \textit{Mutual Information (MI)} at the different levels, in order to quantify the cumulative contributions of the deeper levels of graphs, only. 

\subsection{Test on two-level RB-LFR benchmark}
We first concentrate on the two-level RB-LFR benchmark ensemble. 
The seed LFR benchmark graphs we employ are undirected and unweighed networks with non-overlapping communities. 
The parameters of LFR benchmark are shown in Table \ref{table1}. The number of replicas equals to $R=4$. 

First, we study the accuracy of the community detection methods as a function of the mixing parameter $\mu$. We define three different ground truths: the first ground truth, namely \textit{seed-replica}, corresponds to the hierarchy that should emerge for small mixing parameter (Fig.~\ref{figure4}a); the second ground truth, namely \textit{replica-seed}, corresponds to a larger value of the mixing parameter (Fig.~\ref{figure4}b), and the last ground truth corresponds to a flat structure that there is only one level \cite{fortunato2010community, gregory2010finding}. These three ground truths are represented in black, red, and green color, respectively.

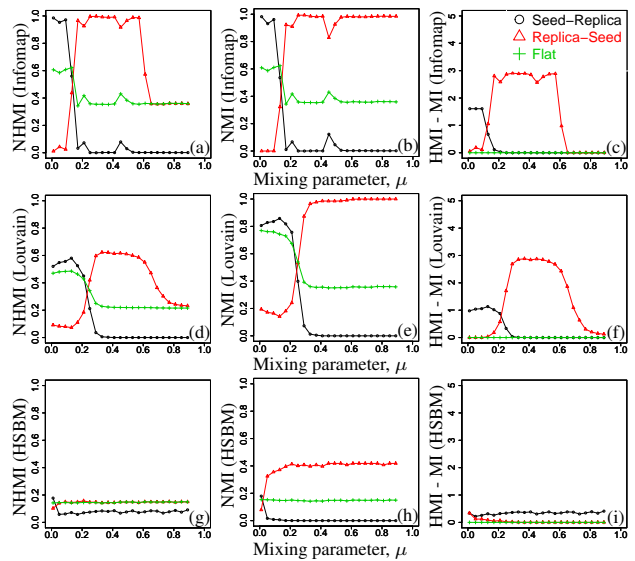
\begin{figure}[ht]
\begin{minipage}{0.15\textwidth}
\begin{tikzpicture}
  \node (img)  {\includegraphics[scale=0.15]{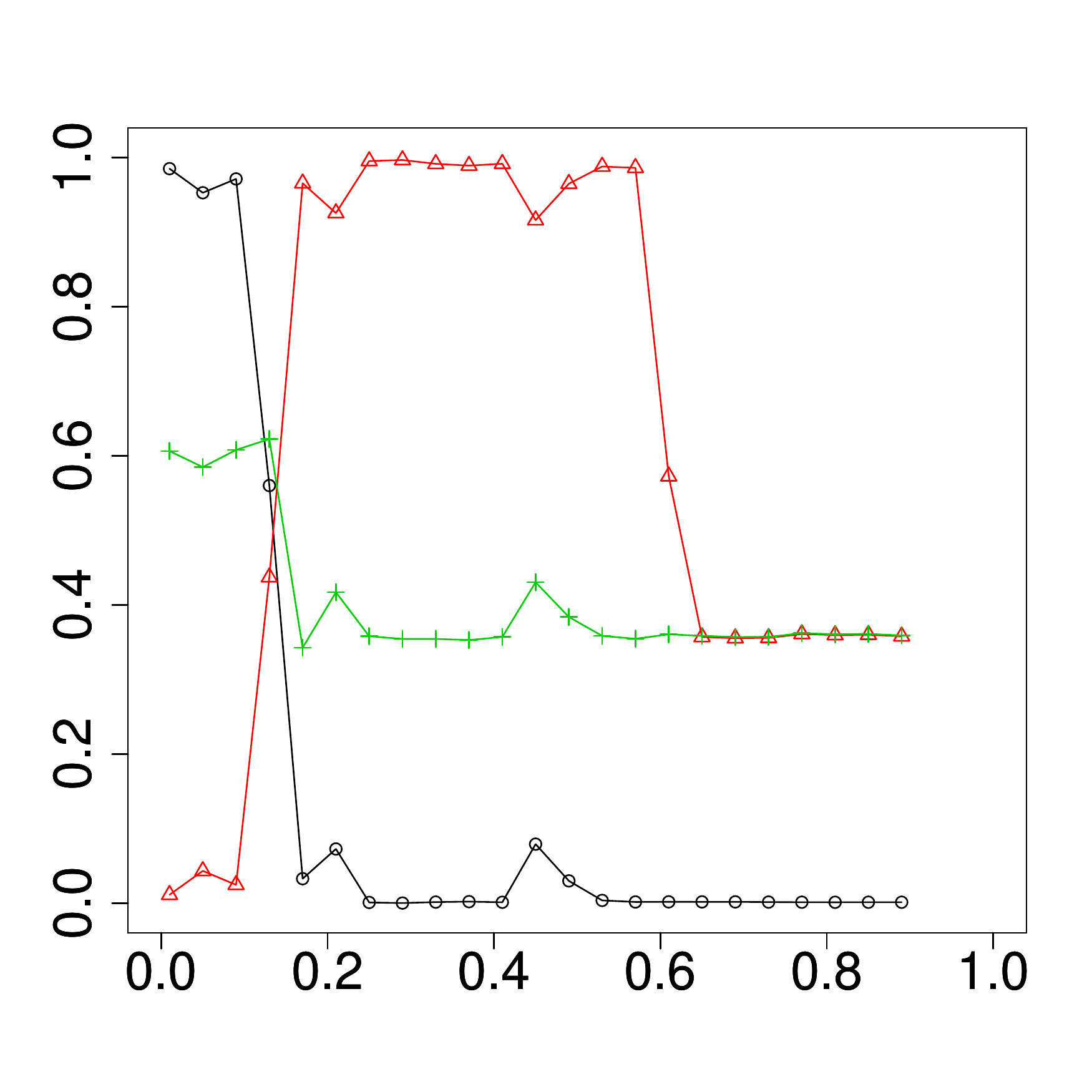}};
  \node[below = of img, node distance=0cm, xshift = 1.03cm, yshift=1.85cm,font=\scriptsize, align = center] {(a)};
  \node[left = of img, node distance=0cm, rotate=90, anchor=center,yshift=-1.1cm,font=\scriptsize , align = center] {NHMI (Infomap)};
 \end{tikzpicture}
\end{minipage} 
\begin{minipage}{0.15\textwidth}
\begin{tikzpicture}
  \node (img)  {\includegraphics[scale=0.15]{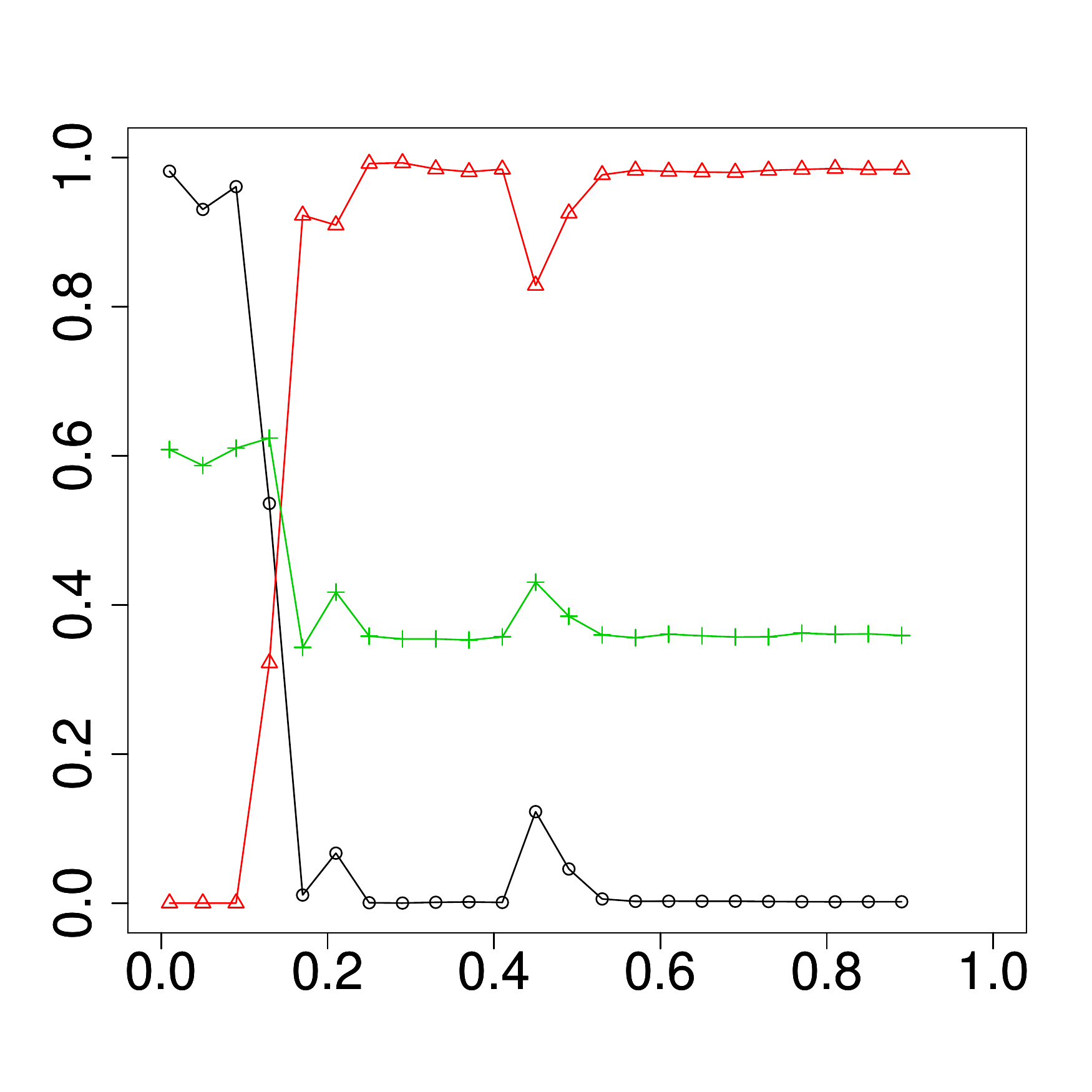}};
  \node[below = of img, node distance=0cm,xshift = 1.03cm, yshift=1.85cm, font=\scriptsize, align = center] {(b)};
    \node[below = of img, node distance=0cm, yshift=1.4cm, font=\scriptsize, align = center] {Mixing parameter, $\mu$};
  \node[left = of img, node distance=0cm, rotate=90, anchor=center,yshift=-1.1cm,font=\scriptsize, align = center] {NMI (Infomap)};
\end{tikzpicture}
\end{minipage}
\begin{minipage}{0.15\textwidth}
\begin{tikzpicture}
  \node (img)  {\includegraphics[scale=0.15]{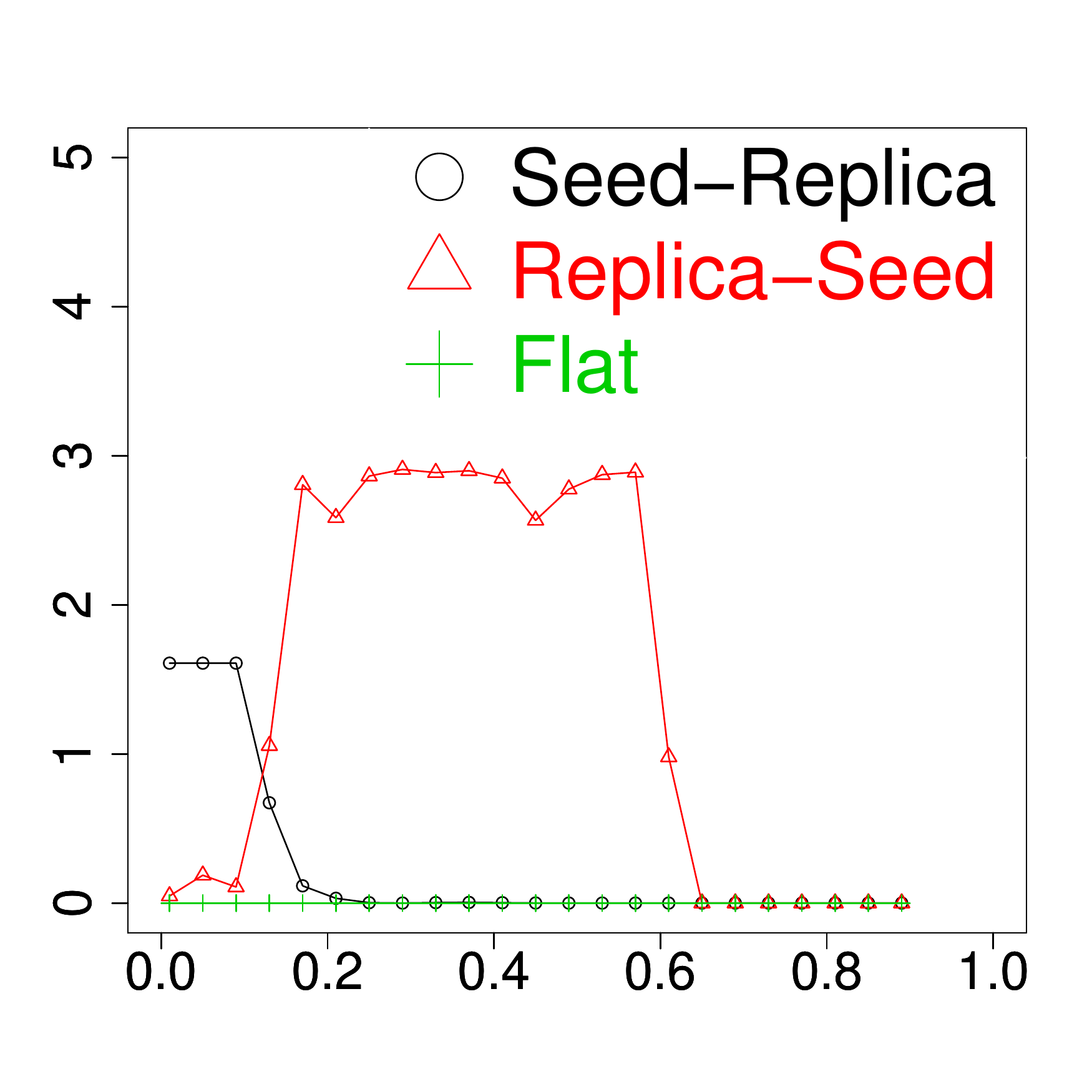}};
  \node[below = of img, node distance=0cm, xshift = 1.03cm, yshift=1.85cm, font=\scriptsize, align = center] {(c)};
  \node[left = of img, node distance=0cm, rotate=90, anchor=center,yshift=-1.1cm,font=\scriptsize, align = center] {HMI - MI (Infomap)};
\end{tikzpicture}
\end{minipage} 

\vspace{-0.5cm}

\begin{minipage}{0.15\textwidth}
\begin{tikzpicture}
  \node (img)  {\includegraphics[scale=0.15]{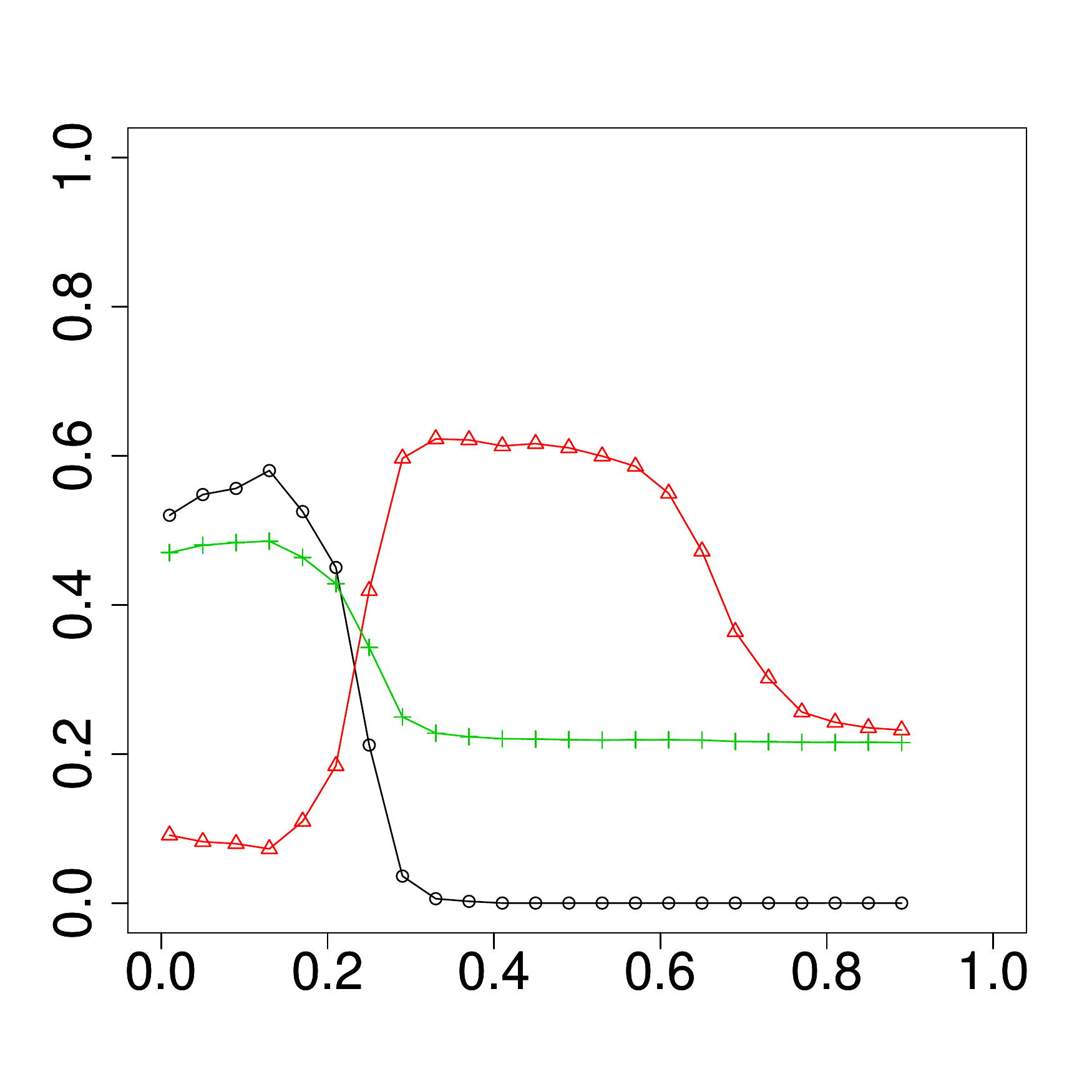}};
  \node[below = of img, node distance=0cm, xshift = 1.03cm, yshift=1.85cm,font=\scriptsize, align = center] {(d)};
  \node[left = of img, node distance=0cm, rotate=90, anchor=center,yshift=-1.1cm,font=\scriptsize , align = center] {NHMI (Louvain)};
 \end{tikzpicture}
\end{minipage} 
\begin{minipage}{0.15\textwidth}
\begin{tikzpicture}
  \node (img)  {\includegraphics[scale=0.15]{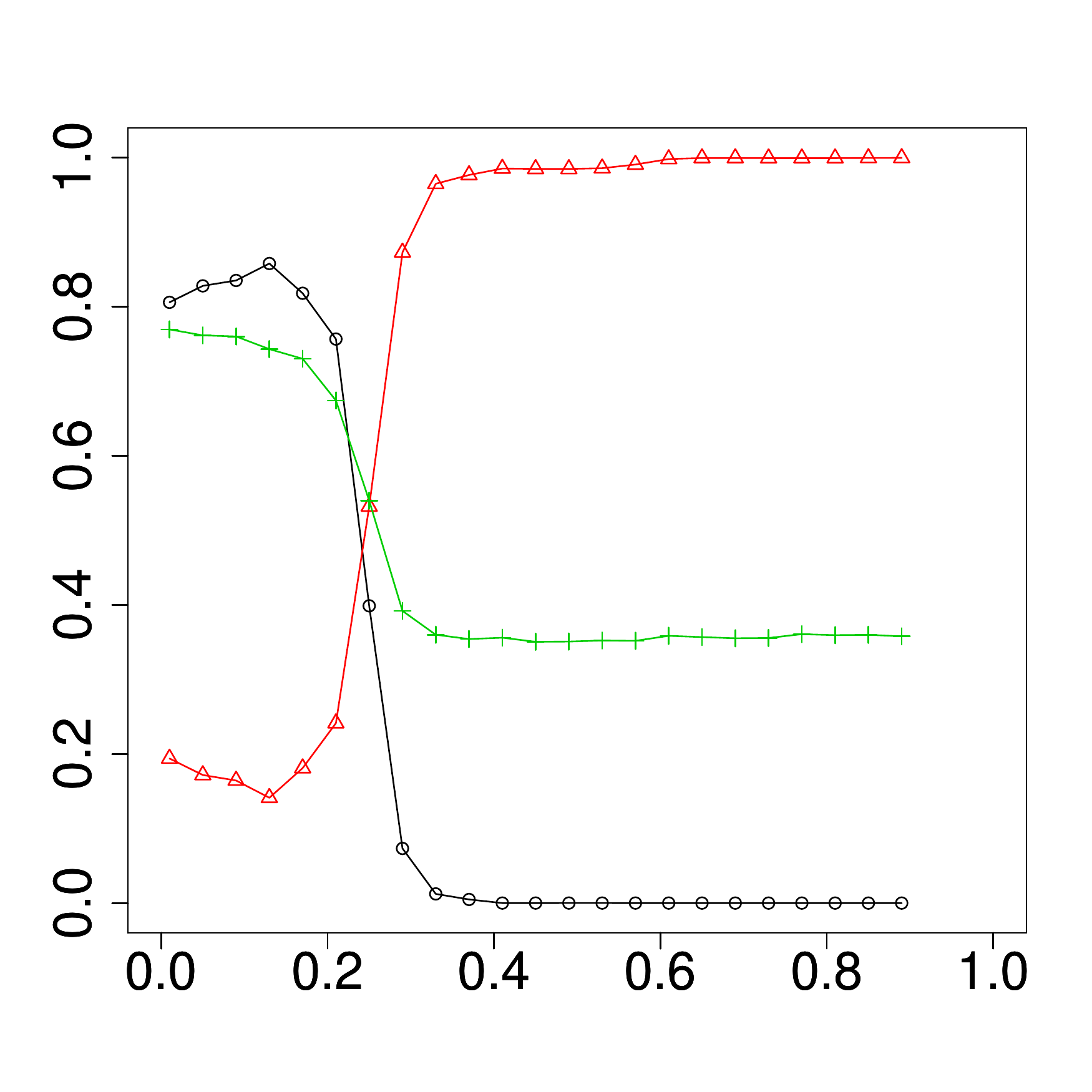}};
  \node[below = of img, node distance=0cm, xshift = 1.03cm, yshift=1.85cm,font=\scriptsize, align = center] {(e)};
      \node[below = of img, node distance=0cm, yshift=1.4cm, font=\scriptsize, align = center] {Mixing parameter, $\mu$};
  \node[left = of img, node distance=0cm, rotate=90, anchor=center,yshift=-1.1cm,font=\scriptsize, align = center] {NMI (Louvain)};
\end{tikzpicture}
\end{minipage}
\begin{minipage}{0.15\textwidth}
\begin{tikzpicture}
  \node (img)  {\includegraphics[scale=0.15]{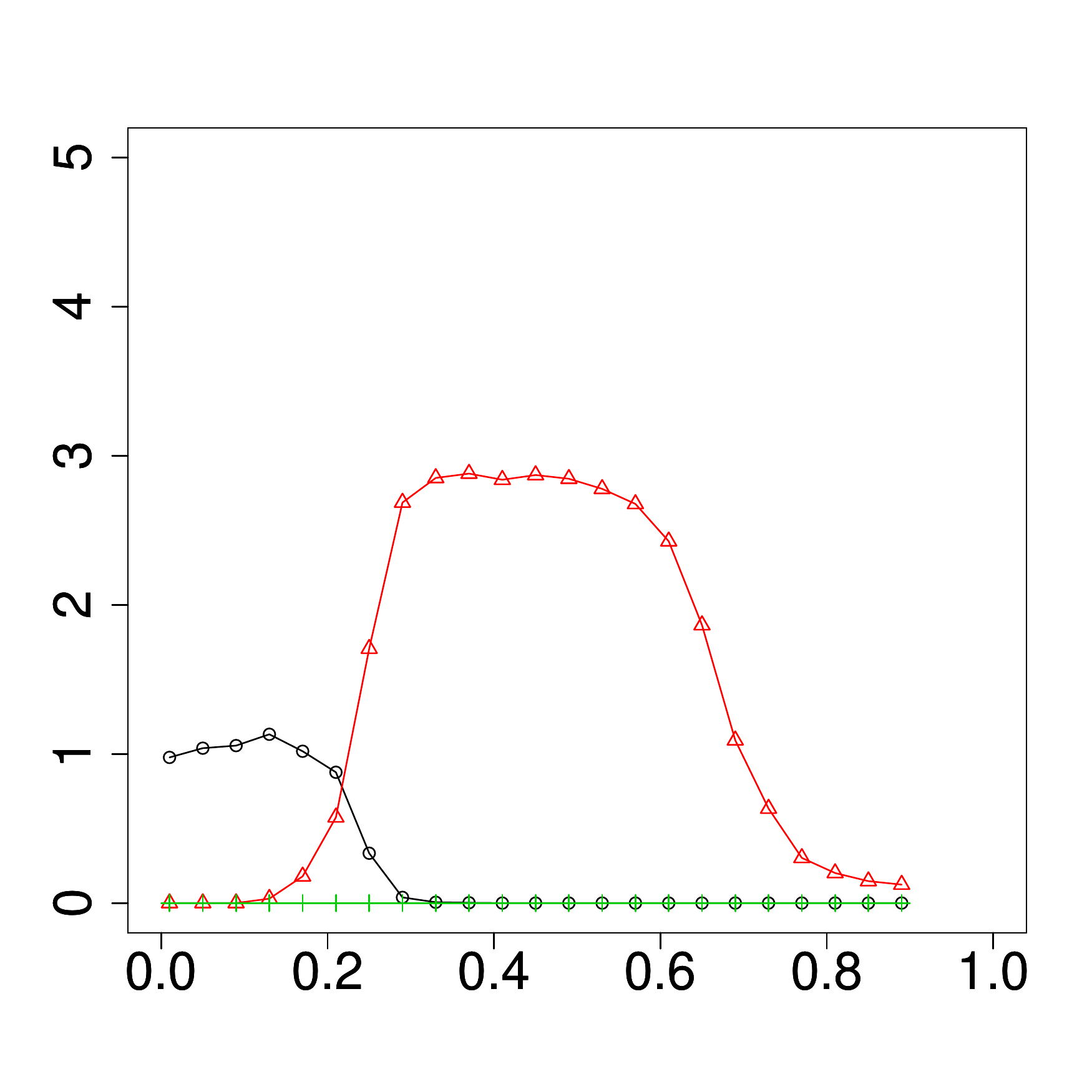}};
  \node[below = of img, node distance=0cm, xshift = 1.03cm, yshift=1.85cm,font=\scriptsize, align = center] {(f)};
  \node[left = of img, node distance=0cm, rotate=90, anchor=center,yshift=-1.1cm,font=\scriptsize, align = center] {HMI - MI (Louvain)};
\end{tikzpicture}
\end{minipage} 

\vspace{-0.5cm}

\begin{minipage}{0.15\textwidth}
\begin{tikzpicture}
  \node (img)  {\includegraphics[scale=0.15]{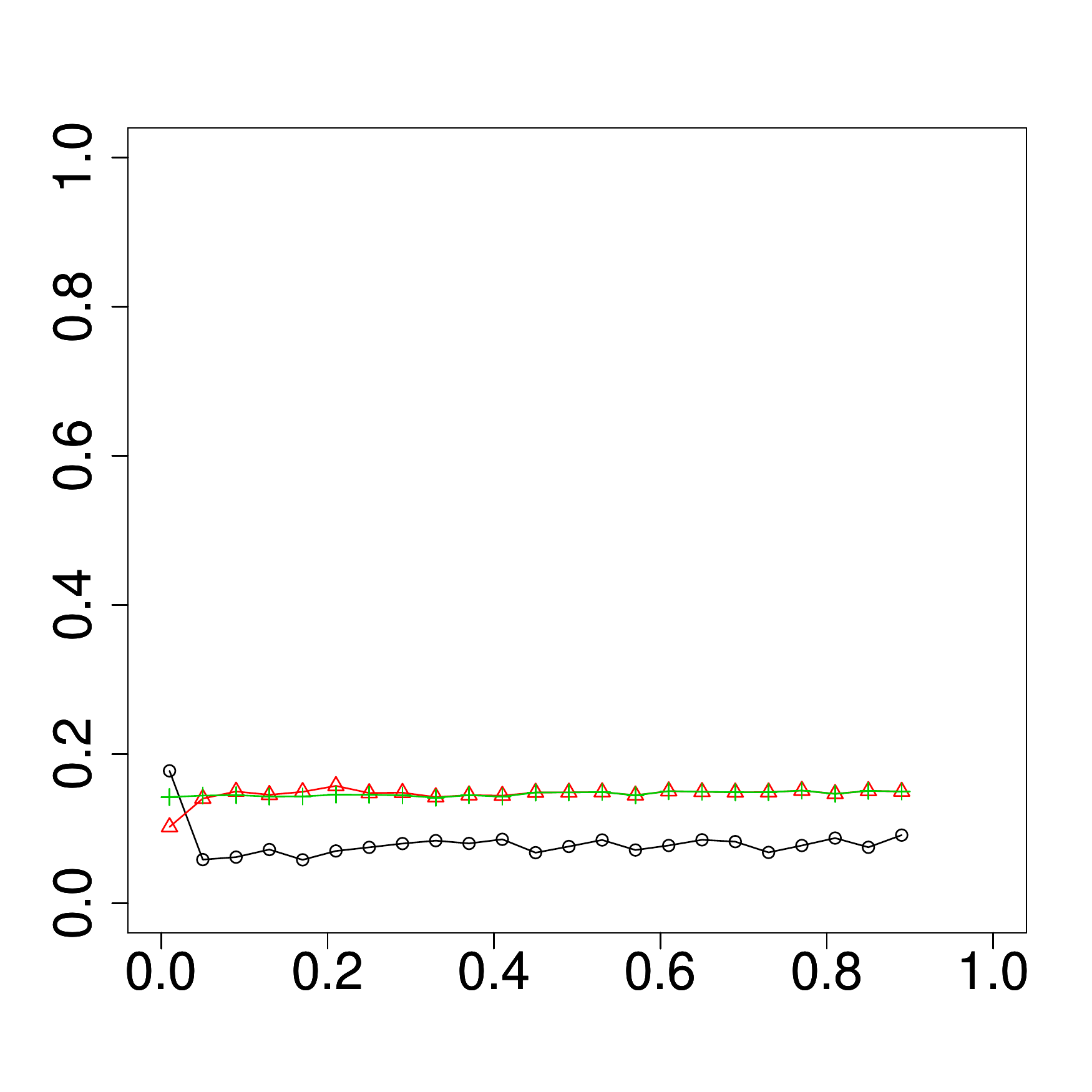}};
  \node[below = of img, node distance=0cm, xshift = 1.03cm, yshift=1.85cm,font=\scriptsize, align = center] {(g)};
  \node[left = of img, node distance=0cm, rotate=90, anchor=center,yshift=-1.1cm,font=\scriptsize , align = center] {NHMI (HSBM)};
 \end{tikzpicture}
\end{minipage} 
\begin{minipage}{0.15\textwidth}
\begin{tikzpicture}
  \node (img)  {\includegraphics[scale=0.15]{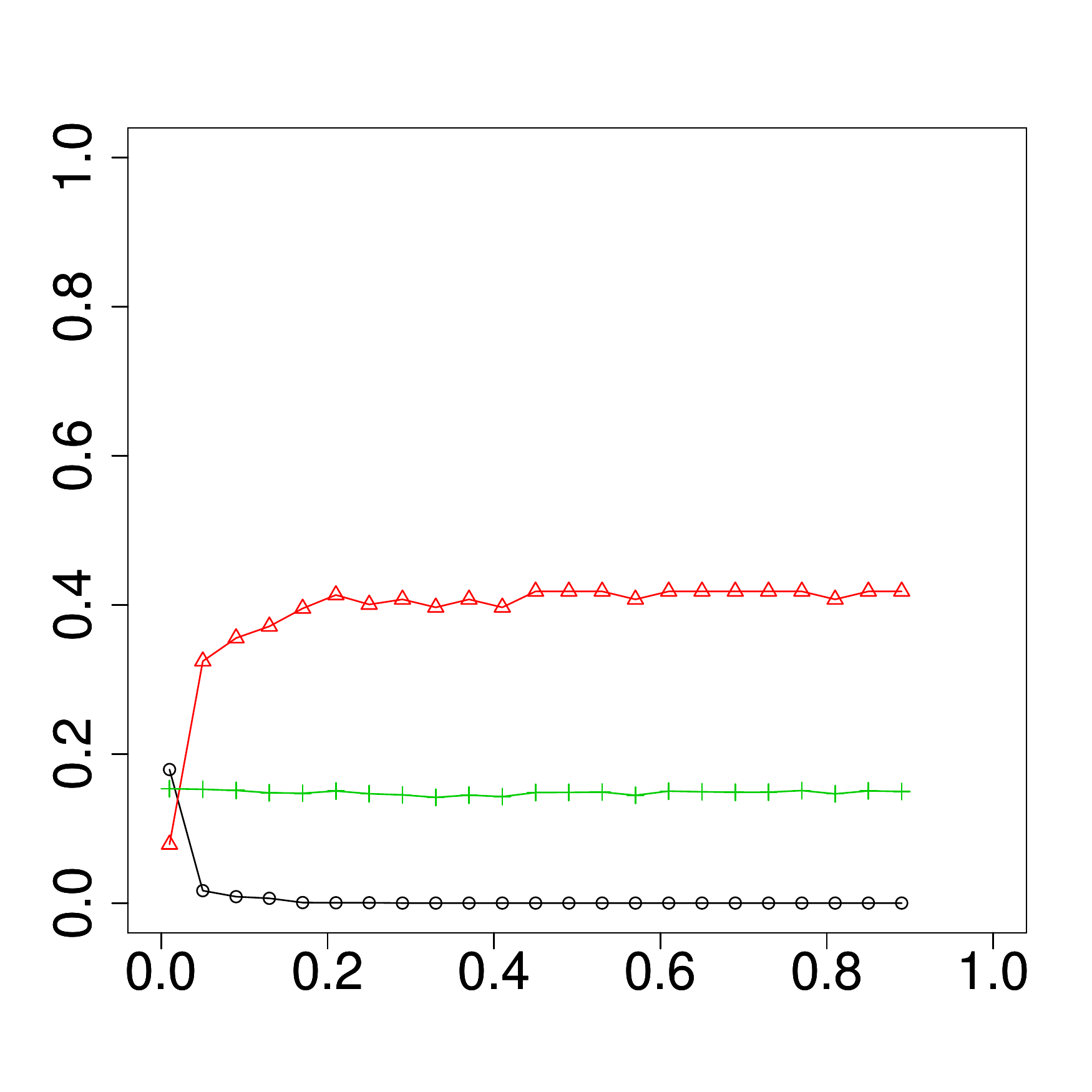}};
  \node[below = of img, node distance=0cm,xshift = 1.03cm, yshift=1.85cm,font=\scriptsize, align = center] {(h)};
        \node[below = of img, node distance=0cm, yshift=1.4cm, font=\scriptsize, align = center] {Mixing parameter, $\mu$};
  \node[left = of img, node distance=0cm, rotate=90, anchor=center,yshift=-1.1cm,font=\scriptsize, align = center] {NMI (HSBM)};
\end{tikzpicture}
\end{minipage}
\begin{minipage}{0.15\textwidth}
\begin{tikzpicture}
  \node (img)  {\includegraphics[scale=0.15]{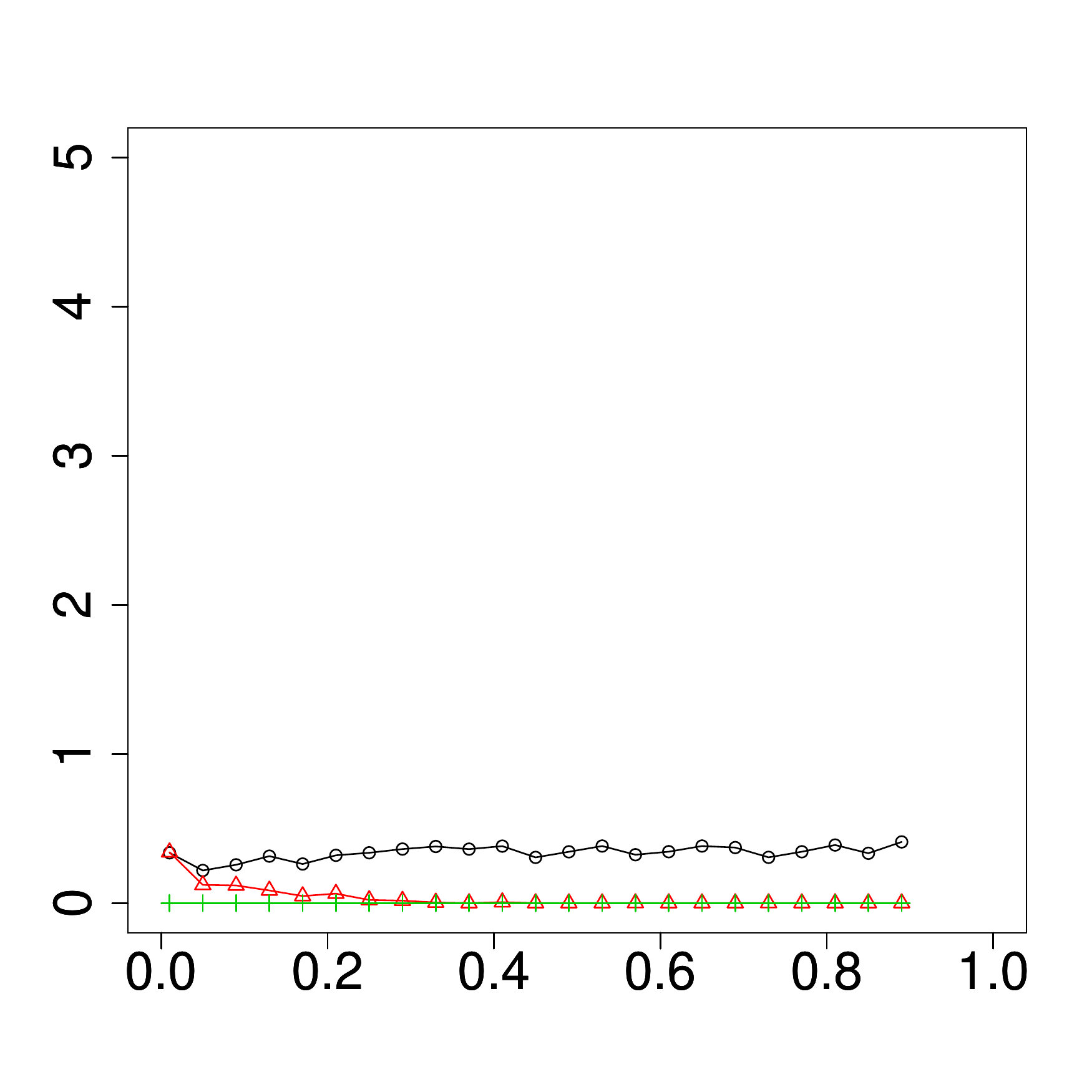}};
  \node[below = of img, node distance=0cm, xshift = 1.03cm, yshift=1.85cm,font=\scriptsize, align = center] {(i)};
  \node[left = of img, node distance=0cm, rotate=90, anchor=center,yshift=-1.1cm,font=\scriptsize, align = center] {HMI - MI (HSBM)};
\end{tikzpicture}
\end{minipage} 
\caption{
Average NHMI, NMI, and (HMI - MI) as a function of the mixing parameter, $\mu$ at the left, middle and right panels, respectively.
Here, the NMI compares partitions at second level of the detected and ground truth hierarchies.
Similarly, the HMI compares full hierarchies while the MI compares partitions at the first level.
From top to bottom, the methods are Infomap, Louvain, and HSBM.
Averages are computed over 10 different network realizations with the same set of parameters of the seed LFR benchmark. The parameters of the seed networks can be found in Table \ref{table1}.
}
\label{figure6}
\end{figure}

The results are shown in Fig.~\ref{figure6}. In the left panels the accuracy of the different community detection algorithms are quantified by  the average value of the NHMI computed between the detected hierarchical community structures and the different ground truths.
In the center column, the similarity is quantified with the average NMI computed between the detected partitions at the second level and those exhibited by the different ground truths. 
In the right panels, the similarity is quantified by the difference HMI - MI between the HMI computed for the full hierarchies and the MI computed for the partitions at the first level. The tested methods are Infomap, Louvain, and HSBM from top to bottom. Taking the top-left panel as an example: Infomap can unveil the community structure until $\mu \approx 0.6$ (with the difference between both ground truths).  For $\mu \lessapprox 0.1$, it detects the first type of ground truth, and for $0.2 \lessapprox \mu \lessapprox 0.6$, it detects the second type of ground truth.
We observe a clear transition between the ground truths for $\mu$ between $\mu = 0.1$ and $\mu = 0.2$; in both regions, the NHMI reaches values close to one making apparent that the algorithm gives a description of the hierarchy very close to the ground truth. For $\mu \gtrapprox 0.6$, Infomap detects a flat community structure. This result showcases that the RB-LFR benchmark shows a clear hierarchical community structure which can be recognized successfully by Infomap. The fact that $\textit{NHMI} = 1$ highlights that this  is indeed non-trivial. 

Comparing panels (a) to (d), and (g) of Fig.~\ref{figure6}, we observe that the new benchmark poses a challenging task that can test the performance of the algorithms: the accuracy of Louvain reaches 0.6 until $\mu \approx 0.6$ but, it still detects some hierarchical community structure until $\mu \approx 0.9$, a far wider range than Infomap. 
The HSBM always has an accuracy smaller than 0.2. We note here that the poor performance of the HSBM is most likely related to its approach, i.e.~a bottom-up approach, while the other two methods are taking the top-down approaches to build the hierarchies~\cite{perotti2015hierarchical}. 

The right panels, Fig.~\ref{figure6}c, f, \& i, which show the difference between the full HMI and MI of the first level, overall giving the contribution that the second level has on the HMI.
In other words, it quantifies how accurately the algorithms detect the second level and how relevant is the corresponding contribution as measured by the HMI. 
For instance, for Infomap, under the second definition of ground truth, the observed value represents 64.7\% of the total value of the HMI when $\mu = 0.37$.
Hence, the contribution of the second level is non-negligible, showing the convenience of \textit{Hierarchical Mutual Information} as a measure for the comparison of hierarchical community structures, when compared to the traditional \textit{Mutual Information}.

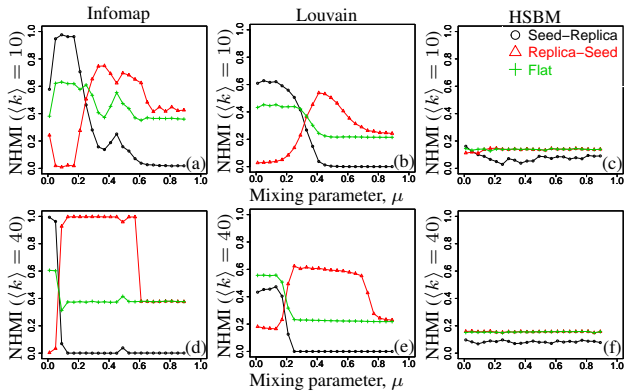
\begin{figure}[ht]
\begin{minipage}{0.15\textwidth}
\begin{tikzpicture}
  \node (img)  {\includegraphics[scale=0.15]{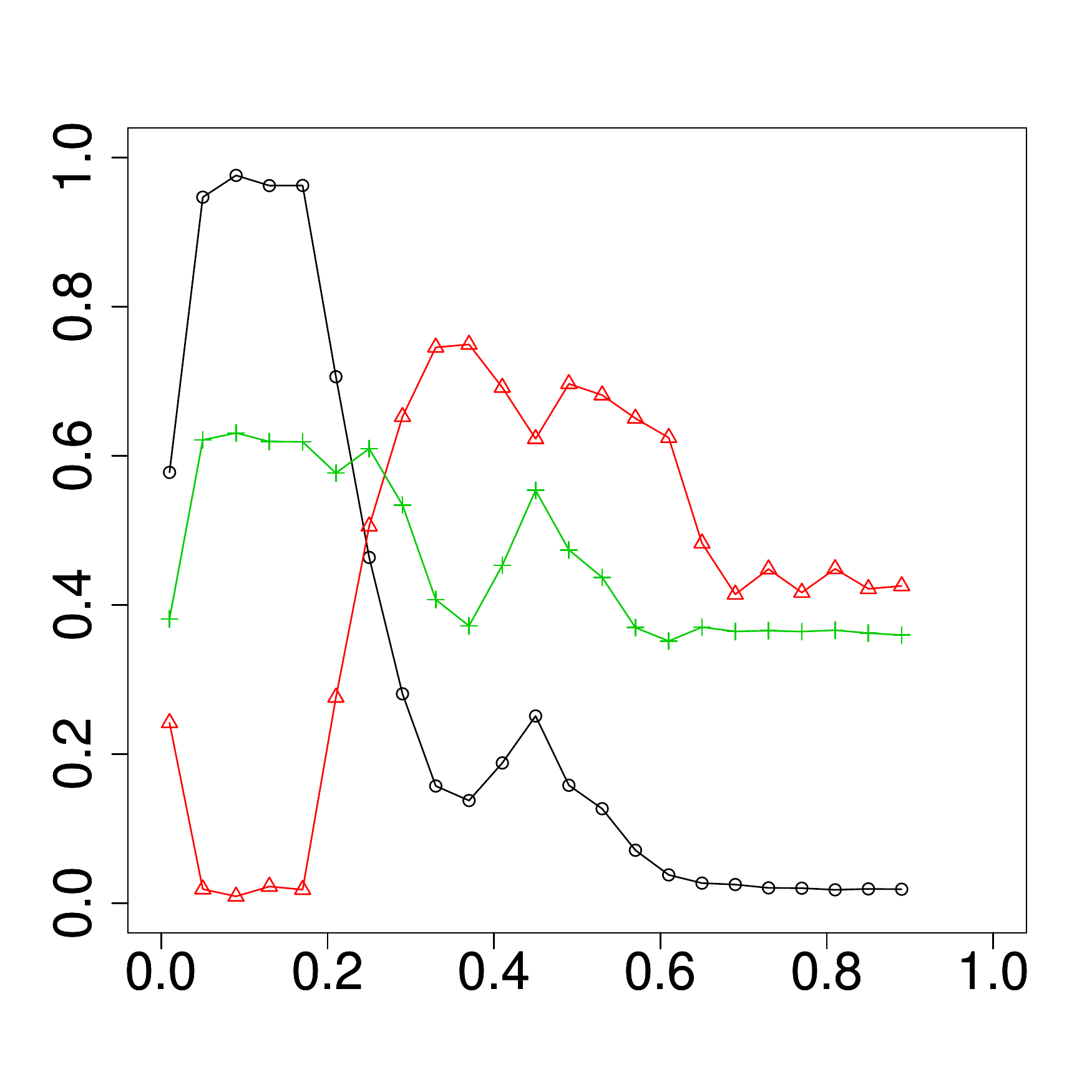}};
  \node[below = of img, node distance=0cm, xshift = 1.03cm, yshift=1.85cm, font=\scriptsize, align = center] {(a)};  
      \node[above = of img, node distance=0cm, yshift=-1.5cm, font=\scriptsize, align = center] {Infomap};  
  \node[left = of img, node distance=0cm, rotate=90, anchor=center,yshift=-1.1cm,font=\scriptsize , align = center] {NHMI ($\langle k\rangle = 10$)};
 \end{tikzpicture}
\end{minipage} 
\begin{minipage}{0.15\textwidth}
\begin{tikzpicture}
  \node (img)  {\includegraphics[scale=0.15]{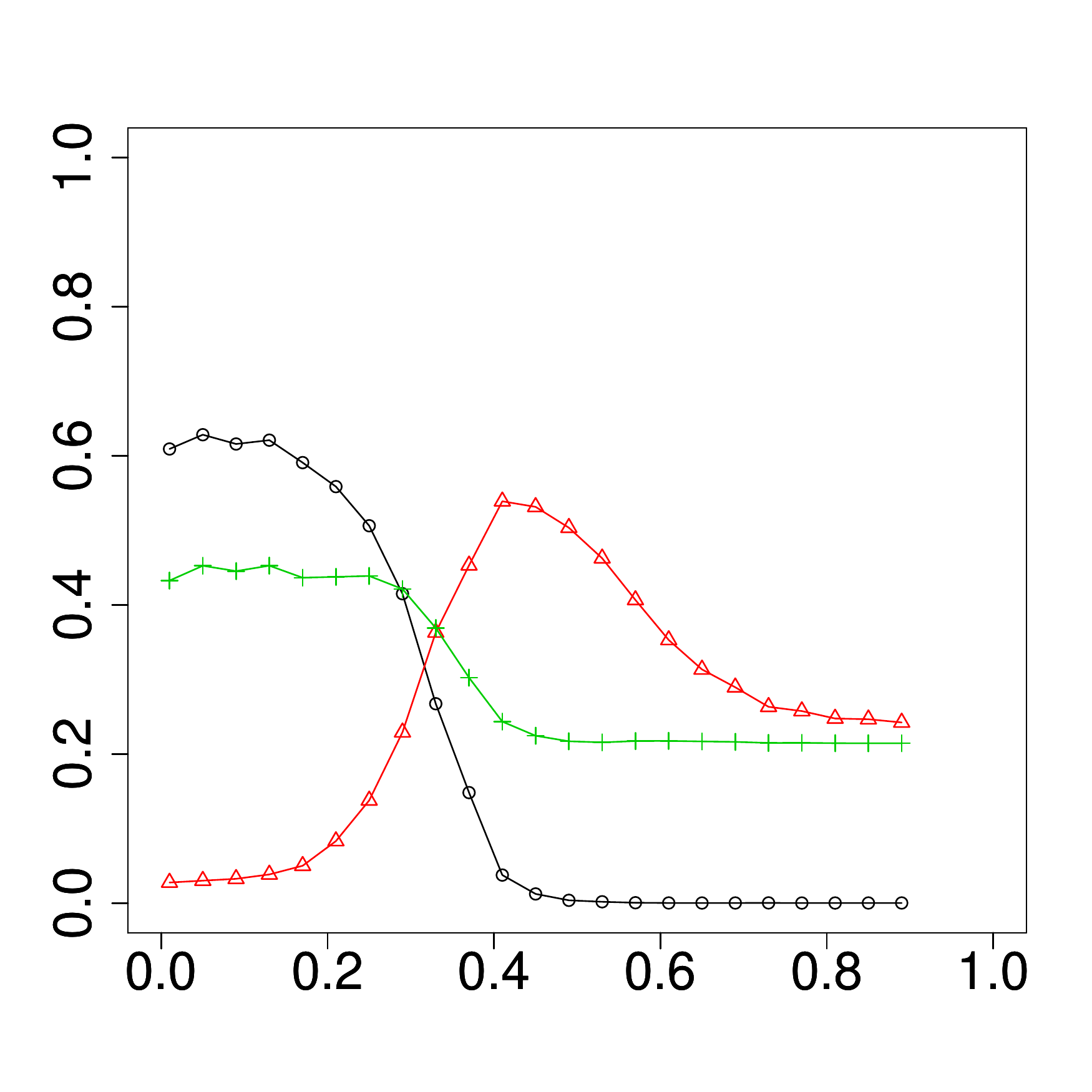}};
  \node[below = of img, node distance=0cm, xshift = 1.03cm, yshift=1.85cm,font=\scriptsize, align = center] {(b)};  
        \node[above = of img, node distance=0cm, yshift=-1.5cm, font=\scriptsize, align = center] {Louvain};  
  \node[below = of img, node distance=0cm, yshift=1.4cm, font=\scriptsize, align = center] {Mixing parameter, $\mu$};
  \node[left = of img, node distance=0cm, rotate=90, anchor=center,yshift=-1.1cm,font=\scriptsize , align = center] {NHMI ($\langle k\rangle = 10$)};
 \end{tikzpicture}
\end{minipage} 
\begin{minipage}{0.15\textwidth}
\begin{tikzpicture}
  \node (img)  {\includegraphics[scale=0.15]{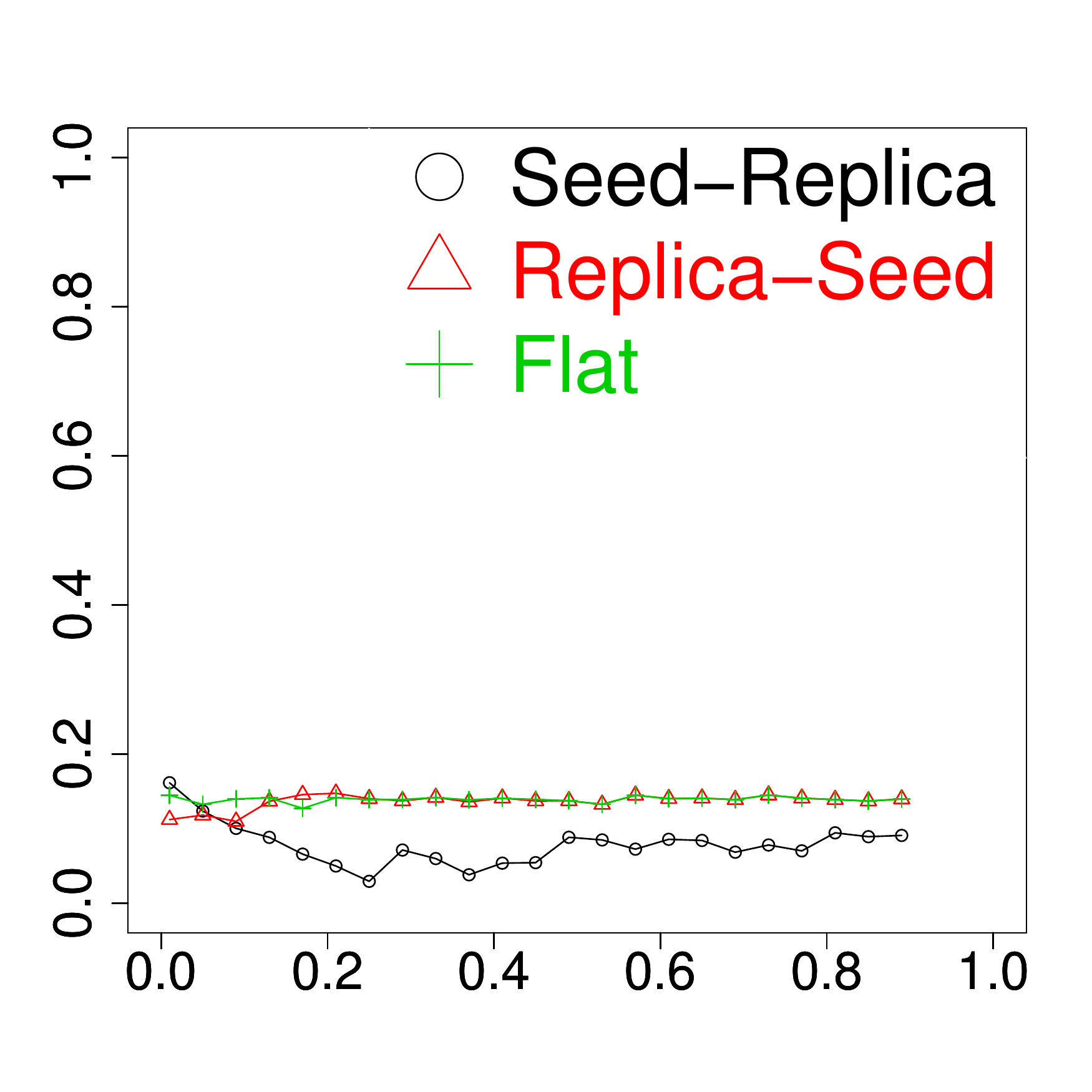}};
  \node[below = of img, node distance=0cm, xshift = 1.03cm, yshift=1.85cm,font=\scriptsize, align = center] {(c)};  
        \node[above = of img, node distance=0cm, yshift=-1.5cm, font=\scriptsize, align = center] {HSBM};  
  \node[left = of img, node distance=0cm, rotate=90, anchor=center,yshift=-1.1cm,font=\scriptsize , align = center] {NHMI ($\langle k\rangle = 10$)};
 \end{tikzpicture}
\end{minipage} 

\vspace{-0.5cm}

\begin{minipage}{0.15\textwidth}
\begin{tikzpicture}
  \node (img)  {\includegraphics[scale=0.15]{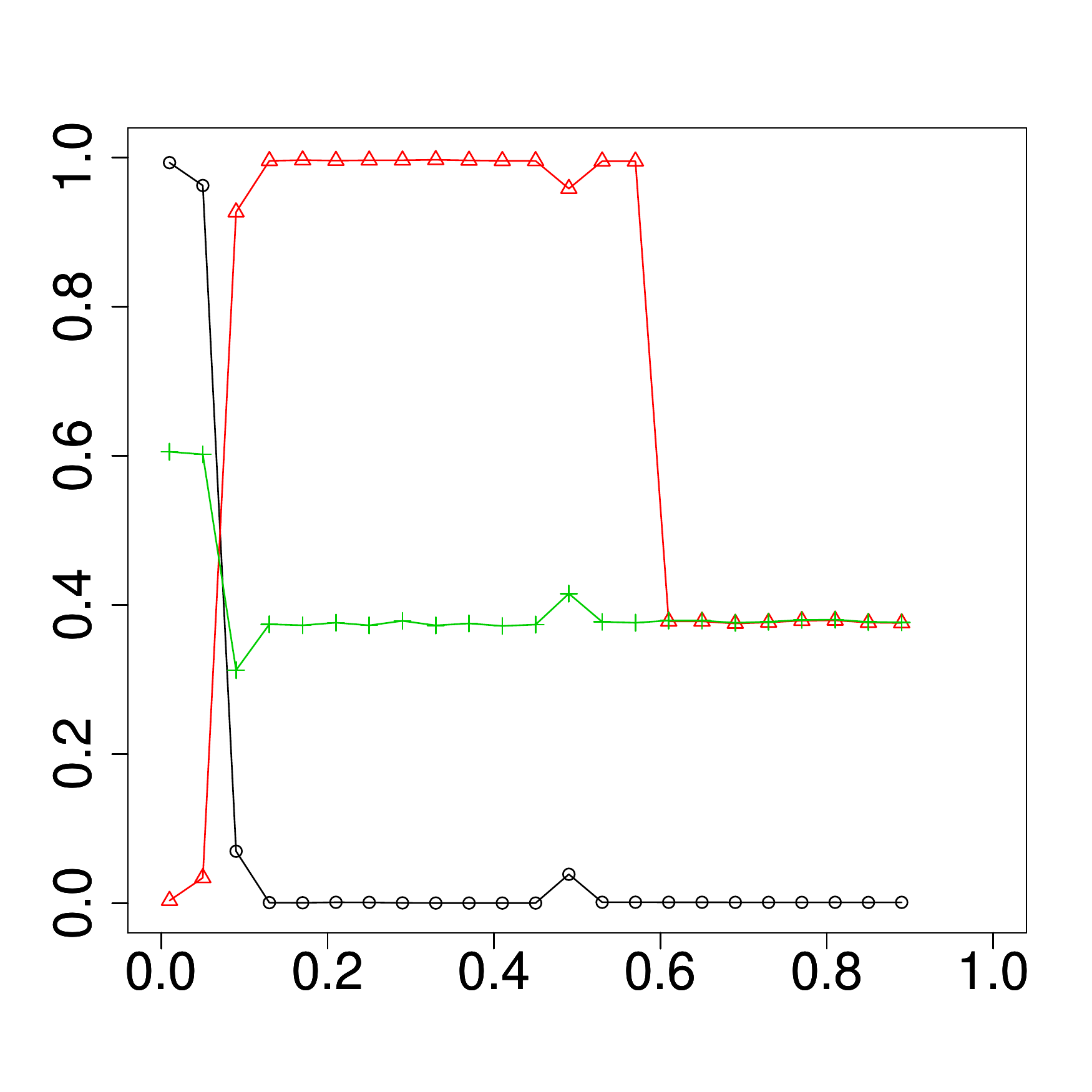}};
  \node[below = of img, node distance=0cm, xshift = 1.03cm, yshift=1.85cm, font=\scriptsize, align = center] {(d)};  
  \node[left = of img, node distance=0cm, rotate=90, anchor=center,yshift=-1.1cm,font=\scriptsize , align = center] {NHMI ($\langle k\rangle = 40$)};
 \end{tikzpicture}
\end{minipage} 
\begin{minipage}{0.15\textwidth}
\begin{tikzpicture}
  \node (img)  {\includegraphics[scale=0.15]{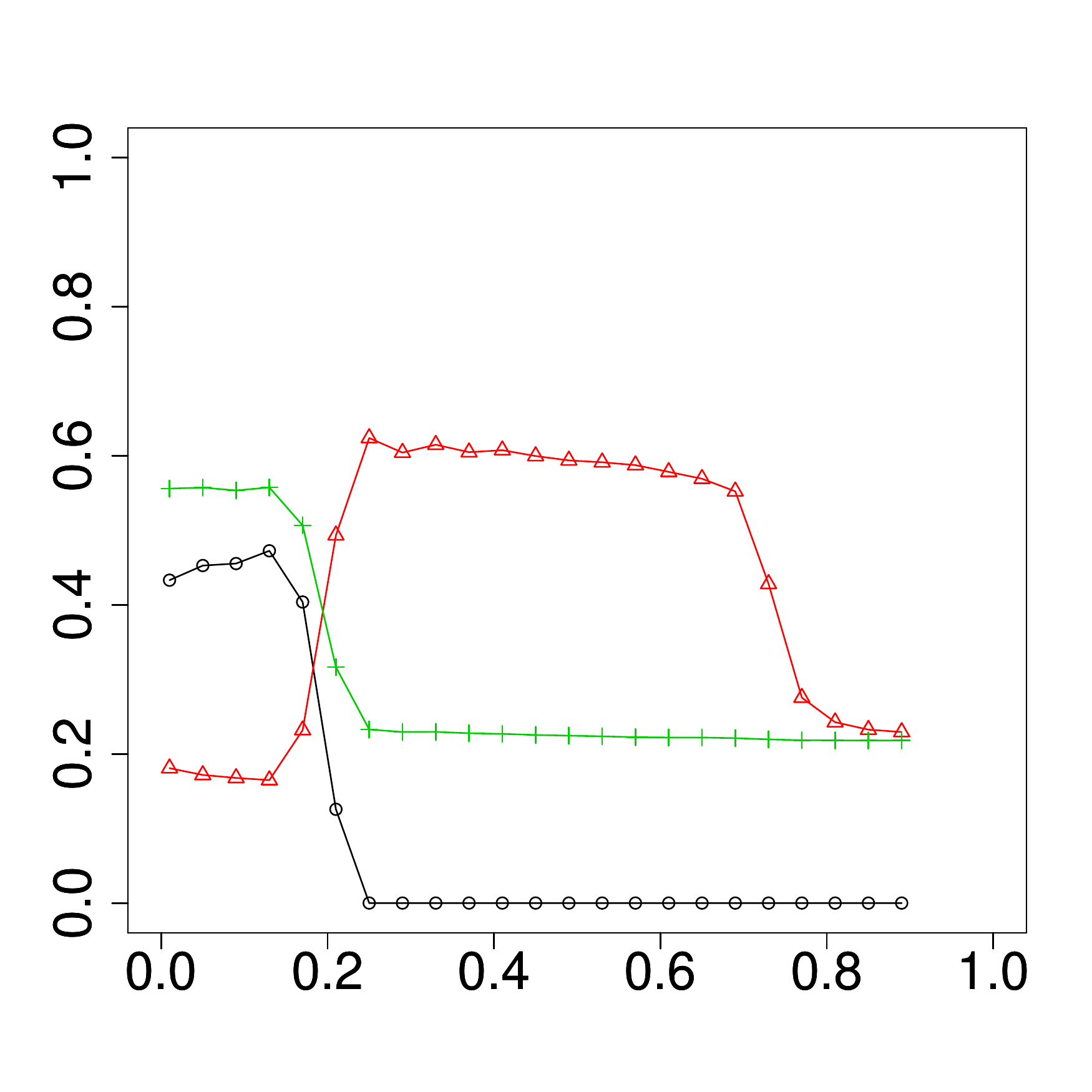}};
  \node[below = of img, node distance=0cm, xshift = 1.03cm, yshift=1.85cm,font=\scriptsize, align = center] {(e)};  
  \node[below = of img, node distance=0cm, yshift=1.4cm, font=\scriptsize, align = center] {Mixing parameter, $\mu$};
  \node[left = of img, node distance=0cm, rotate=90, anchor=center,yshift=-1.1cm,font=\scriptsize , align = center] {NHMI ($\langle k\rangle = 40$)};
 \end{tikzpicture}
\end{minipage} 
\begin{minipage}{0.15\textwidth}
\begin{tikzpicture}
  \node (img)  {\includegraphics[scale=0.15]{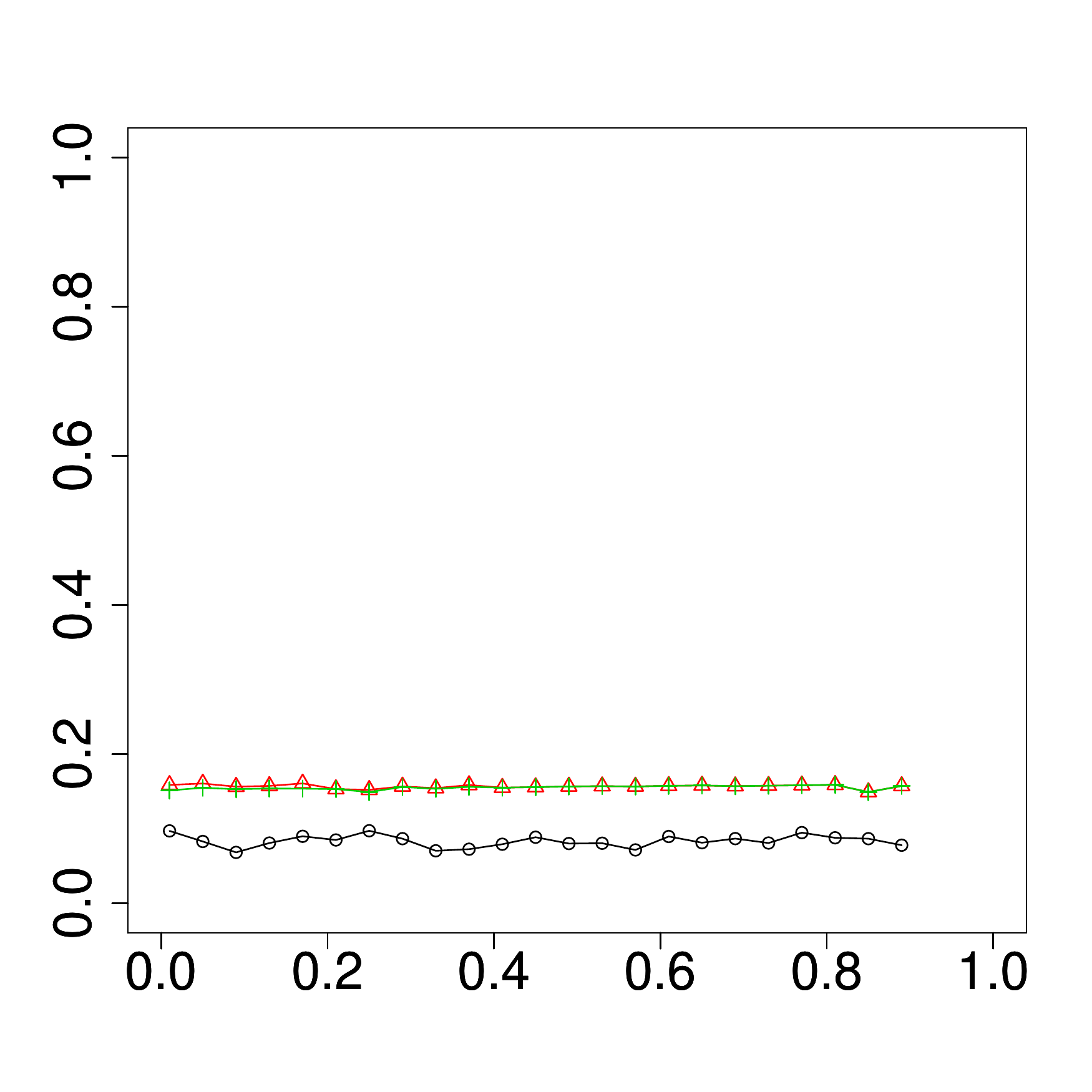}};
  \node[below = of img, node distance=0cm, xshift = 1.03cm, yshift=1.85cm,font=\scriptsize, align = center] {(f)};  
  \node[left = of img, node distance=0cm, rotate=90, anchor=center,yshift=-1.1cm,font=\scriptsize , align = center] {NHMI ($\langle k\rangle = 40$)};
 \end{tikzpicture}
\end{minipage} 
\caption{Average NHMI as a function of the mixing parameter, $\mu$. The top panels correspond to seed LFR benchmaks with average degree $\langle k\rangle = 10$ and the bottom ones to $\langle k\rangle = 40$. From left to right, the methods are Infomap, Louvain, and HSBM.
Averages are computed over 10 different network realizations with the same set of parameters of the seed LFR benchmark.The parameters of the seed networks can be found in Table \ref{table1}.
} 
\label{figure7}
\end{figure}

Now, we measure the effect of the average degree $\langle k\rangle$ on the performance of algorithms. 
We use the NHMI to quantify the accuracies of the algorithms and the results are shown in Fig.~\ref{figure7}. The top panels correspond to $\langle k \rangle = 10$, and the bottom ones correspond to $\langle k\rangle = 40$. 
Comparing panels (a) and (d), and panels (b) and (e), we can observe that for sparse RB-LFR benchmark graphs, the community detection methods have better performance with increasing $\langle k\rangle$. 
This is the result that is typically observed~\cite{fortunato2010community}
and is a reasonable one since, in the sparse regime $\langle k\rangle \ll N_0$, where $N_0$ is the number of nodes in the network, the larger is $\langle k\rangle$ the less important are the sample to sample fluctuations that may affect how well defined the communities are.
Furthermore, we observe a similar pattern to the Fig.~\ref{figure6}: while Infomap exhibits higher accuracy, Louvain is able to detect a hierarchical structure in a wider range of the mixing parameter $\mu$ (Figs.~\ref{figure7}d \& e and Figs.~\ref{figure6}a \& d). 

\begin{figure}[ht]
\begin{minipage}{0.15\textwidth}
\begin{tikzpicture}
  \node (img)  {\includegraphics[scale=0.15]{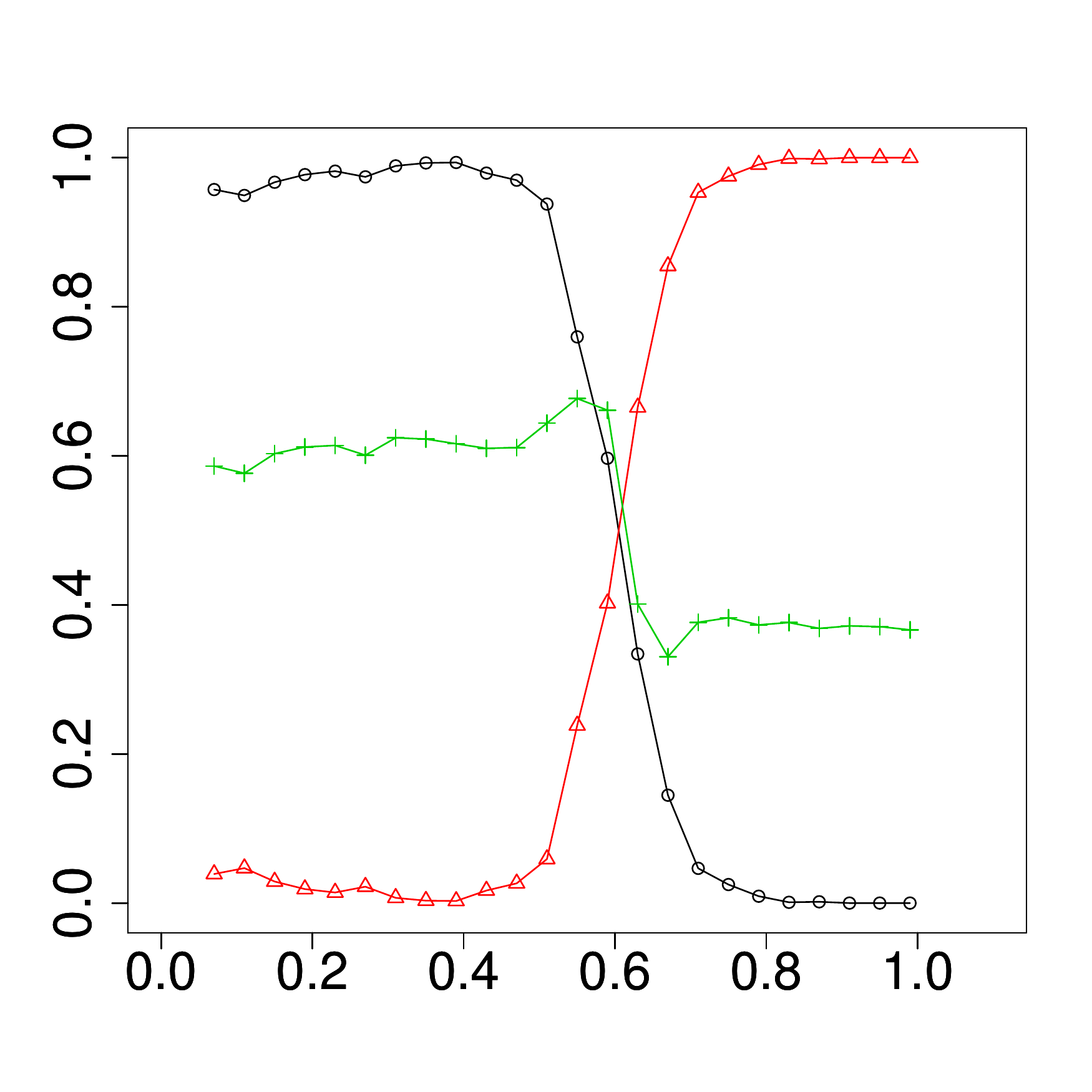}};
  \node[below = of img, node distance=0cm, xshift = 1.03cm, yshift=1.85cm,font=\scriptsize, align = center] {(a)};  
    \node[above = of img, node distance=0cm, yshift=-1.5cm, font=\scriptsize, align = center] {$\mu = 0.05$};  
  \node[left = of img, node distance=0cm, rotate=90, anchor=center,yshift=-1.1cm,font=\scriptsize , align = center] {NHMI (Infomap)};
 \end{tikzpicture}
\end{minipage} 
\begin{minipage}{0.15\textwidth}
\begin{tikzpicture}
  \node (img)  {\includegraphics[scale=0.15]{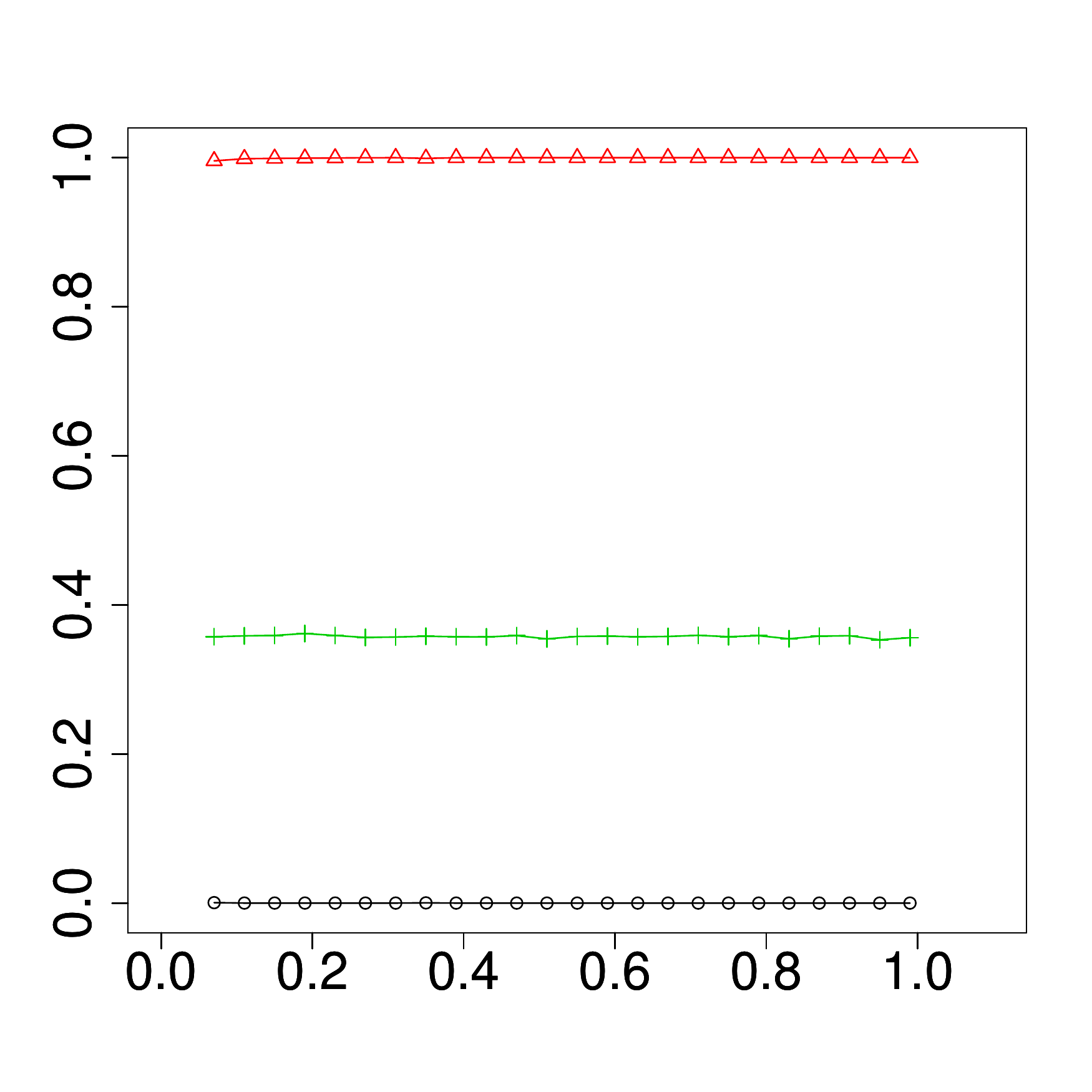}};
  \node[below = of img, node distance=0cm, xshift = 1.03cm, yshift=1.85cm,font=\scriptsize, align = center] {(b)};  
      \node[above = of img, node distance=0cm, yshift=-1.5cm, font=\scriptsize, align = center] {$\mu = 0.3$};  
  \node[below = of img, node distance=0cm, yshift=1.4cm, font=\scriptsize, align = center] {$p$};  
  \node[left = of img, node distance=0cm, rotate=90, anchor=center,yshift=-1.1cm,font=\scriptsize , align = center] {NHMI (Infomap)};
 \end{tikzpicture}
\end{minipage} 
\begin{minipage}{0.15\textwidth}
\begin{tikzpicture}
  \node (img)  {\includegraphics[scale=0.15]{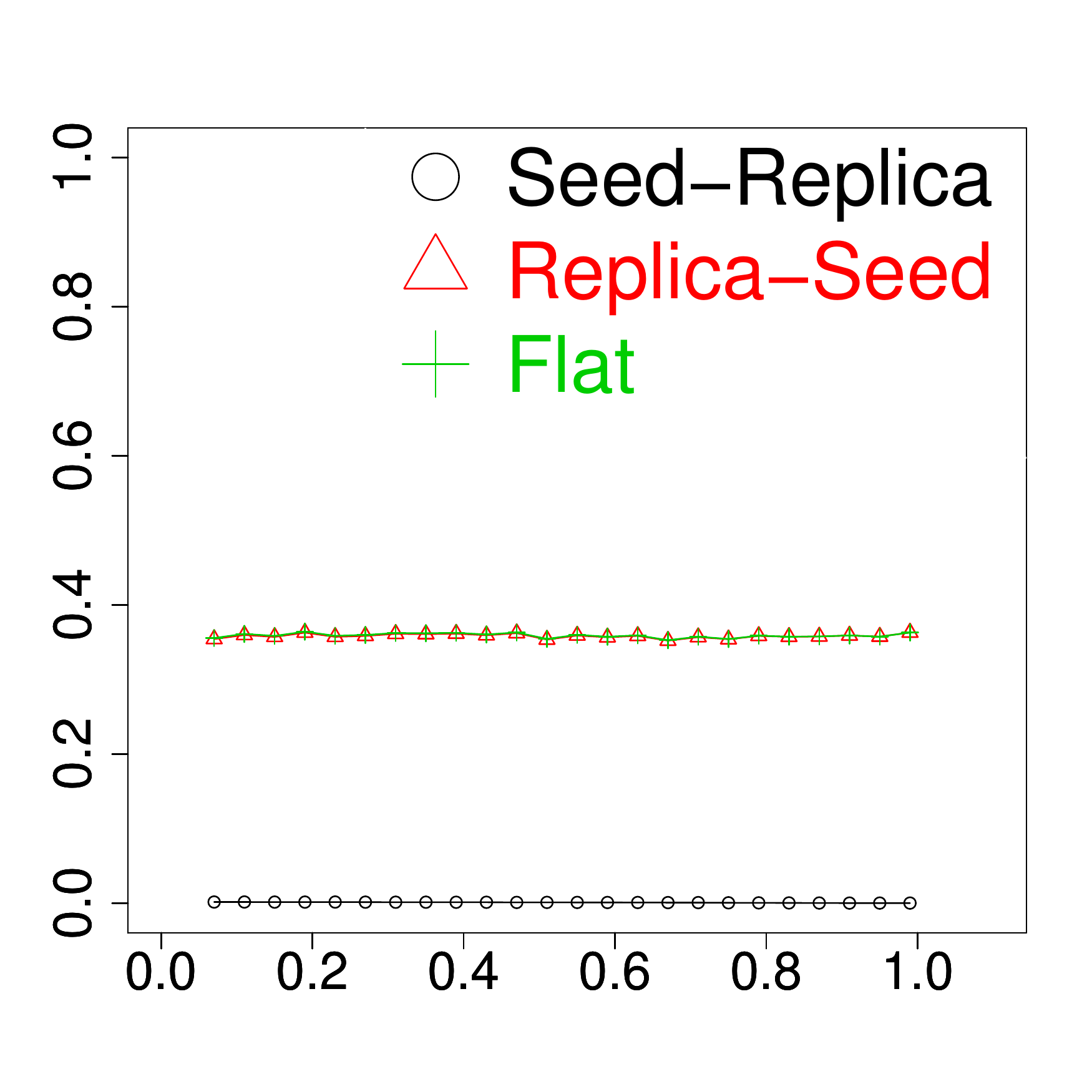}};
  \node[below = of img, node distance=0cm, xshift = 1.03cm, yshift=1.85cm,font=\scriptsize, align = center] {(c)};  
      \node[above = of img, node distance=0cm, yshift=-1.5cm, font=\scriptsize, align = center] {$\mu = 0.7$};  
  \node[left = of img, node distance=0cm, rotate=90, anchor=center,yshift=-1.1cm,font=\scriptsize , align = center] {NHMI (Infomap)};
 \end{tikzpicture}
\end{minipage} 

\vspace{-0.5cm}

\begin{minipage}{0.15\textwidth}
\begin{tikzpicture}
  \node (img)  {\includegraphics[scale=0.15]{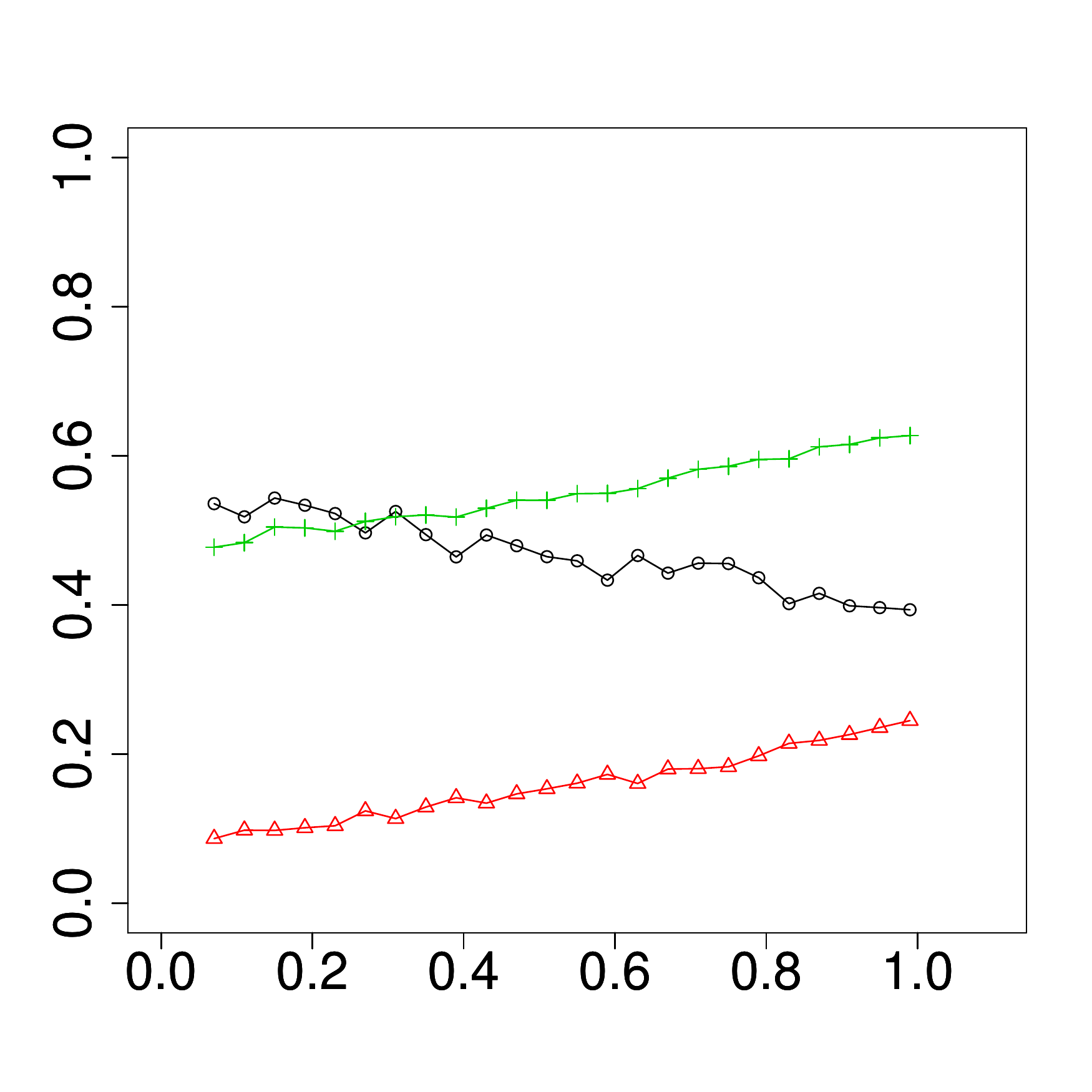}};
  \node[below = of img, node distance=0cm, xshift = 1.03cm, yshift=1.85cm,font=\scriptsize, align = center] {(d)};  
  \node[left = of img, node distance=0cm, rotate=90, anchor=center,yshift=-1.1cm,font=\scriptsize , align = center] {NHMI (Louvain)};
 \end{tikzpicture}
\end{minipage} 
\begin{minipage}{0.15\textwidth}
\begin{tikzpicture}
  \node (img)  {\includegraphics[scale=0.15]{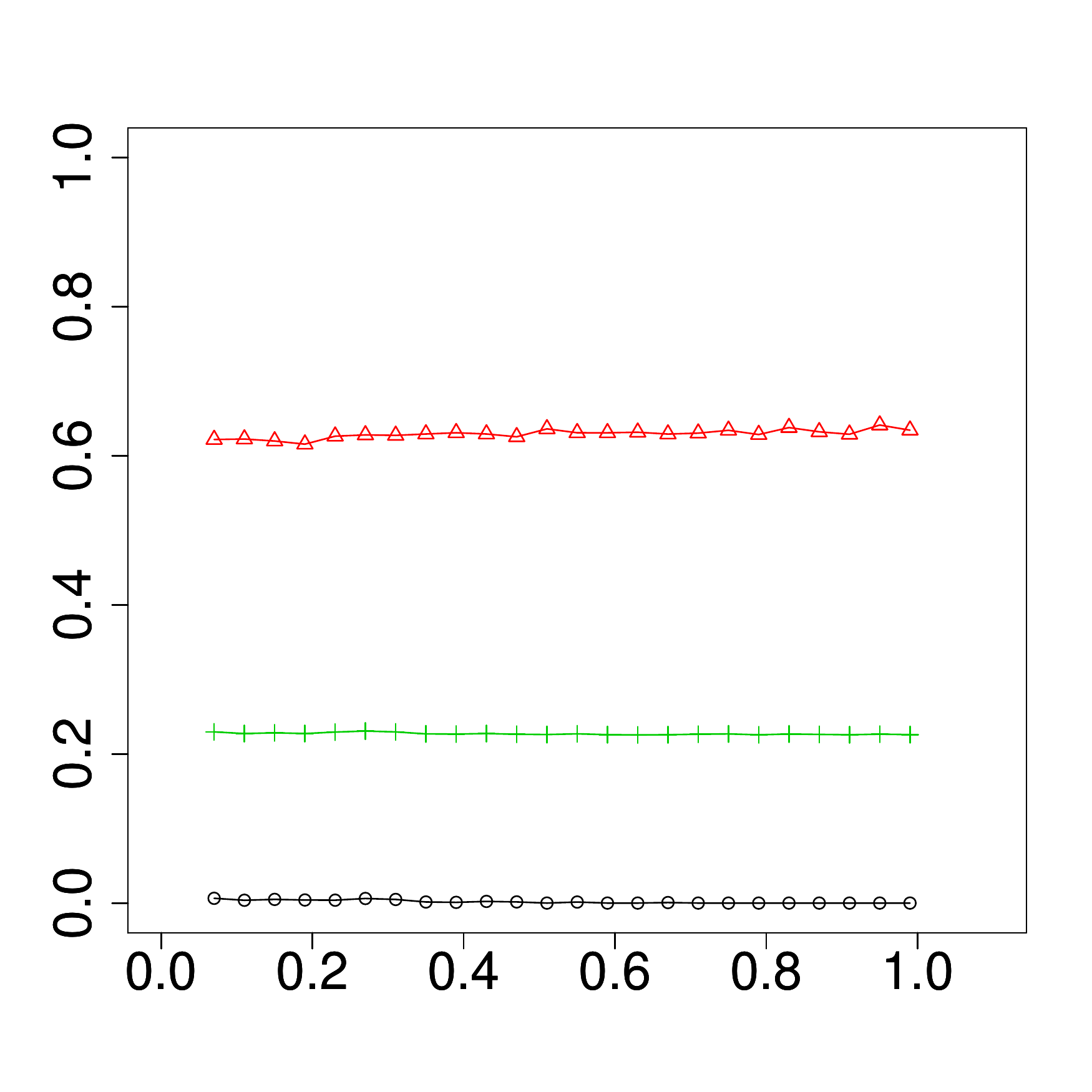}};
  \node[below = of img, node distance=0cm, xshift = 1.03cm, yshift=1.85cm,font=\scriptsize, align = center] {(e)};  
    \node[below = of img, node distance=0cm, yshift=1.4cm, font=\scriptsize, align = center] {$p$};
  \node[left = of img, node distance=0cm, rotate=90, anchor=center,yshift=-1.1cm,font=\scriptsize , align = center] {NHMI (Louvain)};
 \end{tikzpicture}
\end{minipage} 
\begin{minipage}{0.15\textwidth}
\begin{tikzpicture}
  \node (img)  {\includegraphics[scale=0.15]{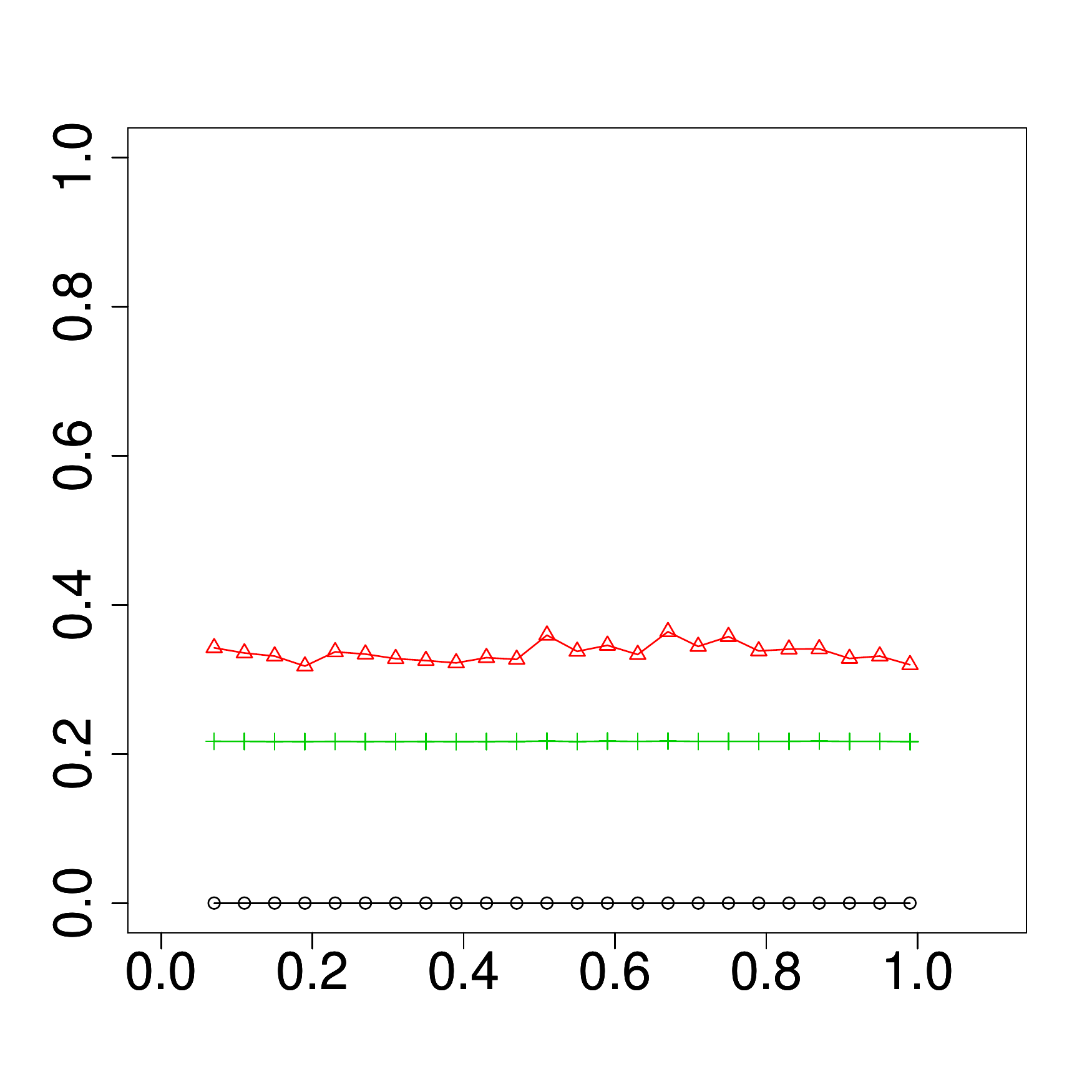}};
  \node[below = of img, node distance=0cm, xshift = 1.03cm, yshift=1.85cm,font=\scriptsize, align = center] {(f)};  
  \node[left = of img, node distance=0cm, rotate=90, anchor=center,yshift=-1.1cm,font=\scriptsize , align = center] {NHMI (Louvain)};
 \end{tikzpicture}
\end{minipage} 

\vspace{-0.5cm}

\begin{minipage}{0.15\textwidth}
\begin{tikzpicture}
  \node (img)  {\includegraphics[scale=0.15]{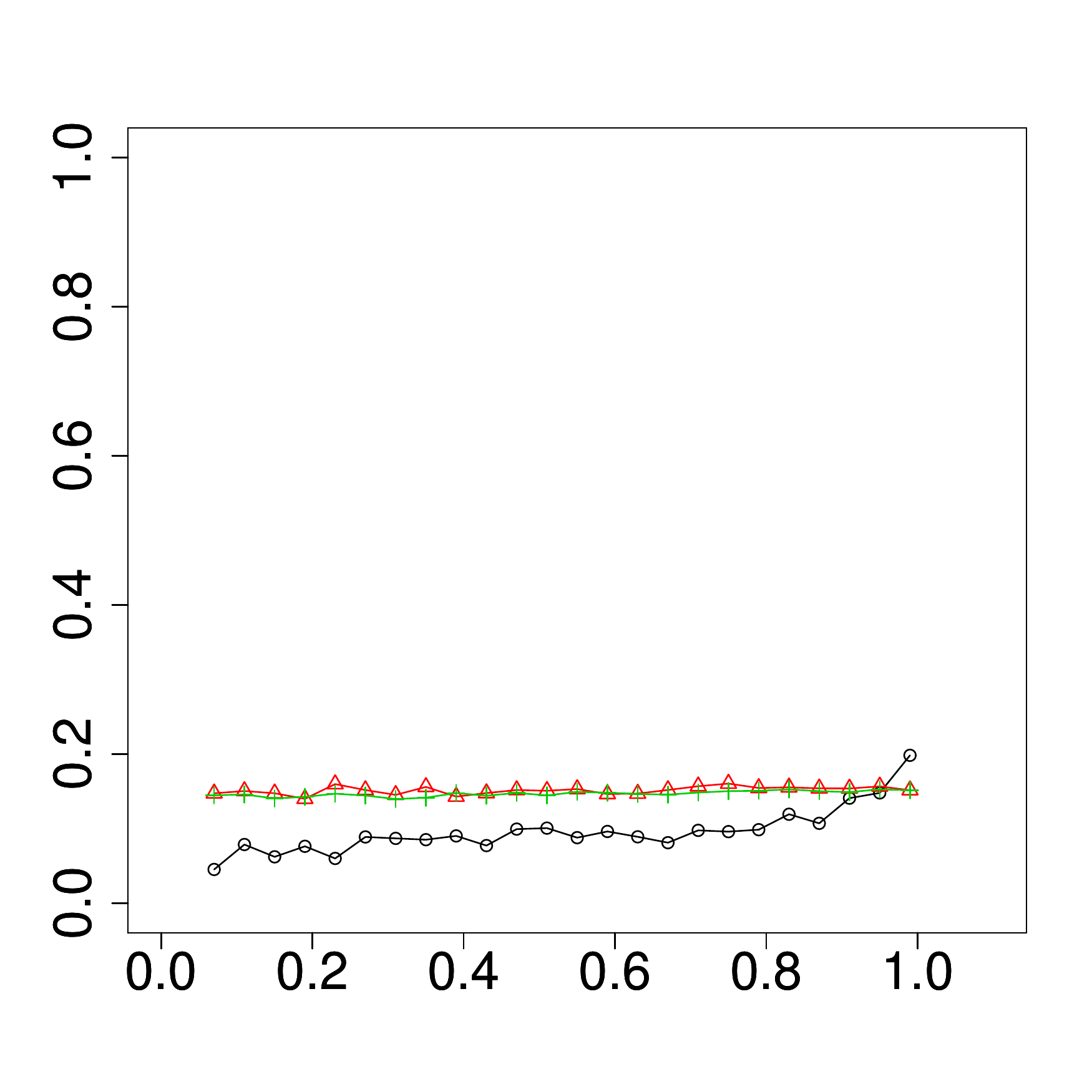}};
  \node[below = of img, node distance=0cm, xshift = 1.03cm, yshift=1.85cm,font=\scriptsize, align = center] {(g)};  
  \node[left = of img, node distance=0cm, rotate=90, anchor=center,yshift=-1.1cm,font=\scriptsize , align = center] {NHMI (HSBM)};
 \end{tikzpicture}
\end{minipage} 
\begin{minipage}{0.15\textwidth}
\begin{tikzpicture}
  \node (img)  {\includegraphics[scale=0.15]{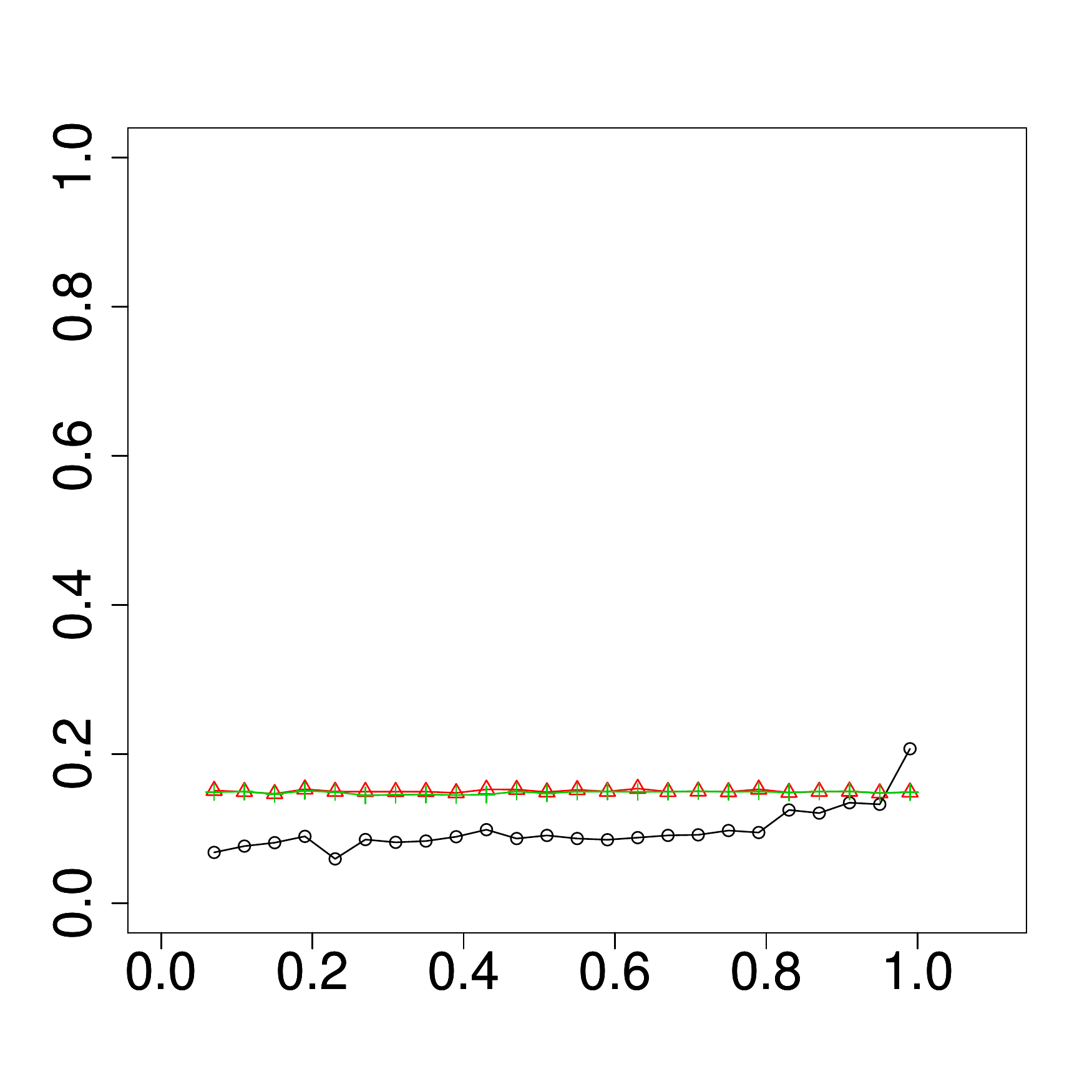}};
  \node[below = of img, node distance=0cm, xshift = 1.03cm, yshift=1.85cm,font=\scriptsize, align = center] {(h)};  
    \node[below = of img, node distance=0cm, yshift=1.4cm, font=\scriptsize, align = center] {$p$};
  \node[left = of img, node distance=0cm, rotate=90, anchor=center,yshift=-1.1cm,font=\scriptsize , align = center] {NHMI (HSBM)};
 \end{tikzpicture}
\end{minipage} 
\begin{minipage}{0.15\textwidth}
\begin{tikzpicture}
  \node (img)  {\includegraphics[scale=0.15]{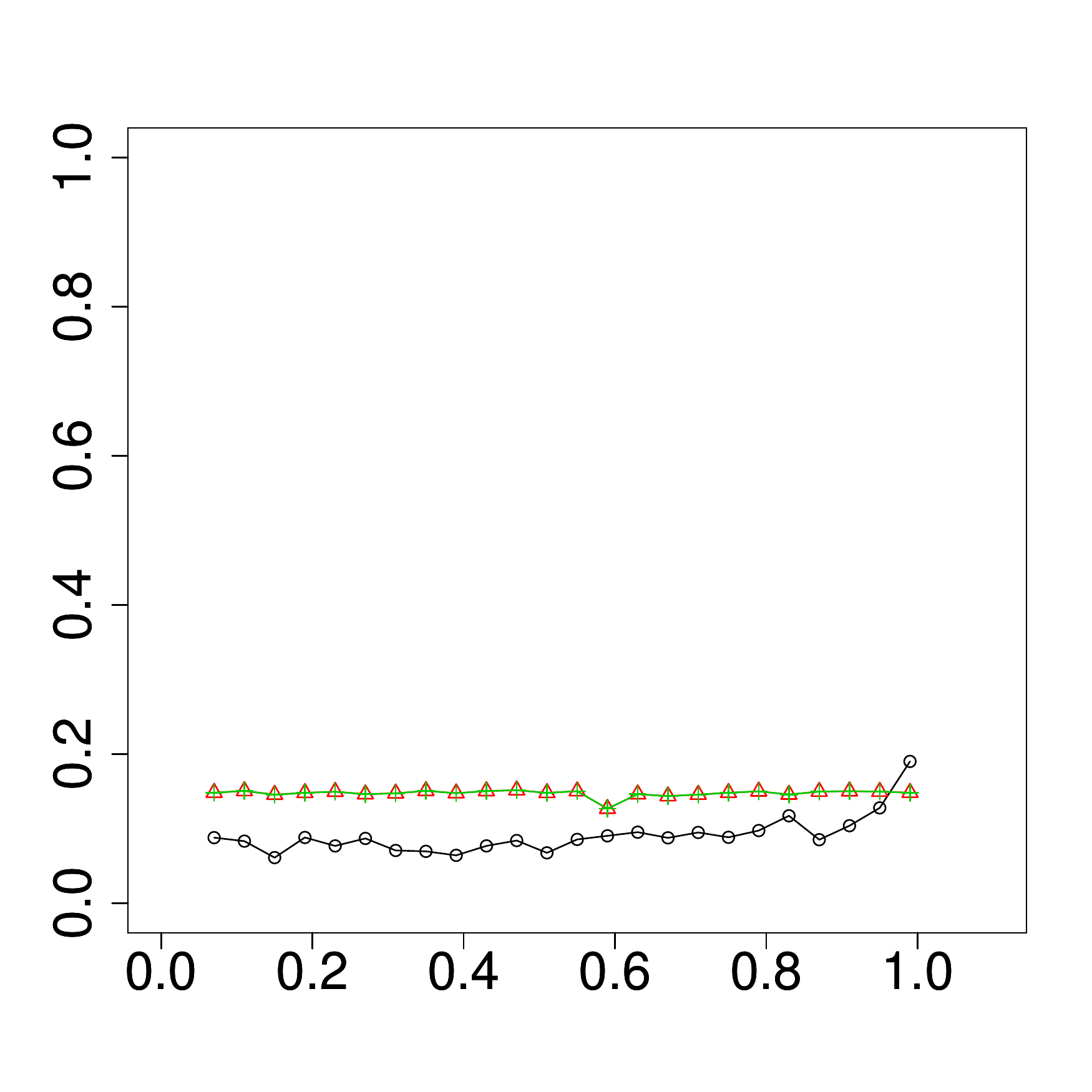}};
  \node[below = of img, node distance=0cm, xshift = 1.03cm, yshift=1.85cm,font=\scriptsize, align = center] {(i)};  
  \node[left = of img, node distance=0cm, rotate=90, anchor=center,yshift=-1.1cm,font=\scriptsize , align = center] {NHMI (HSBM)};
 \end{tikzpicture}
\end{minipage} 
\caption{Average NHMI as a function of the complementary mixing parameter, $p$. From left to right, the mixing parameters are $\mu = 0.05$, $0.3$ and $0.7$, respectively. From top to bottom, the methods are Infomap, Louvain, and HSBM. Averages are computed over 10 different network realizations with the same set of parameters of the seed LFR benchmark. The parameters of the seed networks can be found in Table \ref{table1}}
\label{figure8}
\end{figure}
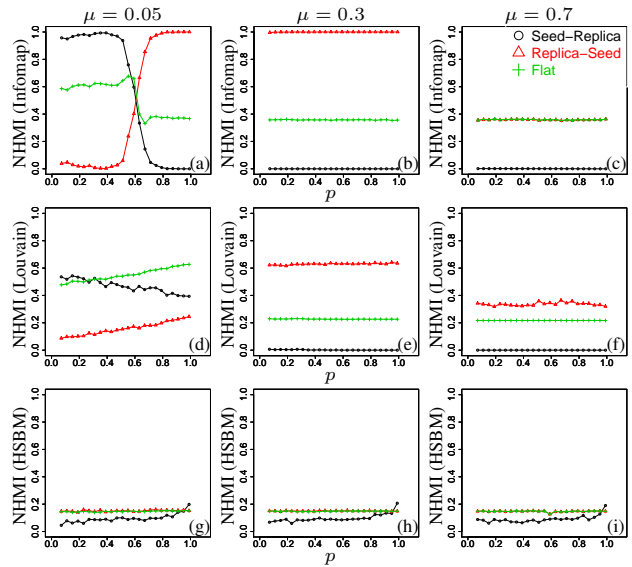

\paragraph{Decimated inter-layer connections}
So far we have considered a highly stylized model where the communities in the seed network are deterministically replicated in deeper layers. In this subsection, we relax this assumption. We note that in these less stylized cases, all the nodes would have more links to their own communities, such that the topologies of the networks would remain the same. 
With this in mind, we introduce a parameter $p$. It specifies the probability of randomly removing connections between the seed communities and the replicas (Fig.~\ref{figure1}d). The decimation procedure associated to $p$ is applied to every pair of seed--replica communities, independently. 
In this way, $p = 0$ means that all connections are kept (the case studied in the previous subsection) and $p = 1$ means  all connections are removed. 
Hence, $p$ is a sort of complementary mixing parameter; while $\mu$ controls the connectivity at the LFR level, $p$ controls the connectivity at the inter-layer level.
We study the accuracy of the community detection methods by plotting the NHMI as a function of $p$. We repeat calculations for three different values of the mixing parameter, $\mu = 0.05$, $0.3$ and $0.7$, i.e.~they represent the three qualitatively different regions for the mixing parameter found in the previous results. 
The findings are shown in Figure \ref{figure8}. 
In Fig.~\ref{figure8}a, a transition between the  two seed-replica and replica-seeds ground truths is observed as $p$ is varied.
This is analogous to what is observed in Fig.~\ref{figure6}a when $\mu$ is varied. 
In other words, the previous result confirm the role of $p$ as a complementary mixing parameter.
The rest of the panels in Fig.~\ref{figure8} essentially show that, when the mixing parameter is large, the number of connections between communities and their replicas is already very small and $p$ cannot have a significant impact on the detected structure. 
Overall, we can conclude that the RB-LFR benchmark graphs are relatively robust to random removal of some connections, a desirable characteristic for a well defined ensemble of benchmark graphs.
Importantly, only the Infomap algorithm is able to unveil such topological transition induced by $p$.
From now on, $p=0$. 

Since the previous results show that Infomap performs well and, in some cases, considerably better than the other options, in what follows we restrict our analysis presenting the results obtained with Infomap, only.

\begin{figure}[h]
\begin{minipage}{0.15\textwidth}
\begin{tikzpicture}
  \node (img)  {\includegraphics[scale=0.15]{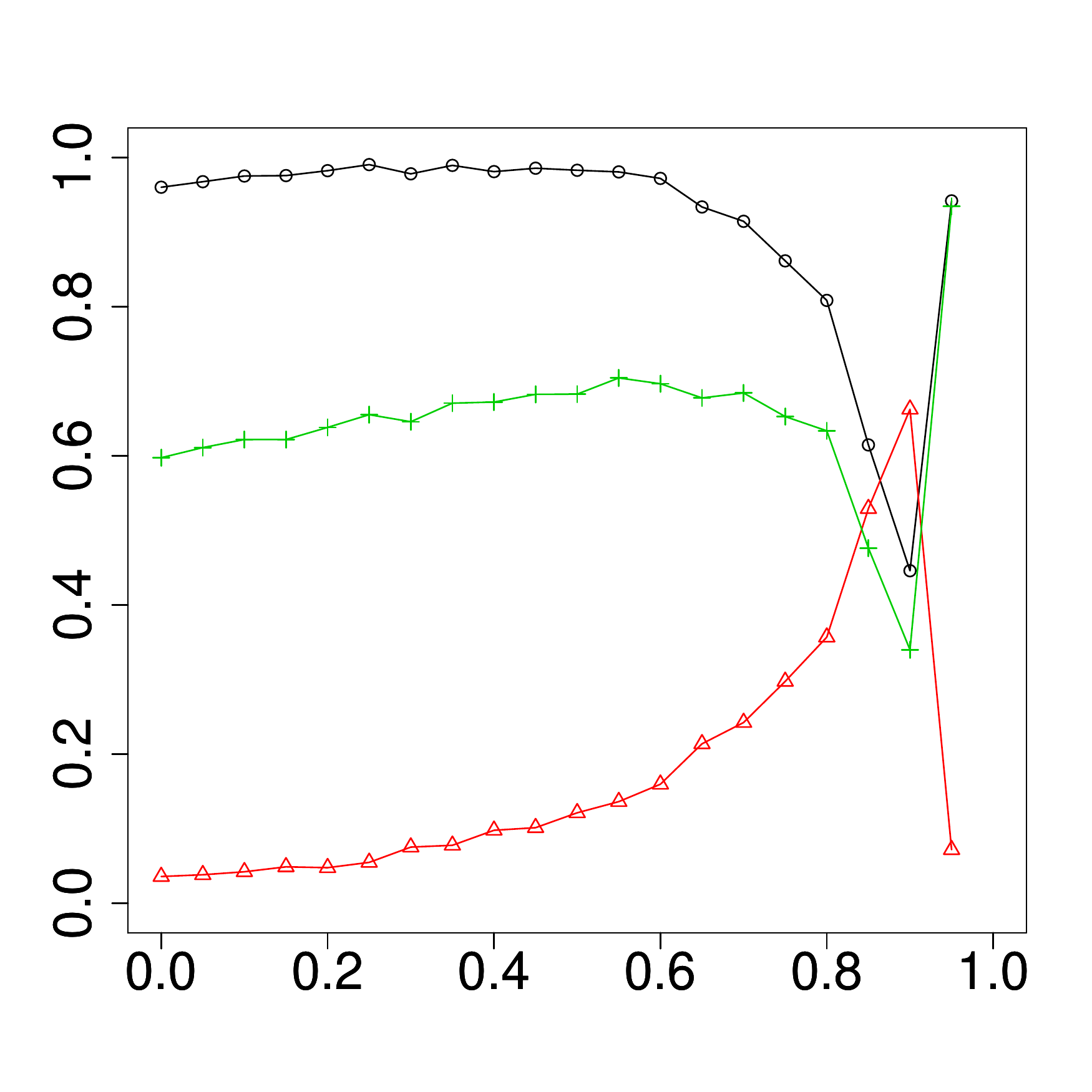}};
\node[below = of img, node distance=0cm,xshift = 1.03cm, yshift=1.85cm,font=\scriptsize, align = center] {(a)};
    \node[above = of img, node distance=0cm, yshift=-1.5cm, font=\scriptsize, align = center] {$\mu = 0.05$};  
  \node[left = of img, node distance=0cm, rotate=90, anchor=center,yshift=-1.1cm,font=\scriptsize , align = center] {NHMI (Infomap)};
 \end{tikzpicture}
\end{minipage} 
\begin{minipage}{0.15\textwidth}
\begin{tikzpicture}
  \node (img)  {\includegraphics[scale=0.15]{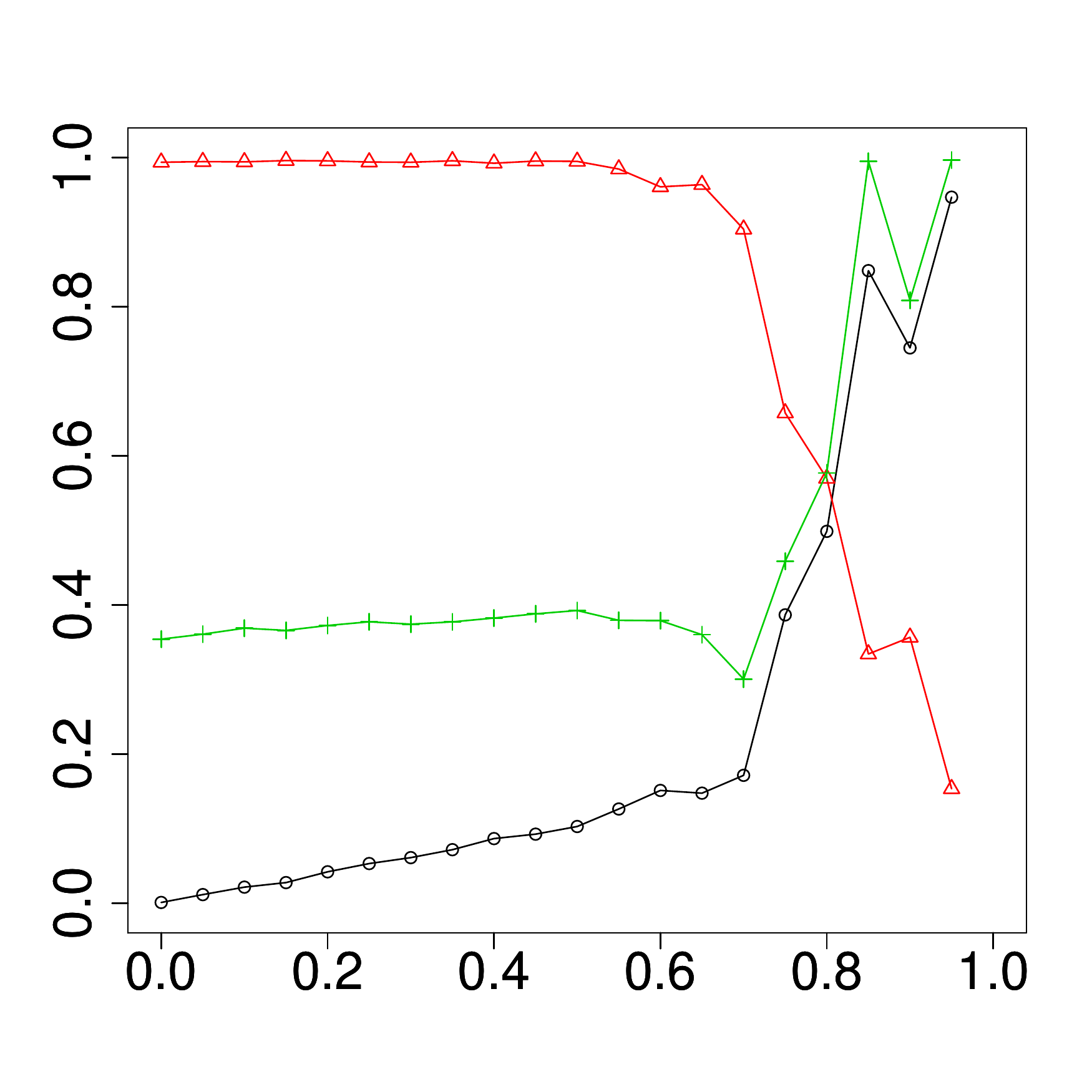}};
\node[below = of img, node distance=0cm,xshift = 1.03cm, yshift=1.85cm,font=\scriptsize, align = center] {(b)};
    \node[above = of img, node distance=0cm, yshift=-1.5cm, font=\scriptsize, align = center] {$\mu = 0.3$};  
\node[below = of img, node distance=0cm, yshift=1.4cm, font=\scriptsize, align = center] {$q$};
  \node[left = of img, node distance=0cm, rotate=90, anchor=center,yshift=-1.1cm,font=\scriptsize , align = center] {NHMI (Infomap)};
 \end{tikzpicture}
\end{minipage} 
\begin{minipage}{0.15\textwidth}
\begin{tikzpicture}
  \node (img)  {\includegraphics[scale=0.15]{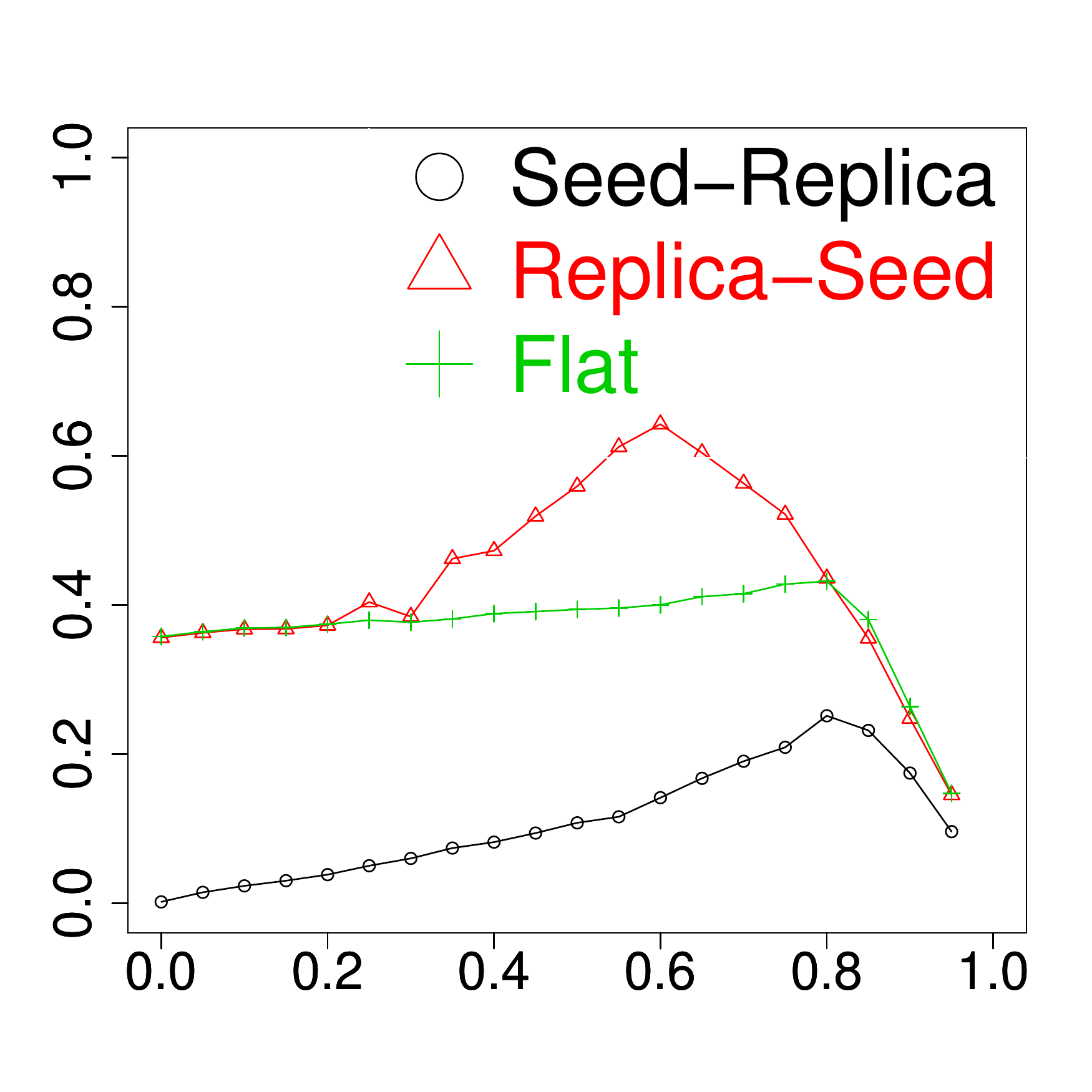}};
\node[below = of img, node distance=0cm,xshift = 1.03cm, yshift=1.85cm,font=\scriptsize, align = center] {(c)};
    \node[above = of img, node distance=0cm, yshift=-1.5cm, font=\scriptsize, align = center] {$\mu = 0.7$};  
  \node[left = of img, node distance=0cm, rotate=90, anchor=center,yshift=-1.1cm,font=\scriptsize , align = center] {NHMI (Infomap)};
 \end{tikzpicture}
\end{minipage} 

\vspace{-0.5cm}

\begin{minipage}{0.15\textwidth}
\begin{tikzpicture}
  \node (img)  {\includegraphics[scale=0.15]{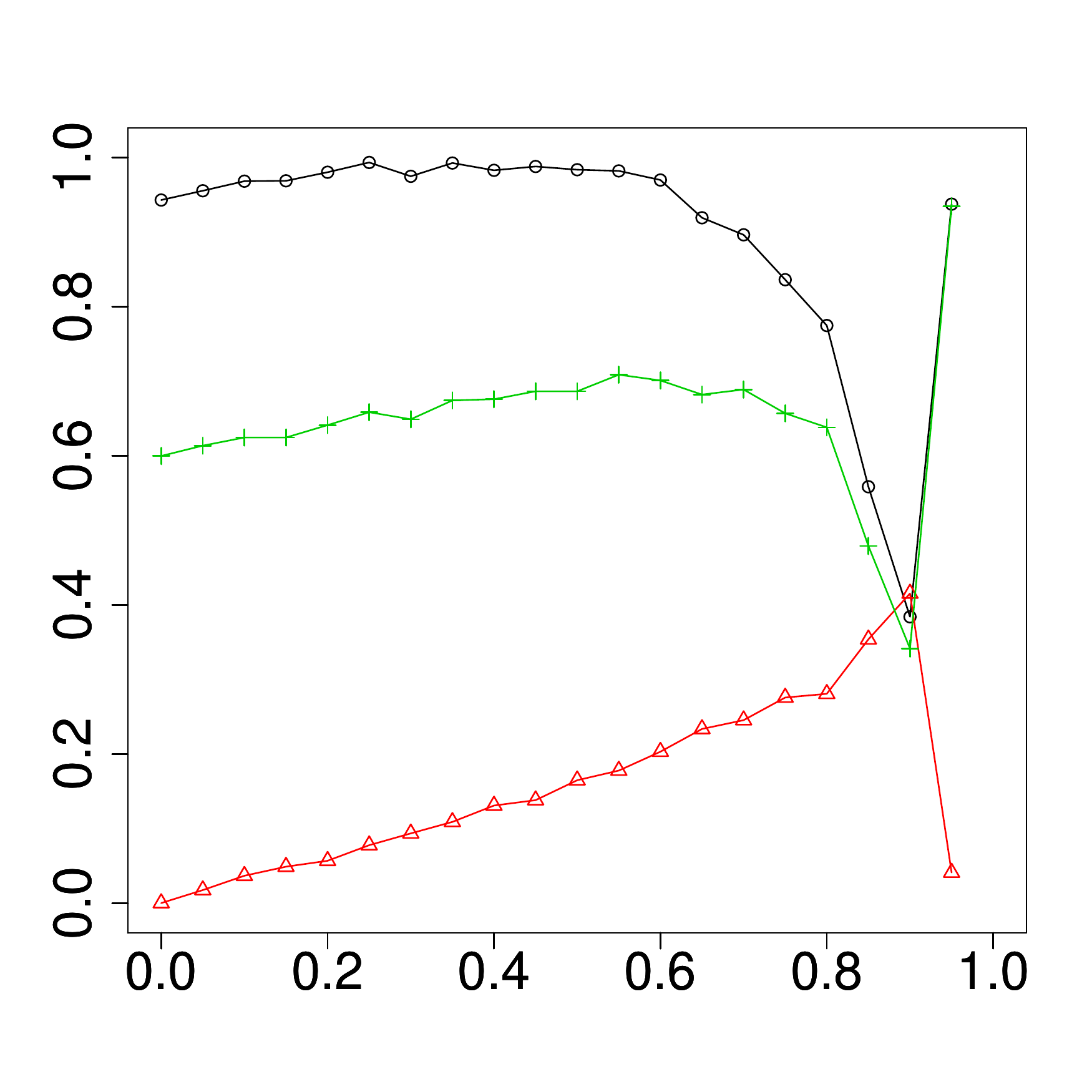}};
\node[below = of img, node distance=0cm,xshift = 1.03cm, yshift=1.85cm,font=\scriptsize, align = center] {(d)};
  \node[left = of img, node distance=0cm, rotate=90, anchor=center,yshift=-1.1cm,font=\scriptsize , align = center] {NMI (Infomap)};
 \end{tikzpicture}
\end{minipage} 
\begin{minipage}{0.15\textwidth}
\begin{tikzpicture}
 \node (img)  {\includegraphics[scale=0.15]{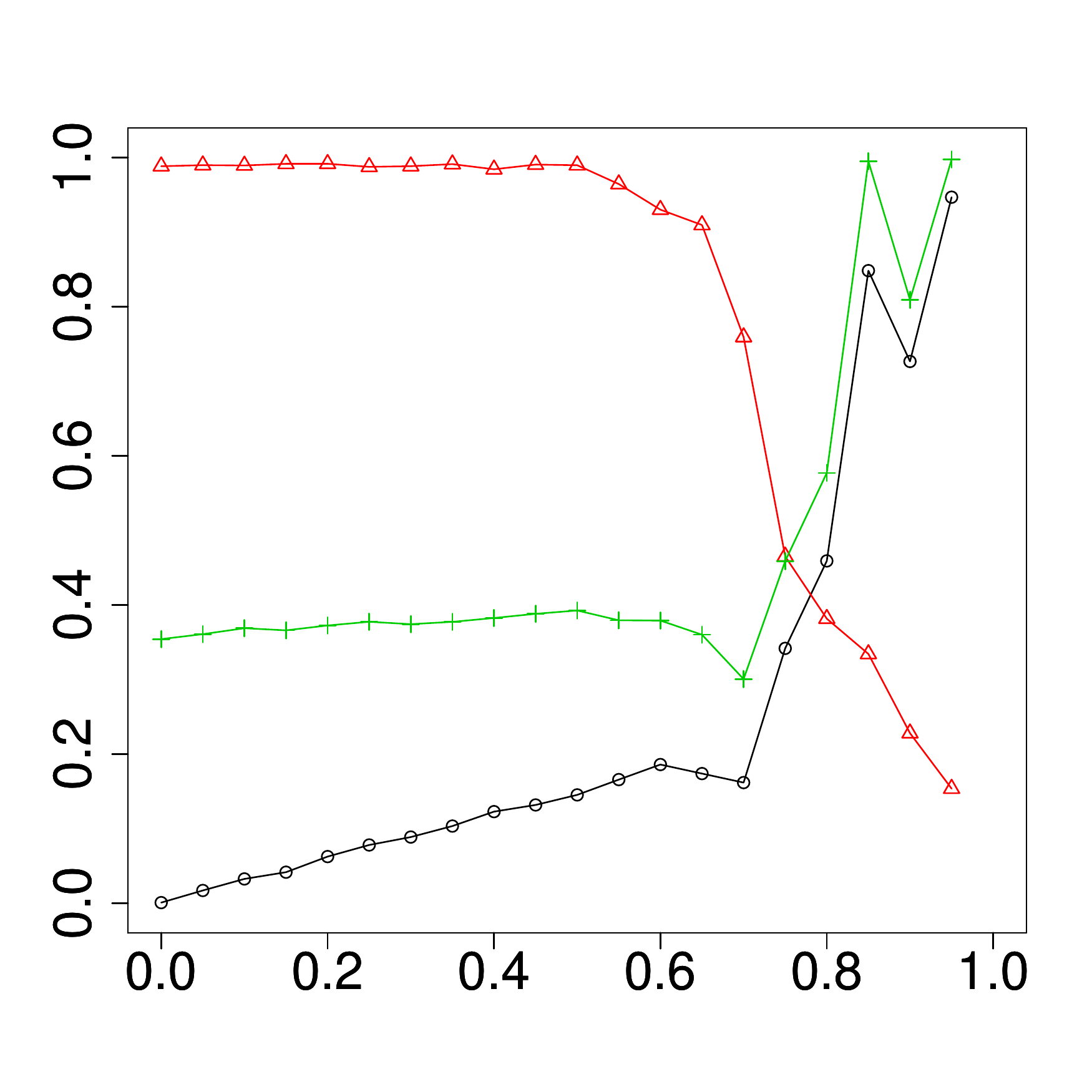}};
\node[below = of img, node distance=0cm,xshift = 1.03cm, yshift=1.85cm,font=\scriptsize, align = center] {(e)};
\node[below = of img, node distance=0cm, yshift=1.4cm, font=\scriptsize, align = center] {$q$};
  \node[left = of img, node distance=0cm, rotate=90, anchor=center,yshift=-1.1cm,font=\scriptsize , align = center] {NMI (Infomap)};
 \end{tikzpicture}
\end{minipage} 
\begin{minipage}{0.15\textwidth}
\begin{tikzpicture}
  \node (img)  {\includegraphics[scale=0.15]{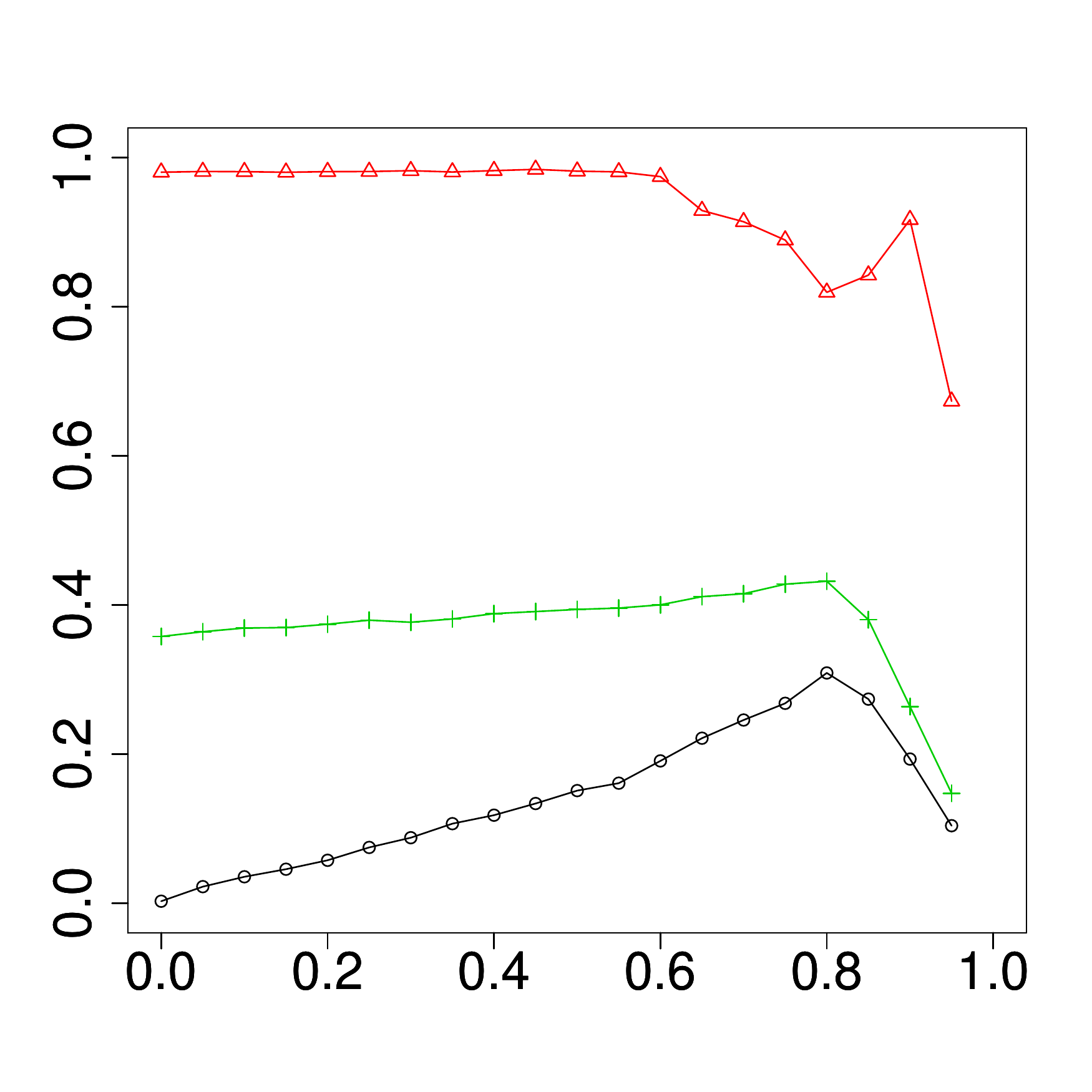}};
\node[below = of img, node distance=0cm,xshift = 1.03cm, yshift=1.85cm,font=\scriptsize, align = center] {(f)};
  \node[left = of img, node distance=0cm, rotate=90, anchor=center,yshift=-1.1cm,font=\scriptsize , align = center] {NMI (Infomap)};
 \end{tikzpicture}
\end{minipage} 
\caption{Average NHMI and NMI (top and bottom, respectively) as a function of $q$, the fraction of replica communities removed from a standard RB-LFR benchmark. From left to right, the mixing parameters is set to $\mu=0.05$, $0.3$ and $0.7$, respectively. Averages are computed over 10 different network realizations with the same set of parameters of the seed LFR benchmark. The parameters of the seed networks can be found in Table \ref{table1}.}
\label{figure11}
\end{figure}
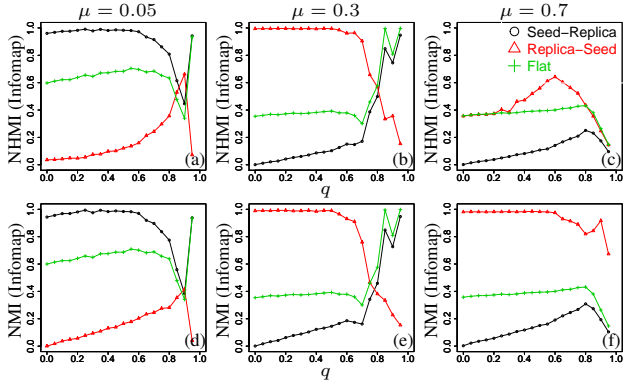

\paragraph{Decimation of replicas}
We now randomly remove a fraction $q$ of the existing replicas---together with all their connections---from a previously generated RB-LFR benchmark graph. 
For $q = 0$ all the replica communities are kept while for $q = 1$ all of them are removed. As before, we use $\mu$ = 0.05, 0.3, \& 0.7 to represent three different regions of the mixing parameter. The results indicate that, in all cases, the RB-LFR benchmark graphs still preserves a relatively stable hierarchical structure even after 60\% of the replicated communities have been removed (Fig.~\ref{figure11}). 
From now on, $q=0$. 

\paragraph{Network sizes}
Then, we have measured the effect of network size on the performance of Infomap, observing that the accuracy of the method mildly decreases as the number of nodes $N_0$ increases.
It only has a mensurable effect when for $\mu \to 0$. 

\paragraph{Number of replicas} In the end, we studied the effect of the number of replicas on the performance of Infomap (going from $R=4$ to $R=9$). We observe that the range of the mixing parameter $\mu$ where the transition between ground truths occur, becomes slightly wider.
Overall, we conclude that the results are robust to variations of the number of replicas.

\subsection{Test on three-level RB-LFR benchmark}

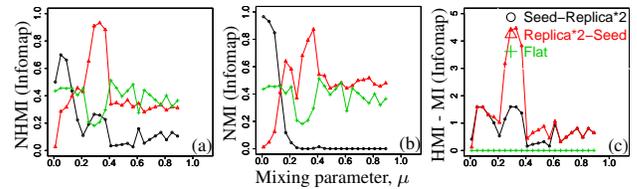
\begin{figure}[h]
\begin{minipage}{0.15\textwidth}
\begin{tikzpicture}
  \node (img)  {\includegraphics[scale=0.15]{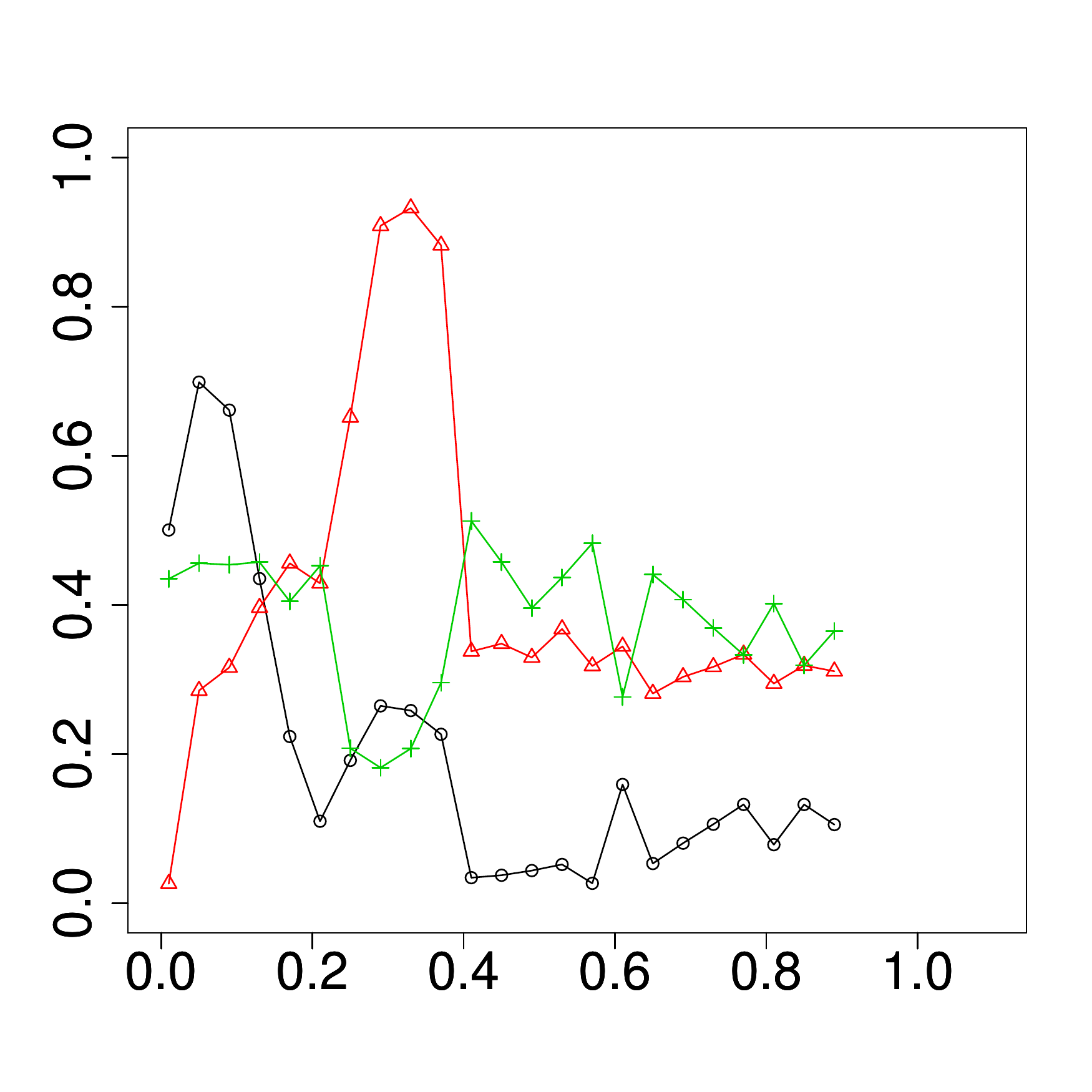}};
\node[below = of img, node distance=0cm,xshift = 1.03cm, yshift=1.85cm,font=\scriptsize, align = center] {(a)};
  \node[left = of img, node distance=0cm, rotate=90, anchor=center,yshift=-1.1cm,font=\scriptsize , align = center] {NHMI (Infomap)};
 \end{tikzpicture}
\end{minipage} 
\begin{minipage}{0.15\textwidth}
\begin{tikzpicture}
  \node (img)  {\includegraphics[scale=0.15]{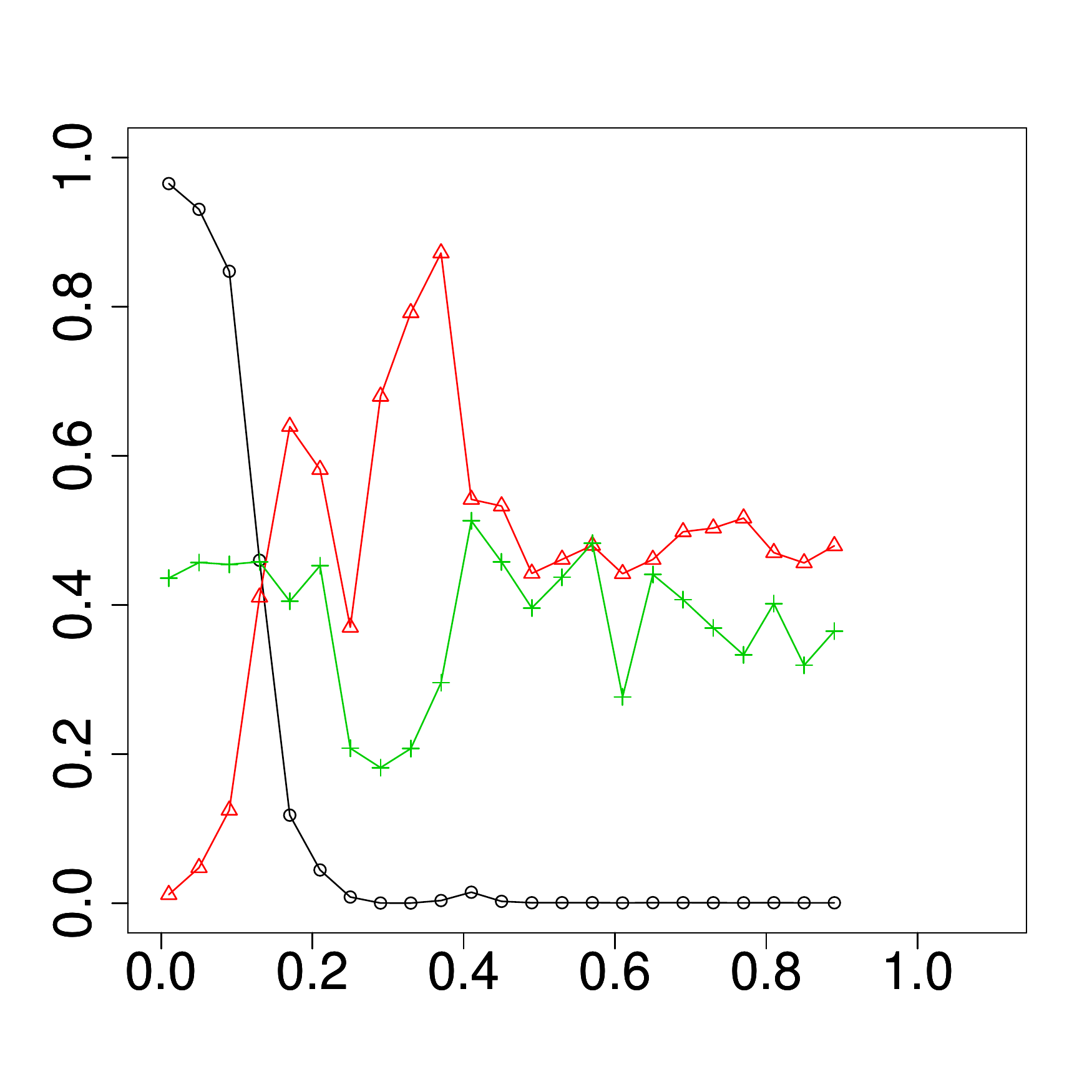}};
\node[below = of img, node distance=0cm,xshift = 1.03cm, yshift=1.85cm,font=\scriptsize, align = center] {(b)};
\node[below = of img, node distance=0cm, yshift=1.4cm, font=\scriptsize, align = center] {Mixing parameter, $\mu$};
  \node[left = of img, node distance=0cm, rotate=90, anchor=center,yshift=-1.1cm,font=\scriptsize , align = center] {NMI (Infomap)};
 \end{tikzpicture}
\end{minipage} 
\begin{minipage}{0.15\textwidth}
\begin{tikzpicture}
  \node (img)  {\includegraphics[scale=0.15]{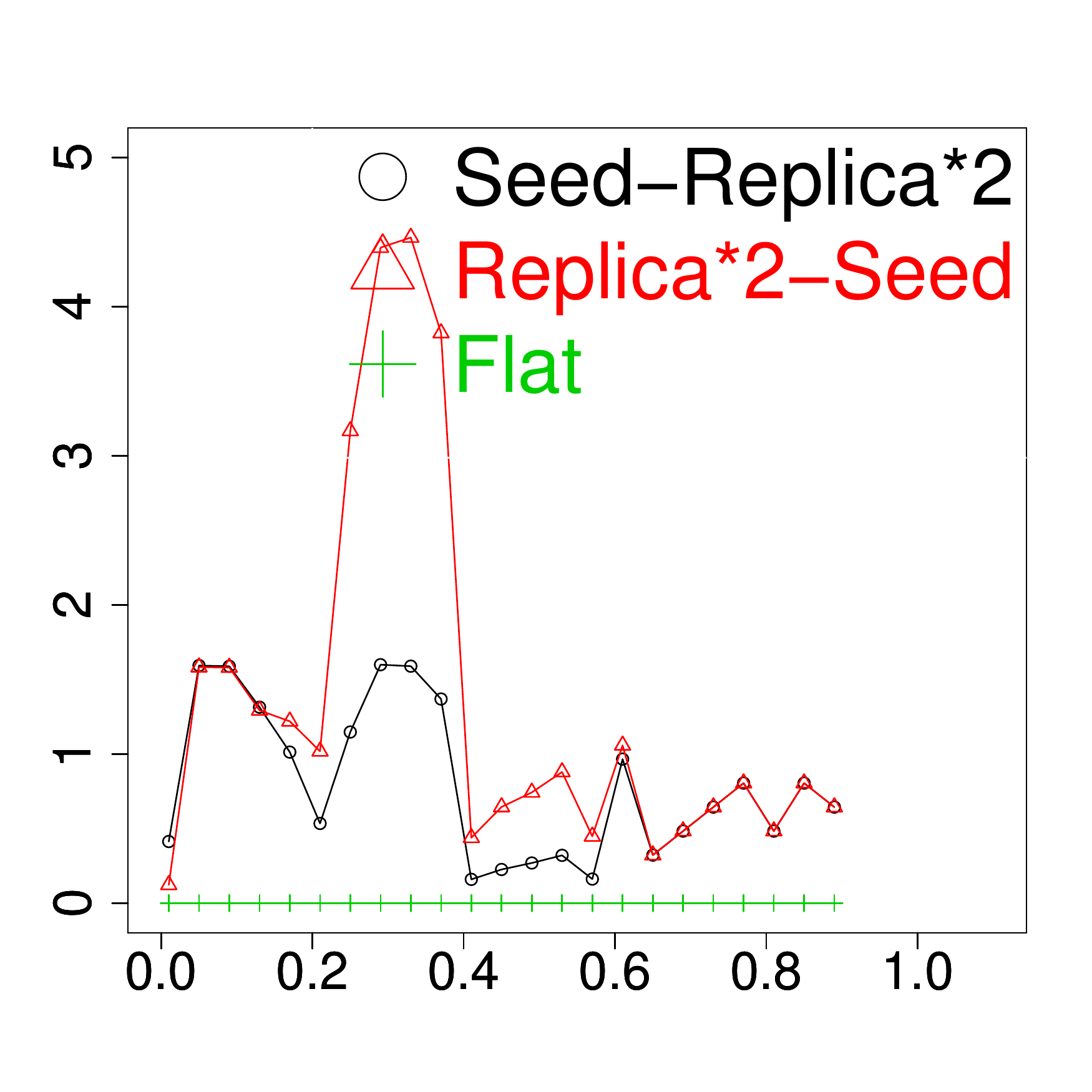}};
\node[below = of img, node distance=0cm,xshift = 1.03cm, yshift=1.85cm,font=\scriptsize, align = center] {(c)};
  \node[left = of img, node distance=0cm, rotate=90, anchor=center,yshift=-1.1cm,font=\scriptsize , align = center] {HMI - MI (Infomap)};
 \end{tikzpicture}
\end{minipage} 
\caption{Average NHMI, NMI, and (HMI - MI) as a function of the mixing parameter $\mu$, at the left, middle and right panels, respectively, for RB-LFR benchmark graphs with three levels. Averages are computed over 10 different network realizations with the same set of parameters of the seed LFR benchmark. The parameters of the seed networks can be found in Table \ref{table1}.}
\label{figure12}
\end{figure}

In the last study, we focus on the three-level RB-LFR benchmark. The setting is the same as those in the first study, i.e. Table~\ref{table1} and Fig.~\ref{figure6}. Under this setting, the first ground truth would be \textit{seed-replica-replica} (Seed-Replica*2), and the second ground truth becomes \textit{replica-replica-seed} (Replica*2-Seed), while the third one remains the same.
We report the accuracy of Infomap as a function of the mixing parameter, $\mu$. The results are shown in Fig.~\ref{figure12}. One could see that the three levels RB-LFR benchmark is a much harder test,  but still Infomap is able to unveil the network structure for certain values of the mixing parameter, $\mu$. 
On the other hand, the accuracies are much worse than those of the two-level benchmark graphs in most of the cases (see Figs.~\ref{figure6}a \& b for a comparison). In Fig.~\ref{figure12}c we show the difference between the full HMI and the MI of the first level. Similar to what we have observed in Fig.~\ref{figure6}c, the second and third levels contribute with an important fraction of the total value of the HMI. 

\begin{figure}[h]
\captionsetup[subfloat]{farskip=-2pt,captionskip=0pt}
\begin{center}
\subfloat[]{\includegraphics[width=.15\textwidth]{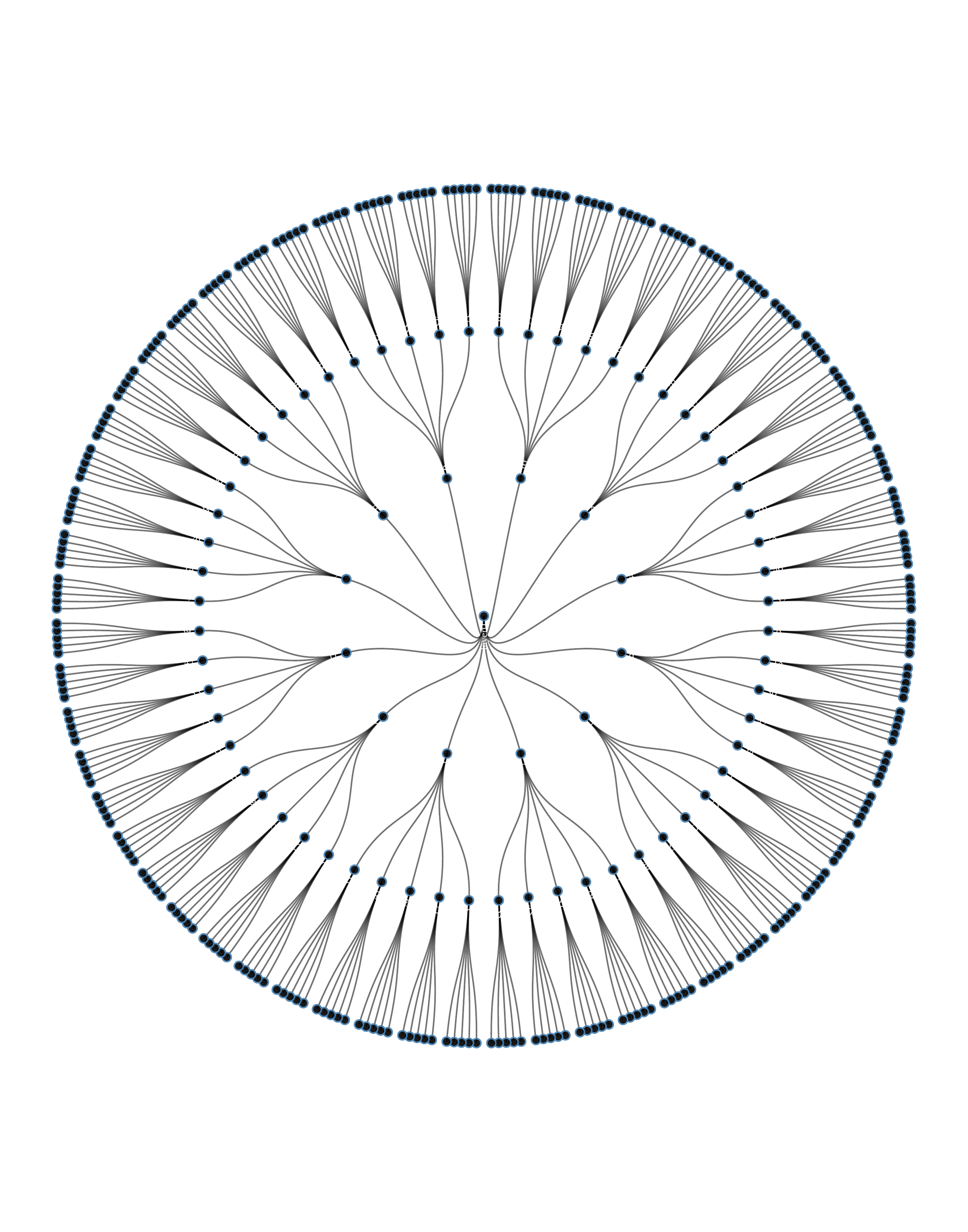}}
\subfloat[]{\includegraphics[width=.15\textwidth]{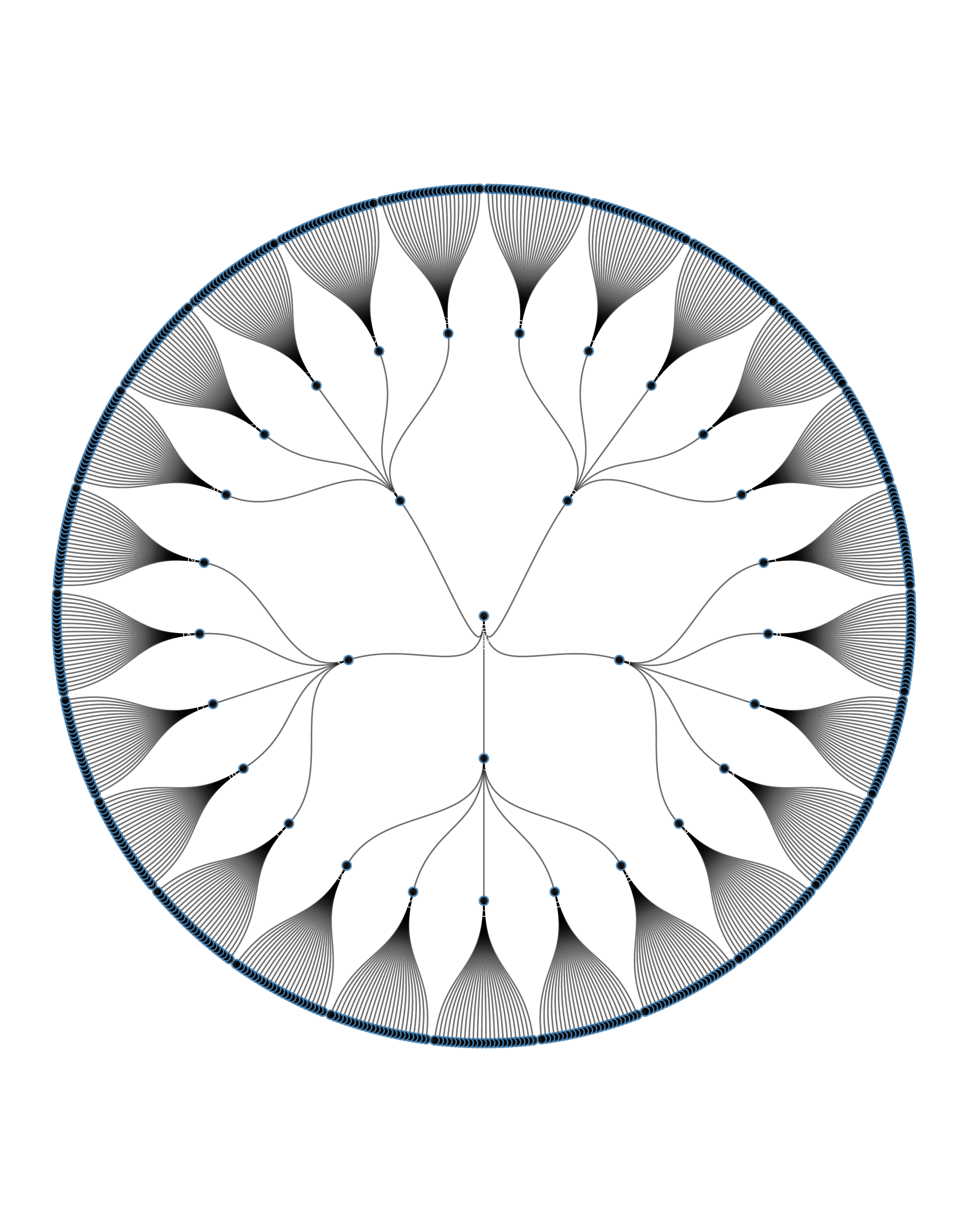}}
\subfloat[]{\includegraphics[width=.15\textwidth]{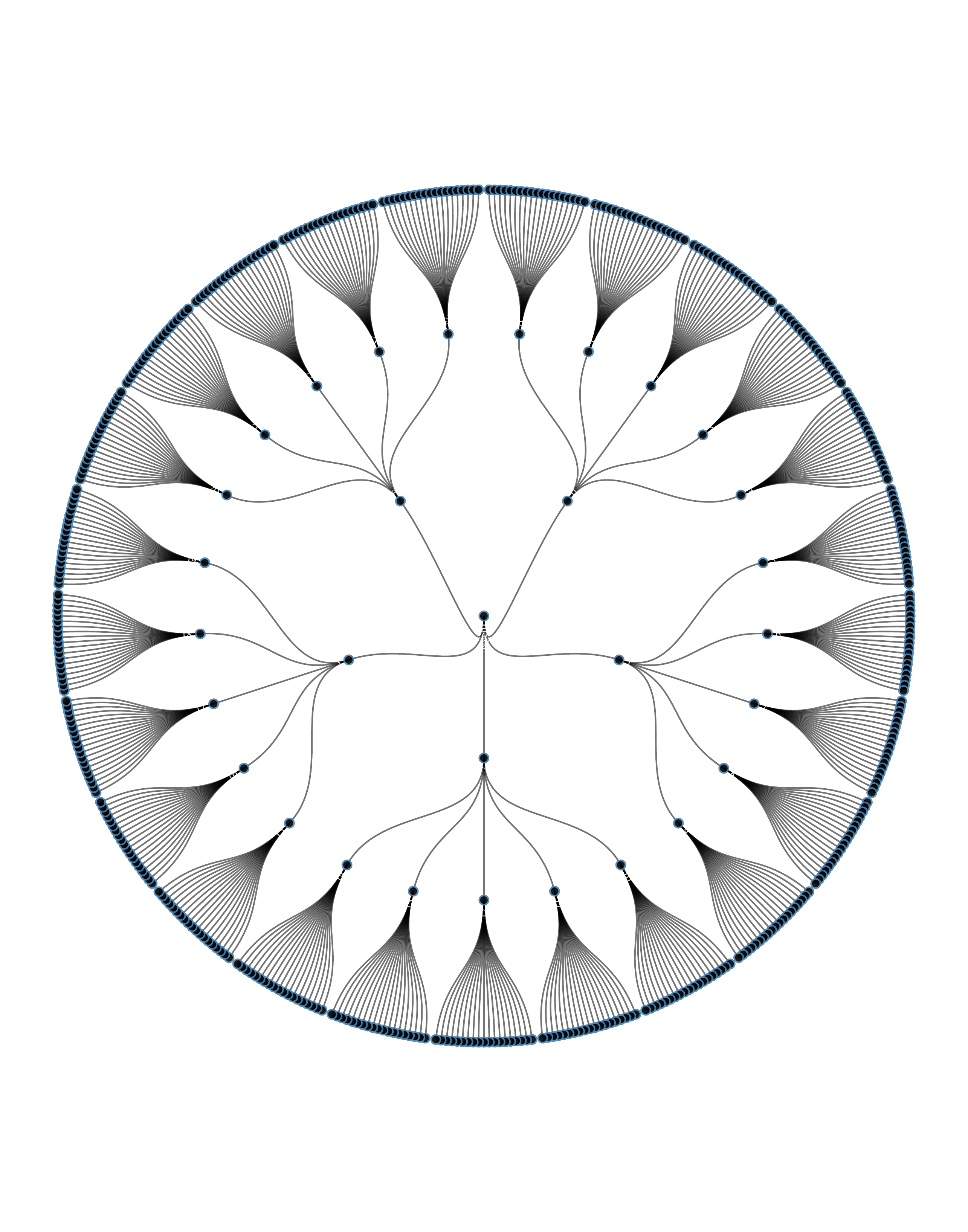}}\\
\subfloat[]{\includegraphics[width=.15\textwidth]{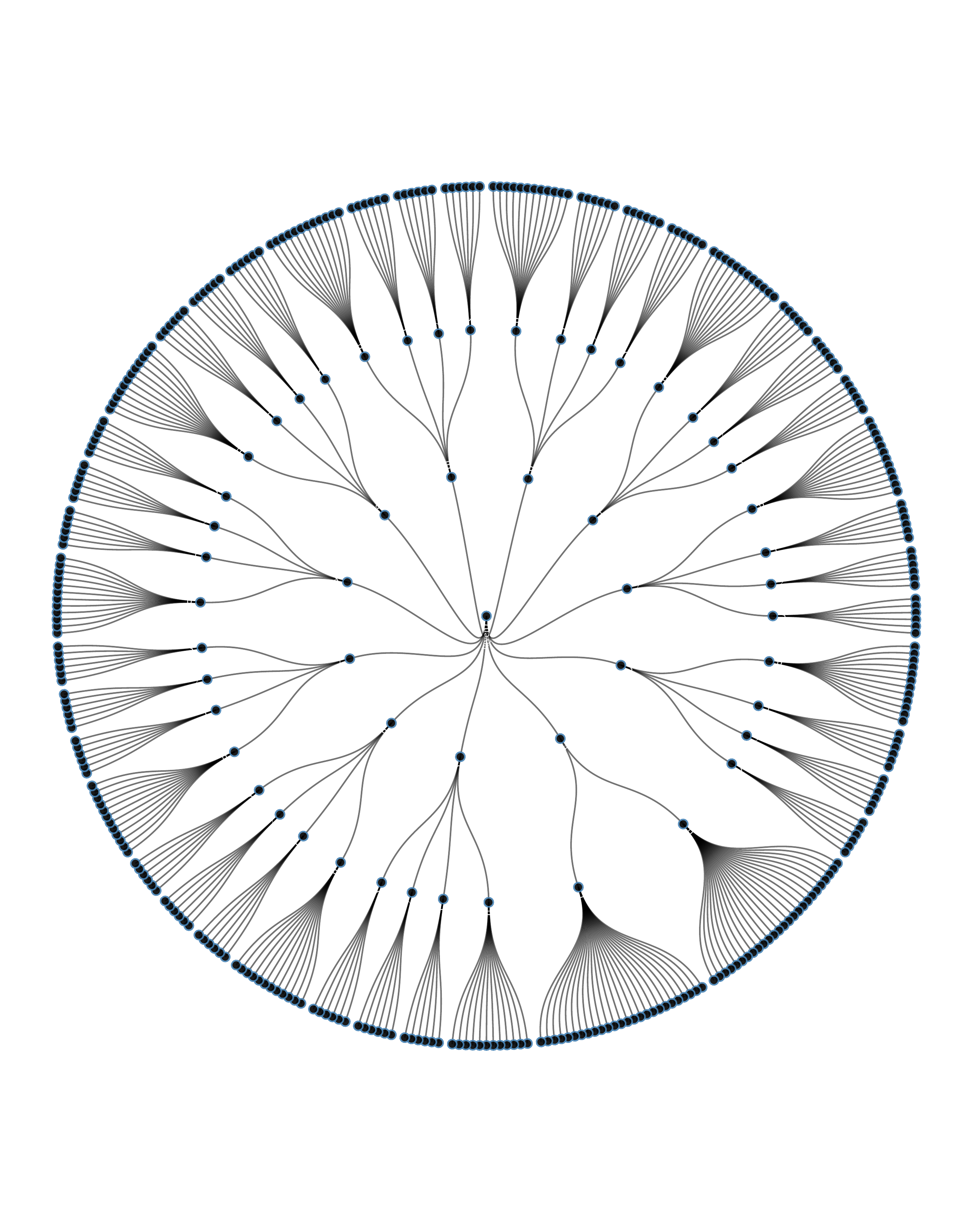}}
\subfloat[]{\includegraphics[width=.15\textwidth]{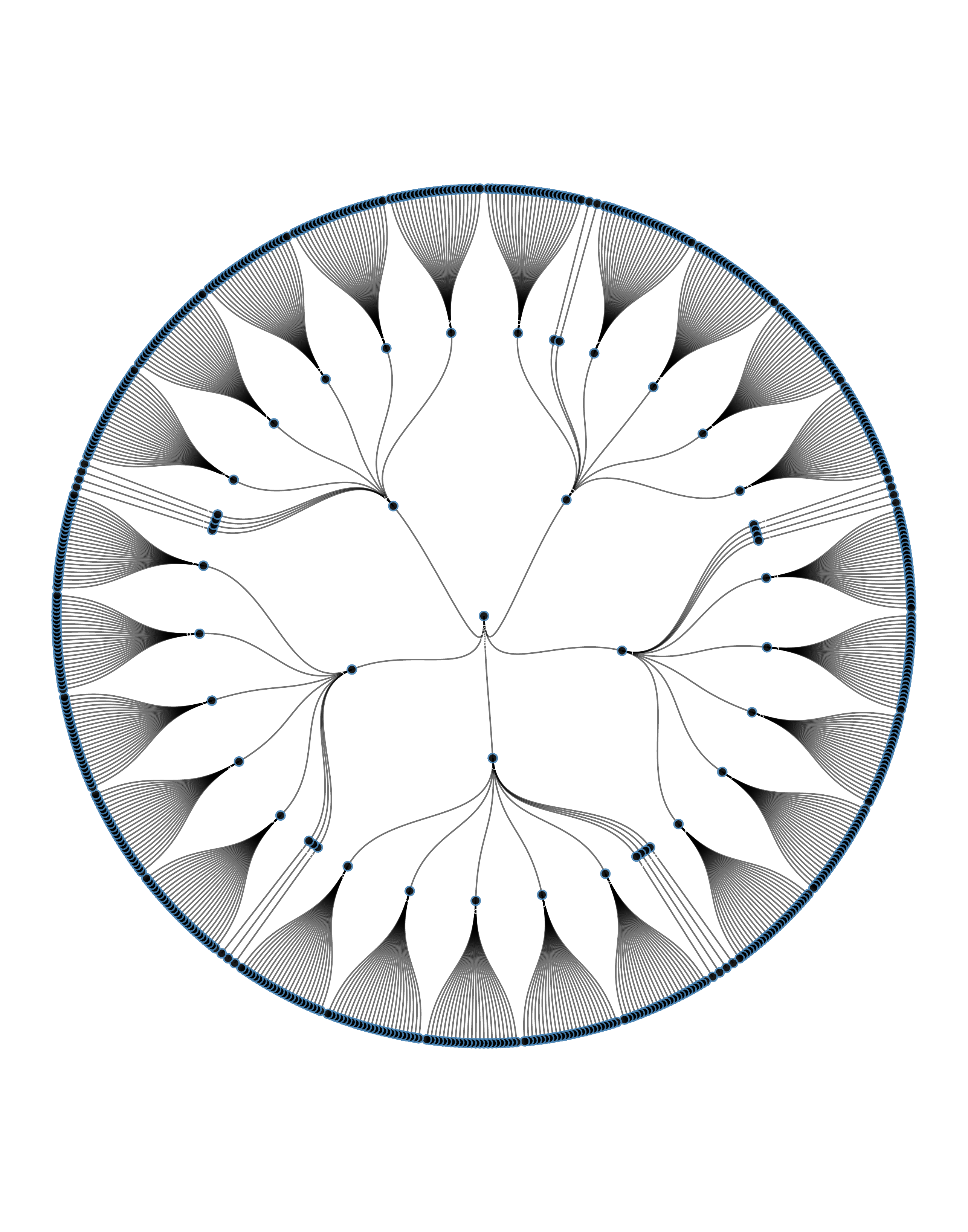}}
\subfloat[]{\includegraphics[width=.15\textwidth]{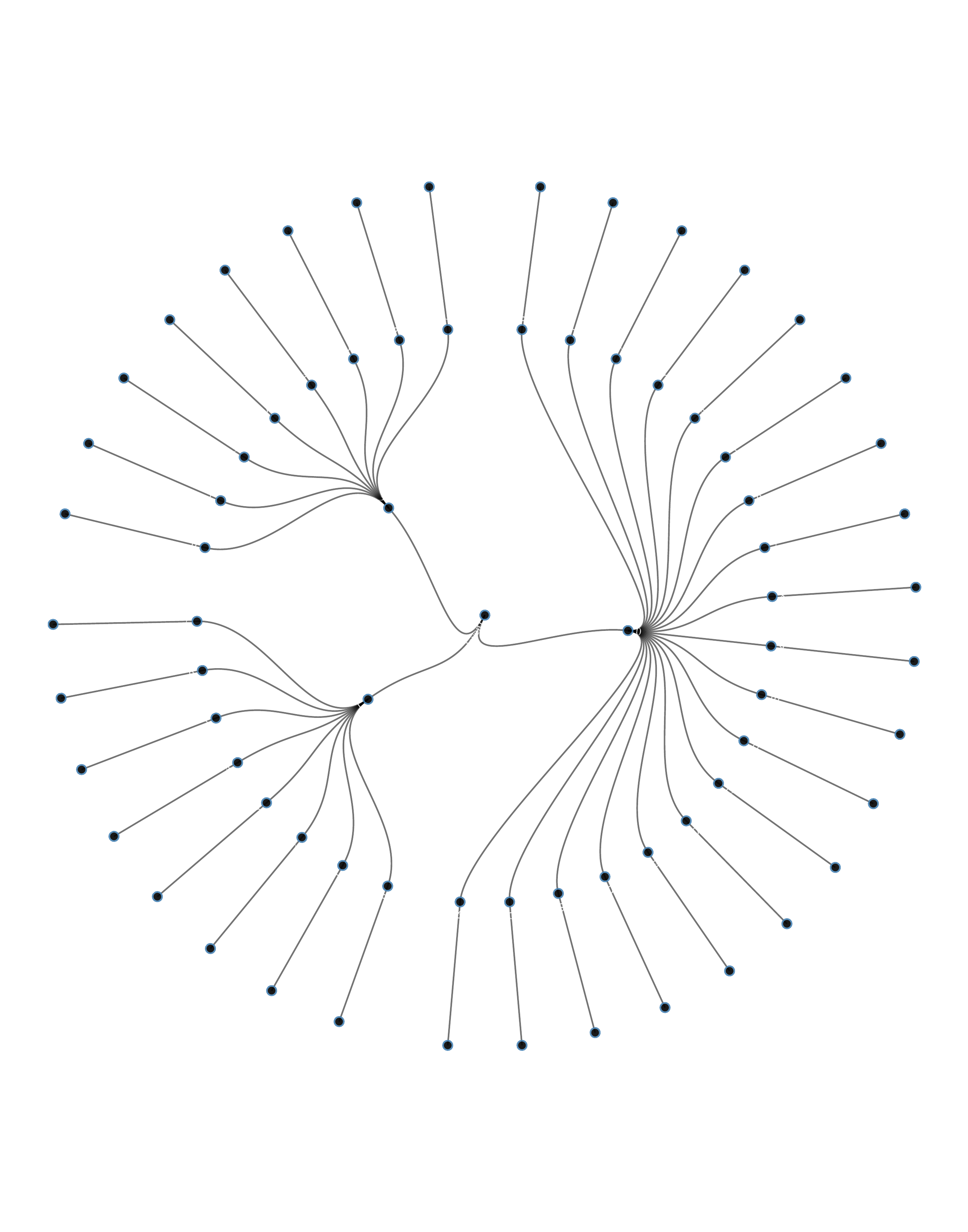}}
\end{center}
\caption{The top panels are the circular representation of the hierarchical structure of three-level RB-LFR benchmark graphs.
The bottom panels are the corresponding hierarchical structures detected by Infomap. 
In cases (a) and (d) the mixing parameters of the seed LFR benchmark is $\mu=0.01$, in cases (b) and (e) $\mu=0.33$ and in cases (c) and (f) $\mu=0.77$. The center of every panel represents the whole network at level 0.
Similar to the $2^{\mathrm{nd}}$ and last level of the two-level RB-LFR, the $3^{\mathrm{rd}}$ and last level of the three-level RB-LFR represents communities that do not contain any sub-communities [see Fig.~3c \& d]. 
} 
\label{figure13}
\end{figure}

Finally, in Fig.~\ref{figure13}, we provide three examples of the ground truth hierarchical structure of different RB-LFR benchmark graphs (top panels) and corresponding hierarchical structures detected by Infomap (bottom panels). The mixing parameters, $\mu$, are 0.01, 0.33, and 0.77 from left to right. 

Panel (a) corresponds to the first type of ground truth. In this case, the mixing parameter is small enough such that the structure of the seed LFR is found on the upper level, and the mechanism of Ravasz-Barab\'asi model is observed in the second and third levels. Panels (b) and (c) correspond to the second type of ground truth. In this case, the mixing parameter is large enough such that the mechanism of Ravasz-Barab\'asi  is observed in levels  1 and 2, while the structure of the seed LFR becomes detected at the third level. In all the cases we have fixed the value of $R, p$, and $q$ to 4, 0, and 0, respectively. Each node on the last level represents a community that doesn't contain any sub-communities [see Fig.~\ref{figure4}c \& d]

Going into the detailed observation of the detected communities, it is possible to compare the structure of the bottom panels with that of the top ones we can see that for $\mu$ = 0.01, Infomap made a mistake in the detection of the first level; two communities have been merged together. 
On the second level, Infomap makes even more mistakes, by merging pairs of communities in several cases (Figs.~\ref{figure13}a \& d).
In the example of $\mu$ = 0.33, Infomap successfully unveils the first level, but it makes mistakes on the second level (Figs.~\ref{figure13}b \& e). 
In the example of $\mu$ is 0.77, Infomap could neither correctly detect the community structure of the first level, nor unveil the structure of the deeper levels. 
In this case, the detected network structure is closed to a flat one: there are three communities on the first level. Each community on the first level contains several sub-communities on the second level, and each community on the second level has only one sub-community, i.e. itself, on the third level (Fig. \ref{figure13}c \& f). 

\section{Summary}
In this study, we have introduced a new class of benchmark graphs to test hierarchical community detection algorithms. These new benchmark graphs combine the LFR benchmark  and the rule for constructing hierarchical network proposed by Ravasz and Barab\'asi,
hence the name of  RB-LFR benchmark.
They integrate the properties of the standard LFR benchmark, i.e.~a power-law degree distribution and community size distribution, while also possess the clear hierarchical structure of the Ravasz-Barab\'asi model, and can be extended to an arbitrary number of levels.

We have found that the newly introduced RB-LFR benchmark graphs pose challenging tests to state-of-the-art hierarchical community detection algorithms. 
In particular, we have seen that the size of the graph and the average degree of nodes have sizeable effect on the accuracies of the methods. 
Our benchmark graphs, while parsimonious, exhibit a rich phenomenology including a variety of topological transitions between co-existing ground truths. 
Furthermore, by introducing two parameters to randomly remove connections and replicas, we have observed that the RB-LFR benchmark exhibits a robust hierarchical community structure. Additionally, our tests have also validated that the recently introduced \textit{Hierarchical Mutual Information (HMI)} suits better for the comparison of hierarchical partitions than the traditional \textit{Mutual Information (MI)} does.

The comparison of the performance of the tested algorithms: Infomap, Louvain, and the Hierarchical Stochastic Block Model (HSBM) against the RB-LFR benchmark, indicates that Infomap produce the best results overall. 
More specifically, the tests on the two-level RB-LFR benchmark graphs indicate that Infomap outperforms the other two methods in terms of accuracy. However, it seems that the three-level RB-LFR benchmark is very challenging for all of the existing algorithms.

Our next step is to conduct a more comprehensive comparison of hierarchical community detection algorithms by evaluating their performance on the RB-LFR benchmark. By doing this, we will gain deeper understanding of the features of the RB-LFR benchmark, and learn more about its limitations and the differences between the RB-LFR benchmark and the real hierarchical systems have. The benchmark introduced in this Paper has a very stylized hierarchical structure, which may be seen as a limitation of the approach. However, existing empirical work on hierarchical community detection has found hierarchies whose complexity is rather limited. Our results highlight that the algorithms for community detection must be vastly improved to ascertain more complex hierarchies. This paper provides the foundation to proceed with this important line of research.

\section*{Acknowledgements}
ZY and CJT acknowledge financial support from the URPP Social Networks at University of Zurich. They are also thankful to the S3IT (Service and Support for Science IT) of the University of Zurich, for providing the support and the computational resources that have contributed to the research results reported in this study. 
JIP acknowledges financial support from grants by CONICET (PIP 112 20150 10028) and SeCyT-UNC (Argentina), and institutional support from IFEG-CONICET.

\end{document}